\newif\ifmnras
\mnrastrue
\ifmnras
	\documentclass[a4paper,fleqn,usenatbib]{mnras}
\else
	\documentclass[apj,numberedappendix]{emulateapj}
\fi

\usepackage{graphics,epsf}
\usepackage{amsmath}                
\usepackage{amsfonts}               
\usepackage{amssymb}                
\usepackage{epsfig}                 
\usepackage{graphicx}
\usepackage{adjustbox}
\def \kms{\rm{km~s^{-1}}}

\def \msyr{~\rm{M_{\odot}}~\rm{yr^{-1}}}
\def \cm{~\rm{cm}}
\def \s{~\rm{s}}
\def \km{~\rm{km}}

\def \g{~\rm{g}}

\def \AU{~\rm{AU}}
\def \erg{~\rm{erg}}

\def \yr{~\rm{yr}}

\def \rmModot{~\rm{M_\odot}}
\def \rmRodot{~\rm{R_\odot}}

\ifmnras
	\def \aap{A\&A}
	
	\def \apj{ApJ}
	\def \apjl{ApJ}
	\def \apjs{ApJS}

	\def \mnras{MNRAS}

	\def \prl{PhRvL}
\fi

\usepackage{xcolor}
\definecolor{redak}{rgb}{0.9,0.15,0.05}

\ifmnras
	\title[Jets in common envelope simulations]{Companion-launched jets and their effect on the dynamics of common envelope interaction simulations} 

	\author[S. Shiber et al.]{
Sagiv Shiber,$^{1}$\thanks{E-mail: \href{shiber@campus.technion.ac.il}{shiber@campus.technion.ac.il}}
Roberto Iaconi,$^{4,2,3}$
\thanks{JSPS International Research Fellow (Graduate School of Science, Kyoto University)}
\thanks{E-mail: \href{roberto.iaconi@mq.edu.au}{roberto.iaconi@kusastro.kyoto-u.ac.jp}}
Orsola De Marco$^{2,3}$
\thanks{E-mail: \href{orsola.demarco@mq.edu.au}{orsola.demarco@mq.edu.au}}
Noam Soker,$^{1}$
\thanks{E-mail: \href{mailto:soker@physics.technion.ac.il}{soker@physics.technion.ac.il}}
\\
$^{1}$Physics Department, Technion -- Israel Institute of Technology, Technion City -- Haifa 3200003, 
Israel\\
$^{2}$Department of Physics and Astronomy, Macquarie University, Sydney, NSW 2109, Australia
\\
$^{3}$Astronomy, Astrophysics and Astrophotonics Research Centre, Macquarie University, Sydney, NSW 2109, Australia
\\
$^{4}$Department of Astronomy, Kyoto University, Kitashirakawa-Oiwake-cho, Sakyo-ku, Kyoto 606-8502, Japan 0000-0002-1940-1950
\\
}

	\date{Accepted XXX. Received YYY; in original form ZZZ}

	\pubyear{2019}
\fi

\begin{document}
\label{firstpage}

\ifmnras
	\pagerange{\pageref{firstpage}--\pageref{lastpage}}
	\maketitle
\else
	\title[Jets in common envelope simulations]{Companion-launched jets and their effect on the dynamics of common envelope interaction simulations}

	\author{Sagiv Shiber}
    \author{Roberto Iaconi}
    \author{Noam Soker}
    \author{Orsola De Marco}
	\affil{Department of Physics, Technion -- Israel
	Institute of Technology, Haifa 32000, Israel; Department of Physics and Astronomy, Macquarie University, Sydney, NSW 2109, Australia; Astronomy, Astrophysics and Astrophotonics Research Centre, Macquarie University, Sydney, NSW 2109, Australia;
    }
\fi

\begin{abstract}
We conduct three-dimensional hydrodynamic simulations of the common envelope binary interaction and show that if the companion were to launch jets while interacting with the giant primary star's envelope, the jets would remove a substantial fraction of the envelope's gas. 
We use the setup and numerical code of an earlier common envelope study that did not include jets, with a 0.88-M$_\odot$, 83-R$_\odot$ red giant star and a 0.3-M$_\odot$ companion. The assumption is that the companion star accretes mass via an accretion disk that is responsible for launching the jets which, in the simulations, are injected numerically.
For the first time we conduct simulations that include jets as well as the gravitational energy released by the inspiraling core-companion system.
We find that simulations with jets unbind approximately three times as much envelope mass than identical simulations that do not include jets, though the total fraction of unbound gas remains below 50\% for these particular simulations. The jets generate high velocity outflows in the polar directions. The jets also increase the final core-companion orbital separation and lead to a kick velocity of the core-companion binary system. 
Our results show that, if able to form, jets could play a crucial role in ejecting the envelope and in shaping the outflow.  
\end{abstract}


\section{INTRODUCTION}
\label{sec:intro}

Many observed close binaries systems that include an evolved, compact star, such as a white dwarf, a neutron star, or a black hole, have orbital separations that are much smaller than a typical radius of the evolved star's progenitor. These binaries must have evolved through at least one common envelope (CE) phase, where the companion inspiralled through the envelope of a giant star (\citealt{Paczynski1976}; for a review see, e.g., \citealt{Ivanovaetal2013}). 
Examples of post-CE binaries are double white dwarfs systems, which may account for a fraction of the progenitors of Type Ia supernovae (e.g., \citealt{ToonenNelemans2013}), or any combination of neutron stars and black holes, that may later merge with the emission of detectable gravitational waves (e.g., \citealt{Abbottetal2016}). 

As the more compact companion orbits inside the CE it loses energy and angular momentum due to tidal interaction and gravitational drag forces, and it inspirals toward the core of the giant star. This interaction transfers orbital energy to the envelope, and is often referred to as CE evolution (CEE). The details of CEE are of great importance because they dictate the outcomes of the binary interaction, in particular the final separation, which in  turn dictates whether a merger takes place inside the CE. The details also determine the morphology of the outflow, hence the shape of the resulting nebula (e.g., \citealt{Ivanovaetal2013}), and the duration of the binary interaction which both play a role in influencing the type of transient lightcurve that would be observed.

Most analytical studies, such as population synthesis simulations (e.g., \citealt{Ablimitetal2016}, \citealt{AblimitMaeda2018}) parameterise CEE using the $\alpha_{\rm CE}$ prescription that is based on energy conservation. This prescription assumes that a fraction $\alpha_{\rm CE}$ of the orbital energy that is released by the spiraling-in secondary star is channeled to unbind the envelope (e.g., \citealt{Webbink1984, LivioSoker1988}). 
What remains to be seen is whether a common efficiency should be used for all types of CEE and whether the orbital energy should be the only source of energy that can be used to unbind the envelope.

Three-dimensional, numerical hydrodynamic simulations of the CE  have been carried out using several different techniques, such as Eulerian, smoothed-particle hydrodynamics (SPH) codes and moving-mesh (e.g., \citealt{LivioSoker1988, RasioLivio1996, Sandquistetal2000, TaamRicker2010, Passyetal2012, RickerTaam2012, Nandezetal2014, Ohlmannetal2016, Staffetal2016MN8, NandezIvanova2016, Kuruwitaetal2016, IvanovaNandez2016, Iaconietal2017, Iaconietal2018, MacLeodetal2018b, Reichardtetal2018}). 
Those simulations that use an ideal-gas equation of state unbind at most 40 percent of the envelope with values of 10-20 percent being more typical. Only simulations that use a tabulated equation of state and thermalise the energy due to recombining gas, can eject the entire envelope under certain circumstances \citep{IvanovaNandez2016}. Just how much of the recombination energy can be harnessed is currently debated \citep{Ivanova2018, Sokeretal2018}.

Questions about when and how the envelope becomes unbound leave the critical issue of the final separation hanging. The final separation in simulations is clearly a function of parameters such as the envelope binding energy and the mass of the companion. There are, however, some doubts about the impact of resolution on the final separation with higher resolution resulting in smaller separations in the simulations of \citet{Iaconietal2018}. It is also not clear whether the initial conditions of the CEE simulations may impact the final separation such as starting at the moment of Roche lobe overflow, instead of starting with a smaller separation (e.g., \citealt{Reichardtetal2018}). Interestingly, simulations that adopt the recombination energy and unbind the envelope do not seem to show a systematically larger or smaller final separation when compared to similar simulations that do not include recombination energy (Reichardt et al., in preparation). This said, it is not clear whether other physical mechanisms that are not yet included in the simulations and that may aid in envelope removal may play a role in the final separation. Hence the problem of envelope unbinding and of final orbital separation are intimately connected.

Several studies proposed the inclusion of additional processes (for a list see \citealt{Soker2017final}). Aside from the inclusion of recombination energy \citep{IvanovaNandez2016}, \cite{Soker2004} and \cite{GlanzPerets2018} considered dust-driven winds after the CE phase, when the envelope is already extended, and show that these winds can further expel envelope material. 
Another energy source could be the accretion of envelope mass onto the more compact companion. The accretion of envelope gas onto the companion releases energy, some of which can be channeled to jets which may contribute to the ejection of the envelope. \cite{ArmitageLivio2000} and \cite{Chevalier2012} examined the ejection of the CE by jets launched from a NS that inspirals inside the giant envelope.
Later studies follow this idea while considering different companions  (e.g., \citealt{Soker2004, Papishetal2015, Soker2015, Soker2016Rev, MorenoMendezetal2017, Shiberetal2017, ShiberSoker2018, Lopezetal2019}).
\begin{table*}
\begin{center}
\begin{adjustbox}{max width=\textwidth}
\begin{tabular}{lccccccccc}
\hline
ID          & Top Grid     & AMR    & Domain       & Max. cell    & Min. cell    & $a_0$            & Jets & Jets mass  & $L_{\rm jets}$ \\
            & resolution   & levels & size         & size         & size         &                  &      & loss rate        &                \\
            & (cells/side) & (\#)   & ($\rmRodot$) & ($\rmRodot$) & ($\rmRodot$) & ($\rmRodot$)     &      & ($\msyr$)   & ($\rmRodot$)   \\   
\hline
\#1         & 64           & 2      & 431          & 6.7            & 1.7          & 83               & N    & N/A         & N/A            \\
\#2         & 64           & 2      & 431          & 6.7            & 1.7          & 83               & Y    & $0.001 $    & 14.4             \\
\#3         & 64           & 2      & 431          & 6.7            & 1.7          & 166              & N    & N/A         & N/A            \\
\#4         & 64           & 2      & 431          & 6.7            & 1.7          & 166              & Y    & $0.001$     & 14.4             \\
\#5         & 128          & 2      & 431          & 3.4            & 0.8          & 83               & N    & N/A         & N/A            \\
\#6         & 128          & 2      & 431          & 3.4            & 0.8          & 83               & Y    & $0.001$     & 7.2              \\
\#7         & 128          & 2      & 431          & 3.4            & 0.8          & 83               & Y    & $0.001$     & 14.4             \\
\#8         & 128          & 2      & 431          & 3.4            & 0.8          & 166              & N    & N/A         & N/A            \\
\#9         & 128          & 2      & 431          & 3.4            & 0.8          & 166              & Y    & $0.001$     & 7.2              \\
\#10        & 128          & 2      & 431          & 3.4            & 0.8          & 166              & Y    & $0.001$     & 14.4             \\
\hline
\#11        & 128          & 3      & 862          & 6.7            & 0.8          & 166              & Y    & $0.003$     & 3.6            \\
\#12        & 128          & 3      & 862          & 6.7            & 0.8          & 166              & Y    & $0.003$     & 7.2              \\
\#13        & 128          & 3      & 862          & 6.7            & 0.8          & 166              & Y    & $0.003$     & 14.4             \\
\hline
\end{tabular}
\end{adjustbox}
\end{center}
 \begin{quote}
  \caption{\protect\footnotesize{The initial parameters for the simulations of the present study. In all simulations the mass of the companion main sequence star is $M_2=0.3$~$M_\odot$ and the initial mass, core mass and radius of the RGB star are $M_1=0.88$~$M_\odot$,  $M_c=0.39$~$M_\odot$, and $R_1=83$~$R_\odot$, respectively.}} \label{tab:simulation_parameters}
\end{quote}
\end{table*} 

Whether accretion plays a role in the envelope's dynamics, rests on how much mass the companion can accrete  during the dynamical inspiral phase. We do not yet know whether the accretion rate is close to the Bondi-Hoyle-Lyttleton (BHL; \citealt{Bondi1952}) value, or if it is much smaller. \cite{RickerTaam2012} have found in their CE simulations that the accretion rate is two orders of magnitudes below the BHL value. \cite{Murguiaetal2017} further argued that disk formation in the CEE is a rare event. However, observational evidence points to accretion taking place at some stage during the CEE: main sequence companions in post-CE binaries in the planetary nebulae (PNe) Hen~2-155 and Hen~2-161 are inflated, something that can be explained by accretion \citep{Jonesetal2015}; the companion in the post-CE central star of the Necklace PN is polluted by carbon-rich material \citep{Miszalskietal2013}, implying accretion of giant envelope gas; several companions in post-CE central stars of PNe are spun-up, possibly by having accreted angular momentum \citep{Montezetal2010, Montezetal2015}. Additionally, several PN harboring post-CE binaries show jets that can be shown to have kinematic ages that are almost coincidental with the ejection of the envelope \citep{Tocknelletal2014} implying accretion phases.  Finally \cite{BlackmanLucchini2014} show that the momenta of the outflow of bipolar pre-PN cannot be explained by any plausible mechanism except by accretion during the CEE, if the accretion rates are those implied by studies such as that of \citet{RickerTaam2012}. These observations may support the feasibility of substantial accretion during CEE.

\cite{Shiberetal2016} argued that the launching of jets from an accretion disk around the mass-accreting companion removes mass, angular momentum, and pressure and therefore makes accretion at a high rate possible. Indeed, in a recent paper \cite{Chamandyetal2018a} find high accretion rates when they use a subgrid `valve mechanism' that removes energy, hence reduces pressure, from the vicinity of the accreting companion. Jets can remove energy as in the subgrid prescription of \cite{Chamandyetal2018a}. 

The launching of the jets must also affect the ejecta's morphology. Bipolar outflows are frequently seen in PNe and in pre-PNe. These outflows are usually attributed to a binary interaction (e.g., \citealt{Soker1994, BalickFrank2002, Wittetal2009}), although the precise relationship between the stunning range of axi-symmetric morphologies and the driving interaction is still unclear. CEE simulations, for example, usually yield outflows that are concentrated around the equatorial plane. The fast spherical wind expected from the post-CE primary ploughs into the CE ejecta resulting in a very collimated structure \citep{Garciaetal2018, Franketal2018}. In this case the equatorial outflow is on the ``outside" of the bipolar flow.
On the other hand, the CE jets simulated by \cite{Shiber2018} would result in a very collimated CE ejection, at least in the early inspiral, while the later inspiral may still produce an equatorial enhancement, with rings and arcs. Although a PN resulting from this type of outflow has not been simulated yet, it is clear that a CE with jets would be quite a different morphological phenomenon.

In this paper we continue to explore the role that jets play in the CEE.  \cite{Shiberetal2016} and \cite{ShiberSoker2018} studied the effect of jets on CEE focusing on cases where the jets efficiently remove the envelope from the beginning of the in-spiraling phase, i.e., where the system experiences the grazing envelope evolution (GEE; \citealt{Soker2015}). In the GEE wider final binary separations can result, or the CEE can be interrupted or delayed. Here, on the other hand, we specifically focus on mass removal with substantial in-spiral, and include the gravity of all components in the system. Namely, for the first time  we include the gravity of the secondary star and the self-gravity of the envelope in simulating jets in a CEE. This implies that both the orbital energy and the accretion energy carried by jets contribute to envelope removal and to the shaping of the outflow. 
   
We describe our numerical scheme in Section \ref{sec:numerics}, and in Section \ref{sec:results} we present our results. We discuss our numerical method limitations in section \ref{sec:numerical_limitations}, primarily the effect of the jets injection length, $L_{\rm jets}$, on our results. In Section \ref{sec:summary} we discuss and summarize our results.

\section{NUMERICAL SET-UP}
\label{sec:numerics}

The CEE simulations in this work use a modified version of {\sc Enzo} (\citealt{O'Sheaetal2004}, \citealt{Bryanetal2014}). The CE specific features have been implemented by \citet{Passyetal2012} and primarily consist on the ability to map 1D stellar models into the {\sc Enzo} 3D domain and in the creation of a new gravity solver for point-mass particles (\citealt{PassyBryan2014}), which allows for improved energy and angular momentum conservation. 
The above papers contain all details on the numerical techniques and the equations the code solves. We here mention only the treatment of the `particles'. In our simulations we treat the core of the giant star and the companion star as point-mass particles. From now on by `particles' we refer to both the core and companion.
The point-mass particles used in {\sc Enzo} CE simulations are dimensionless particles interacting only gravitationally with the gas and the other point-masses in the computational domain, they have a fixed mass (i.e., they are not allowed to modify their mass value by accreting gas from the grid), 
and in their surroundings gravity is smoothed according to the prescription by \citet{Ruffert1993}. 

Several studies have used {\sc Enzo} for the simulation of the CEE and well tested it in different setups and at different resolutions (e.g.,    
\citealt{Passyetal2012, Staffetal2016MN, Staffetal2016MN8, Kuruwitaetal2016, Iaconietal2017, Iaconietal2018}), with the study of \citet{Iaconietal2018} using the adaptive mesh refinement (AMR) capability of {\sc enzo} alongside the new solver of \cite{PassyBryan2014}. Energy and angular momentum are conserved to an acceptable level (\citealt{Iaconietal2017}, \citealt{Iaconietal2018}) and simulation outcomes have been compared with other codes (\citealt{Passyetal2012}, \citealt{Iaconietal2017}) showing reasonable agreement.

In order to simulate jet launching from the companion we implemented a new particle type that acts gravitationally as a point-mass particle, but that is additionally launching jets. This particle is the companion star. We launch jets in the same manner as in \cite{Shiberetal2017} and \cite{ShiberSoker2018}. The salient points are as follows. We numerically insert the jets in two opposite cones around the z-axis, the axis perpendicular to the orbital plane, with a half-opening angle of $30^{\circ}$. 
At each time step we set the density in each cell inside the two cones to have a value corresponding to freely expanding jets with the prescribed velocity and mass loss rate. Namely, the density in the cone decreases as $r^{-2}$. A cell is included in the cone only if its centre belongs to the cone.
The cells where the companion resides in are not inside the cones. Namely, the jets are not inserted in the equatorial plane itself, but from one cell above and below it and up to a distance $L_{\rm jets}$. 
The cones extend from the particle (the companion) position to a length of $L_{\rm jets}$. We want $L_{\rm jets}$ to be large enough to keep the conical geometry, but still small. In simulations \#6, \#7, \#9, \#10, $L_{\rm jets}$ is between two to four times the maximal cell length. In simulations \#11-13, $L_{\rm jets}$ is between half to two the maximal cell length. The jets have two velocity components: a radial one from the location of the particle with a magnitude of $400 \km \sec^{-1}$, and a velocity component equals to the companion velocity. The jets' velocity corresponds to the escape velocity of a low mass main sequence star. 

In this work we carry out thirteen simulations. We utilize AMR in all our simulations, where a cell is split into two in each direction based on a density threshold criterion. We use 2 or 3 levels of refinement, depending on the simulation, as we describe below. For maximum energy conservation we use a smoothing length for the gravitational potential of the point-mass particles equal to three times the length of a cell at the deepest refinement level (\citealt{Staffetal2016MN}).

All the simulations use the same 1D primary model as the one  \citet{Passyetal2012} used, namely an RGB star with a total mass of $M_1 = 0.88 \rmModot$, a core mass of $M_c = 0.39 \rmModot$ and a radius of $R_1 = 83 \rmRodot$. This allows us to compare directly simulations with and without jets.
We map the giant into our 3D computational domain and stabilize it using the procedure described in \citet{Passyetal2012}, utilizing a point-mass particle to represent the core of the giant, that cannot be resolved in the {\sc Enzo} grid.
The companion star is instead a $0.3 \rmModot$ main sequence star, which, similarly to the primary's core, cannot be resolved and it is therefore represented by a point-mass particle.
The particles are initialized with velocities resulting in Keplerian circular orbits appropriate for the respective initial orbital separations $a_0$. Note that in the case of the primary star both the point-mass core and the envelope are given the same, uniform velocity.

Using this binary we build several different initial setups, varying the resolution, AMR levels, domain size, initial orbital separation, presence of jets, jets' length of injection, and mass rate carried by the jets. We summarize our different simulation setups in Table~\ref{tab:simulation_parameters}. 

The initial orbital separation chosen corresponds to either the initial radius of our RGB star ($ 83$~$R_\odot$; i.e., the companion is placed on the surface of the primary) or double that value. In the first case the primary is greatly overflowing its Roche lobe at the beginning of the simulation   ($R_{\mathrm{RL},83} = 39~\rmRodot$), while in the second case the primary is only slightly overflowing its Roche lobe radius ($R_{\mathrm{RL},166} = 79\rmRodot$, only slightly smaller than the radius of the primary of $83~\rmRodot$; \citealt{Eggleton1983}). $a_0$ is measured from the center of the giant star. 

The rate of mass carried by the two opposite jets is a few percent of the BHL accretion rate when the companion is on the giant star surface. The reason for reducing the jet mass-loss rate compared to the BHL accretion rate is the indication that accretion rates in CEE might be lower (e.g., \citealt{RickerTaam2012}) and the assumption that jets carry about $10$ per cent of the accreted mass, as observed in, e.g., young stellar objects. For a $0.3 \rmModot$ companion orbiting with a Keplerian velocity on the surface of a giant star with a mass of $0.88 \rmModot$ and a radius of $83 \rmRodot$, the BHL accretion rate is $0.15 \msyr$. We therefore assume two values for the mass-loss rates in the jets, $\dot{M}_{\rm jets}$: $0.7$ and $2$ per cent of this value, i.e., $0.001 \msyr$ and $0.003 \msyr$ respectively.

\begin{figure*}
\centering
\includegraphics[trim= 0.4cm 0.4cm 0.2cm 0.3cm,clip=true,width=0.99\textwidth]{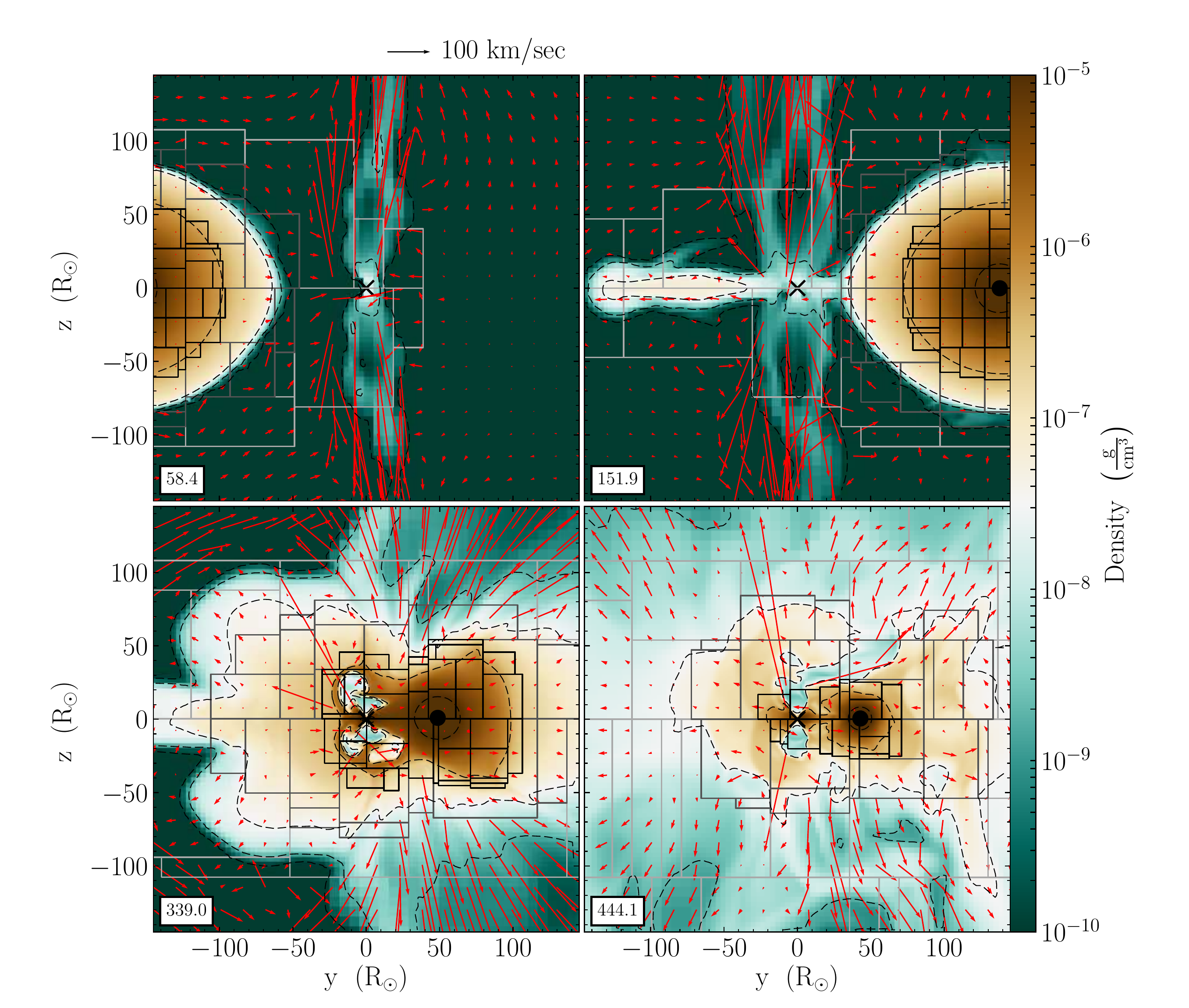} 
\caption{Density maps, AMR grids, and velocity vectors in the meridional $y-z$ plane that is perpendicular to the orbital plane of simulation \#13 at four different times, given in days, at which the companion crosses the x-axis. An 'X' symbol marks the location of the companion and a black circle marks the giant's core. The length of the arrows is proportional to the gas speed, with scaling of $100 \km \s^{-1}$ as indicated with the arrow above the first panel. The first, second, and third level of refinement grids are colored in light gray, dark gray and black, respectively. The dashed contours show equal density surfaces.
}
\label{fig:jets_grid}
\end{figure*}

To illustrate the shape of the jets within the grid with different refinements we show in Fig. \ref{fig:jets_grid} density maps, AMR grids, and velocity arrows in the $y-z$ plane of simulation \#13 at four different times. This simulation has 3 levels of refinement. The first, second, and third level of refinement grids are colored in light gray, dark gray and black, respectively. As said, in order for the jets to preserve their conical shape it is important that $L_{\rm jets}$ is more than the length of two cells.
As the refinement is automatically determined by the density and the companion starts outside the giant, at early times the code injects the jets in less refined regions. However, as the companion dynamically in-spirals through the primary's envelope the jets quickly move to regions with the highest level of refinement. This ensures that they are properly resolved during the crucial part of the CEE.

\section{RESULTS}
\label{sec:results}
  
We summarize in Table \ref{tab:simulation_results} the final properties of the 13 simulations that we have conducted. The second column lists the simulation times. We note that the orbital period of an $M_2=0.3\rmModot$ stellar companion that orbits a $M_1=0.88\rmModot$ red giant branch star on its surface is 81 days. We stop the simulation either at about 584 days, or when the distance of the giant's core to the boundary of the grid becomes smaller than the initial radius of the giant star, which ever comes first. Each simulation produces vast amounts of data. In subsection \ref{subsec:flow_structure} we focus on simulation \#13 and show in detail the flow structure of this simulation. In \ref{subsec:compare_simulations} we discuss the differences between simulations with and without jets. Finally, we mention some energy considerations of our simulations in subsection \ref{subsec:Energy}.
\begin{table*}
\begin{center}
\begin{adjustbox}{max width=\textwidth}
\begin{tabular}{lcccccc}
\hline
ID   & $t_{\mathrm{f}}$ & $M_{\mathrm{out}}$  & $M_{\mathrm{out}}^{\mathrm{unbound}}$ & $M_{\mathrm{gas,\;in}}$ & $a_{\mathrm{f}}$ & $e_{\mathrm{f}}$ \\
     & (days)           & ($\rmModot$)       & ($\rmModot$)                         & ($\rmModot$)       & ($\rmRodot$)     &       \\
\hline
\#1  & 584             &  0.2                & 0.03                                 & 0.26               & 5.9              & 0.11  \\
\#2  & 254             &  0.29               & 0.14                                 & 0.09               & 35.5             & 0.46  \\
\#3  & 581             &  0.17               & 0.04                                 & 0.3                & 7.9              & 0.15  \\
\#4  & 494             &  0.32               & 0.17                                 & 0.07               & 37.1             & 0.55  \\
\#5  & 584             &  0.27               & 0.03                                 & 0.2                & 4.6              & 0.07  \\
\#6  & 289             &  0.33               & 0.15                                 & 0.09               & 20.3             & 0.32  \\
\#7  & 342             &  0.29               & 0.13                                 & 0.1                & 19.7             & 0.16  \\
\#8  & 584             &  0.18               & 0.06                                 & 0.3                & 8.3              & 0.1   \\
\#9  & 549             &  0.33               & 0.16                                 & 0.1                & 20.9             & 0.16  \\
\#10 & 459             &  0.32               & 0.18                                 & 0.09               & 26.8             & 0.4   \\
\hline
\#11 & 584             &  0.2                & 0.13                                 & 0.25               & 20               & 0.04  \\
\#12 & 584             &  0.24               & 0.16                                 & 0.2                & 21.5             & 0.14  \\
\#13 & 584             &  0.3                & 0.24                                 & 0.1                & 26.1             & 0.56  \\
\hline
\end{tabular}
\end{adjustbox}
\end{center}
 \begin{quote}
  \caption{\protect\footnotesize{ Some final properties of the simulations of the present study. Table \ref{tab:simulation_parameters} presents the initial parameters of each simulation. Simulations \#1, \#3, \#5, and \#8 have no jets, as is evidence from the low value of unbound mass that was lost. The rest do have jets.
  $t_f$: the final time; $M_{\rm out}$: the mass that left through a sphere at the boundary of the grid with radius $R_{\rm out}=L/2$ where L is the domain size; $M^{\rm unbound}_{\rm out}$: the unbound mass that left the system (i.e., with positive energy); $M_{\rm gas,\;in}$: the gas mass resides in a sphere with radius $R_{\rm out}$; $a_f$ and $e_f$: the final semi-major axis and eccentricity of the core-companion binary system, respectively.}}
\label{tab:simulation_results}
\end{quote}
\end{table*} 

\subsection{Flow structure}
\label{subsec:flow_structure}

Earlier simulations (e.g., \citealt{ShiberSoker2018}) of the GEE and early CEE with jets show the complicated and highly asymmetrical flow structures. We find similar structures in all of our simulations that include jets. Here we focus on simulation \#13, which has a high resolution and a large domain. 

Simulation \#13 has a domain size of $862 \rmRodot$ and 3 levels of refinement. The companion continuously launches jets with a velocity of $v_{\rm jets}=400 \km \s^{-1}$ and a half-opening angle of $\theta_{\rm jets}=30^{\circ}$ as is the case in all of our simulations. The jets carry mass at a rate of $\dot{M}_{\rm jets}=0.003 \msyr$ and the jets' injection length is $L_{\rm jets}=14.4 \rmRodot$.

\begin{figure*}
\centering
\includegraphics[trim= 4cm 0.4cm 0.2cm 0.3cm,clip=true,height=0.9\textheight]{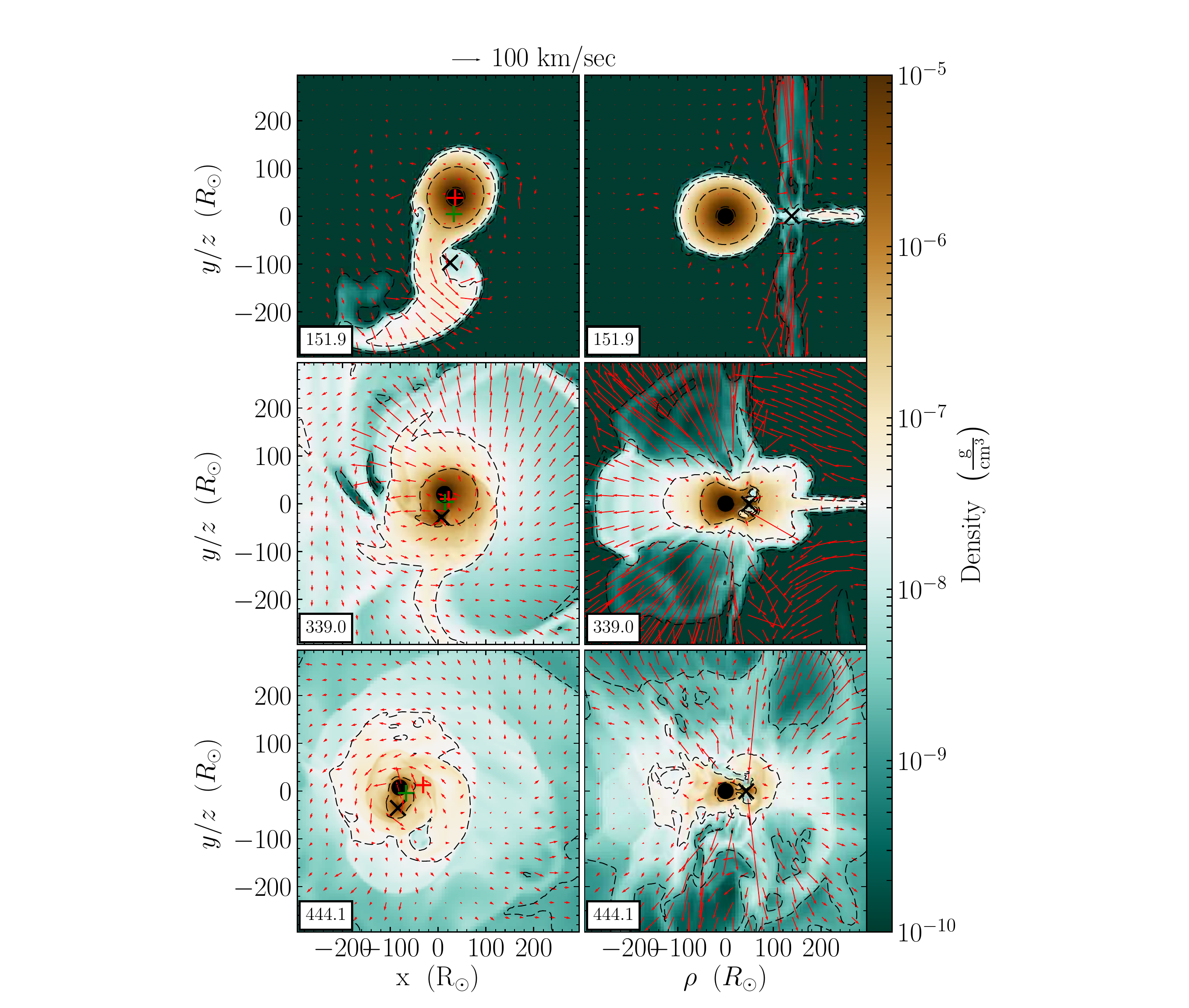} 
\caption{Density maps and velocity vectors of simulation \#13 at three different times given in days. Left: the equatorial plane, $x-y$. The green crosses mark the position of the center of mass (gas + particles), while the red crosses mark the position of the center of mass of the gas (without the particles). `Particles' refer here and in what follows to the core and companion. Right: the meridional, $\rho-z$, plane that traces the companion movement, $\rho \equiv \pm \sqrt {(x-x_c)^2+(y-y_c)^2}$, where $(x_c,y_c)$ is the momentary position on the orbital plane of the giant's core. An 'X' symbol marks the location of the companion and a black circle marks the giant's core. The length of the arrow is proportional to the gas speed, with scaling of $100 \km \s^{-1}$ as indicated with the arrow above the first panel.
}
\label{fig:fiducial_dense}
\end{figure*}

Fig. \ref{fig:fiducial_dense} presents density maps and velocity arrows of simulation \#13 at three times as indicated on the panels in days. At these three times the companion completed about $0.75$, $2.75$, and $6.75$ orbits, respectively. In the left panels we present the flow in the equatorial (orbital), $x-y$, plane, and in the right panels we present it in the meridional, $\rho-z$, plane that traces the companion, namely, the plane that is perpendicular to the orbital plane and contains both the core and the companion particles ($\rho \equiv \pm \sqrt {(x-x_c)^2+(y-y_c)^2}$, where $(x_c,y_c)$ is the momentary position on the orbital plane of the giant's core; right panels). The companion orbits counterclockwise in the $x-y$ plane, and its instantaneous location is marked with 'X', while the giant's core position is marked with a black circle. The length of the arrows is proportional to the gas speed. The green crosses on the left panels mark the position of the center of mass (gas + particles), while the red crosses mark the position of the center of gas mass only (without the particles).  

Based on the flow (and outflow) structure, we can divide the evolution into three different phases. Each of the three times in Fig.~\ref{fig:fiducial_dense} represents a different phase. During the first phase, starting from the beginning of the simulation, the gravity of the companion attracts the mass in the outer shells of the giant. As a result, a thin fan of envelope gas comes out of the giant star and starts trailing the companion in the orbital plane. This particular initial flow morphology is commonly observed in CEE simulations starting from separations larger than the initial radius of the primary (see, e.g., \citealt{RasioLivio1996}, \citealt{RickerTaam2012}, \citealt{Iaconietal2017}, \citealt{MacLeodetal2018a}, \citealt{Reichardtetal2018}). During this early stage, a large fraction of the jets material flows uninterrupted out of the grid in the polar directions, while a small fraction of the jets material accumulates at the equatorial plane around the companion. The reason for this accumulation is that the companion moves around the primary, so the jets continuously encounter fresh circumstellar gas, even if initially of very low density. When jets are shocked, high pressure pushes gas toward the equatorial plane \citep{AkashiSoker2008}.

Simulations \#1-2 and \#5-7 that start with the companion on the giant surface, i.e., an initial orbital separation of $a_0=R_1$, on the other hand, do not develop a fan that trails the companion, and the spiraling-in phase starts almost instantaneously. Therefore, in these cases the division into three phases is more dubious as the first phase does not exist. 

Due to dynamical friction drag forces, the companion starts spiraling-in towards the giant envelope and releases gravitational energy. This is the second phase. The interaction inflates the envelope. Material from the fan trailing the companion outflows through the two outer Lagrange points (the second and then the third). The jets collide with the denser regions inside the envelope. The envelope diverts the jets toward the equatorial direction and stops their propagation altogether, cocooning them inside the giant. 

During the third phase, the companion is deep in the envelope and the spiraling-in rate declines to a relatively low rate. The shocked jet material inflates two opposite hot bubbles around the companion. The bubbles eventually escape from the envelope more or less along the polar directions as we will discuss later.  
The jets remove a substantial amount of mass from the envelope that flows out in all directions, but not in a spherical manner. The rapid non-spherical mass ejection and momentum conservation cause the remaining binary system and bound gas to move away from the center of the grid. The final core-companion binary orbit stabilizes with a fairly eccentric orbit of $e_f=0.56$. 

We record the mass loss and kinetic energy loss rates per unit solid angle as a function of direction, $\dot m(\theta, \phi)$ and $\dot e_k (\theta, \phi)$, respectively, through a sphere of radius $R_{\rm out} = 2 \AU$ around the center of the grid. From these we calculate the momentary outflow velocity magnitude, $\sqrt{v^2}=\sqrt{2\dot{e_k}/\dot{m}} $, the total mass loss, $m(\theta, \phi)$, the total kinetic energy loss, $e_k(\theta, \phi)$, and the average outflow velocity $\sqrt{\langle v\rangle}=\sqrt{2e_k/m} $. 

\begin{figure*}
\centering
\includegraphics[trim= 1cm 0.4cm 0.2cm 0.3cm,clip=true,width=0.45\textwidth]{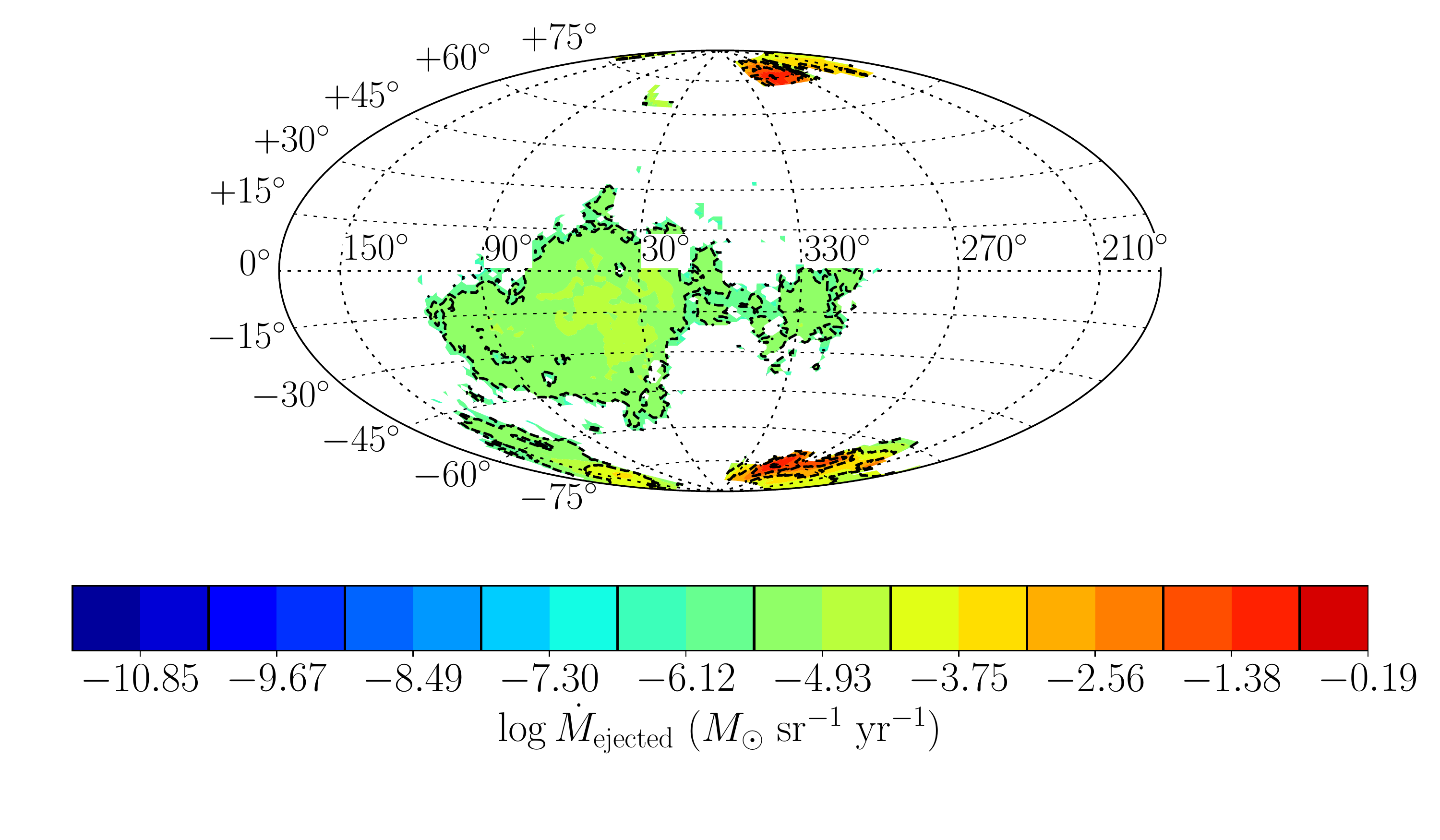}
\includegraphics[trim= 1cm 0.4cm 0.2cm 0.3cm,clip=true,width=0.45\textwidth]{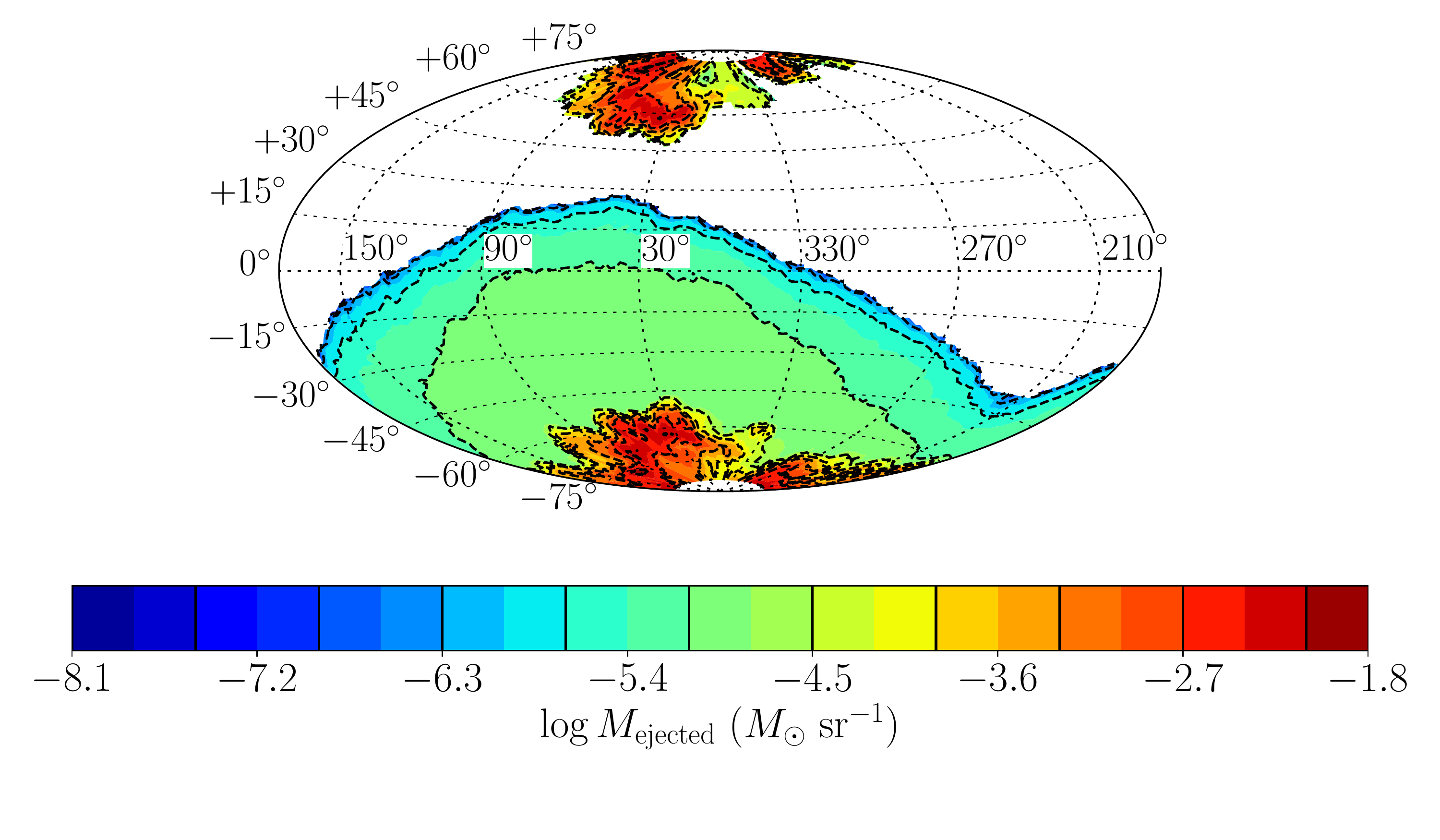}
\hspace*{0.4cm}
\vspace*{0.4cm}
\includegraphics[trim= 1cm 0.4cm 0.2cm 0.3cm,clip=true,width=0.45\textwidth]{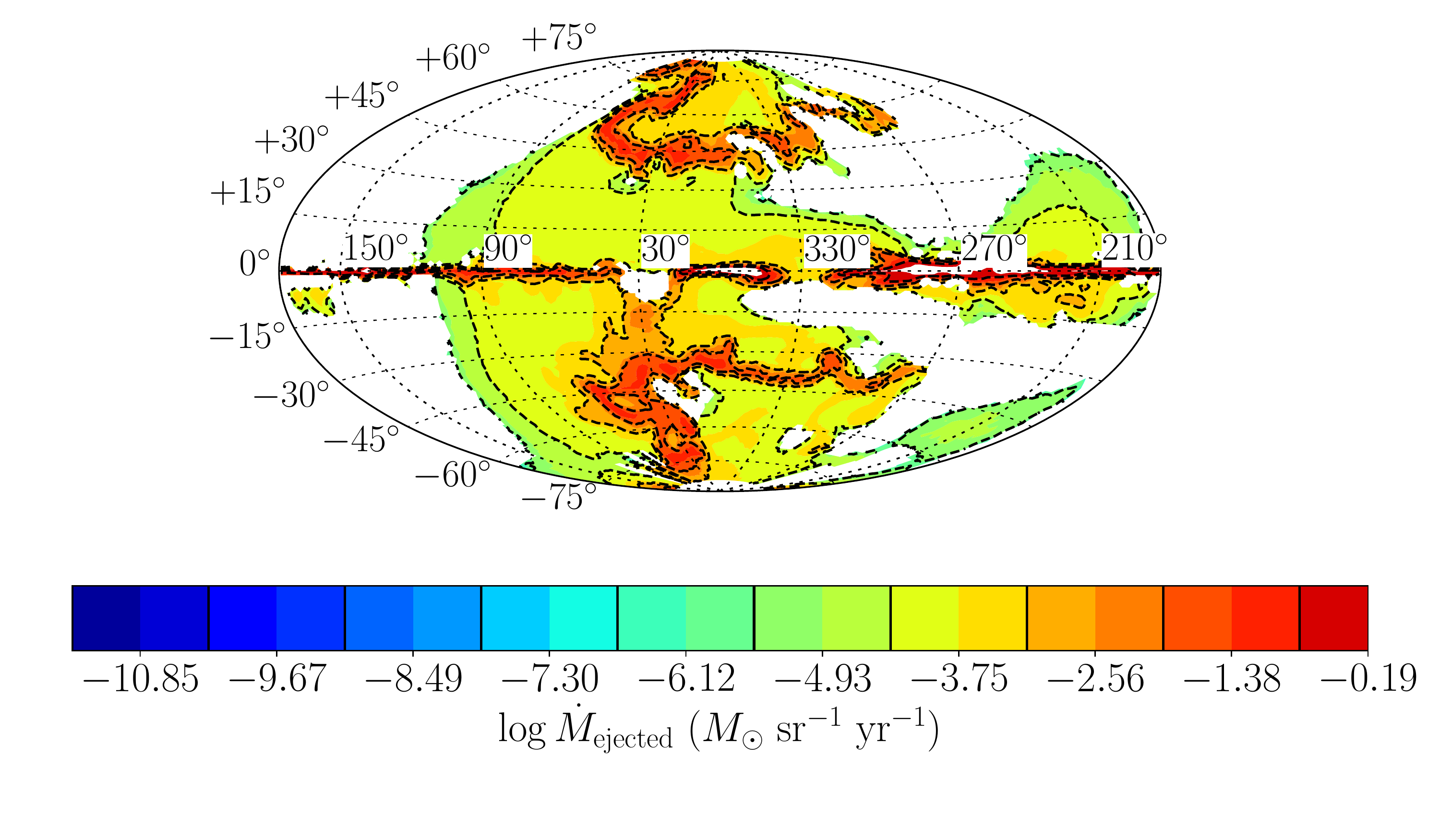}
\includegraphics[trim= 1cm 0.4cm 0.2cm 0.3cm,clip=true,width=0.45\textwidth]{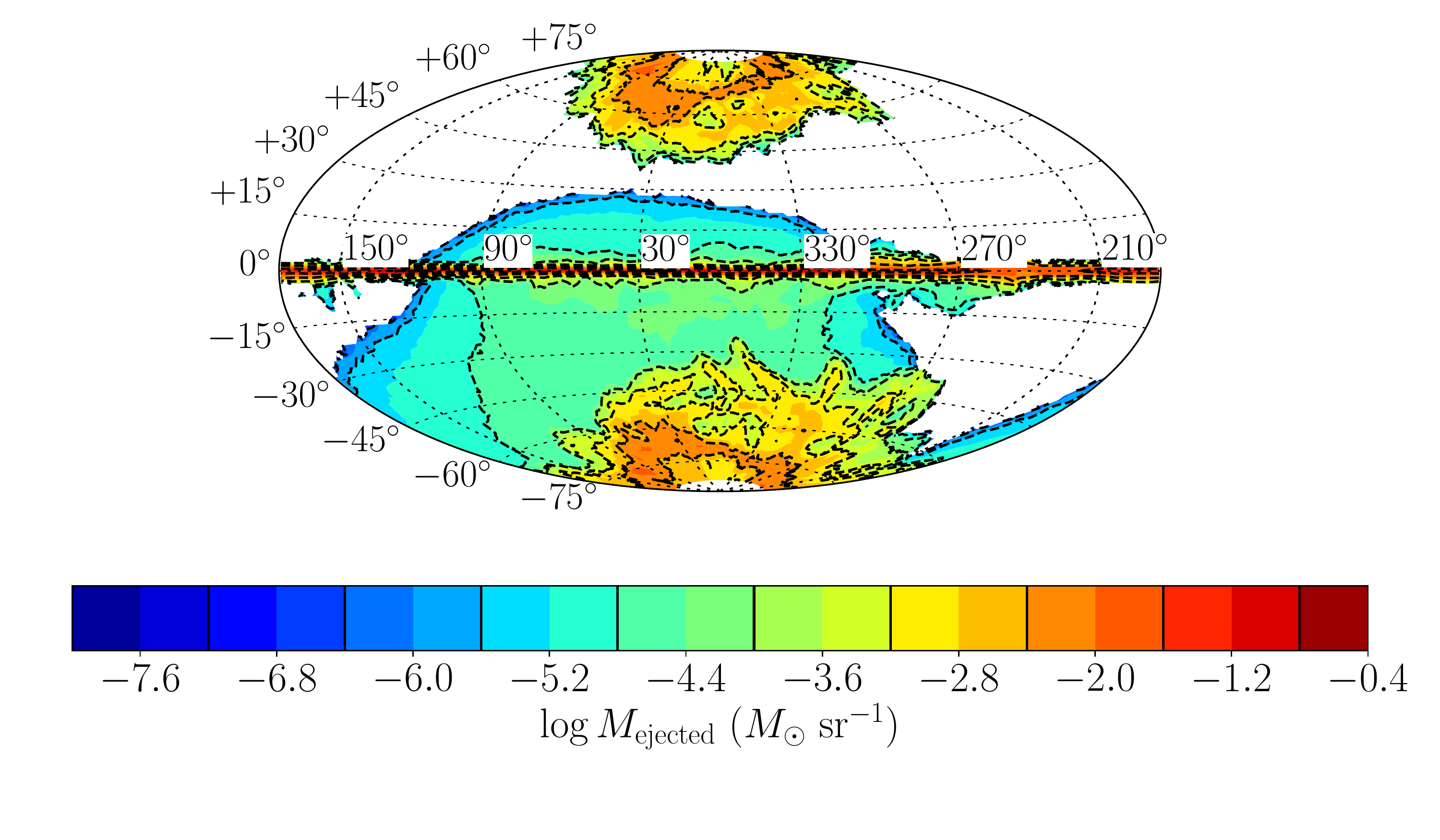}
\hspace*{0.4cm}
\includegraphics[trim= 1cm 0.4cm 0.2cm 0.3cm,clip=true,width=0.45\textwidth]{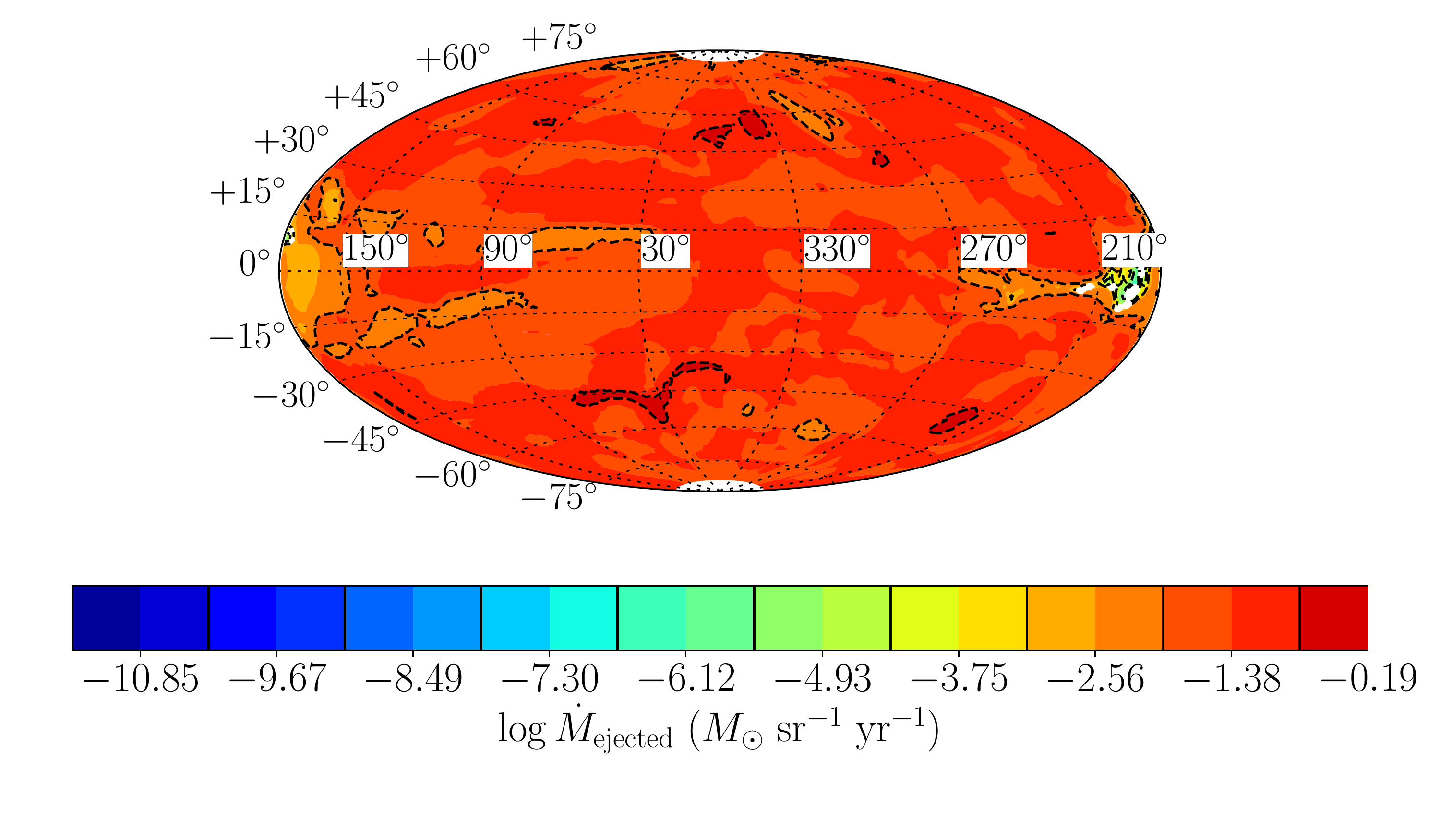}
\includegraphics[trim= 1cm 0.4cm 0.2cm 0.3cm,clip=true,width=0.45\textwidth]{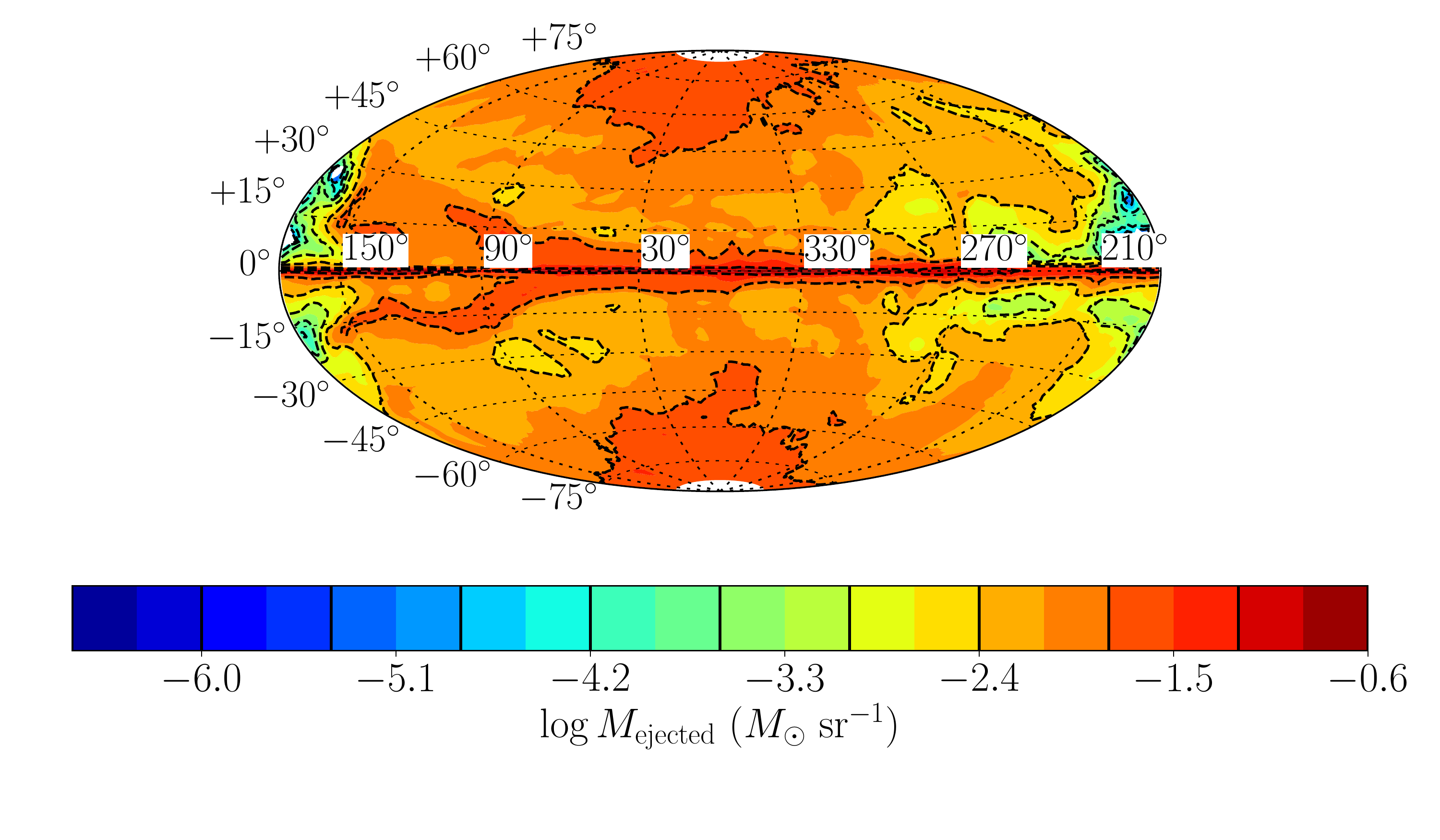}
\hspace*{0.4cm}
\caption{Mass loss rate per unit solid angle maps (left panels; in $M_\odot ~ {\rm sr}^{-1} \s^{-1}$) and total mass loss per unit solid angle maps (right panels;  in  $M_\odot ~ {\rm sr}^{-1}$) of a spherical shell with a radius of $R_{\rm out} = 2 \AU$ of simulation \#13 at three times. The times are (from top to bottom) 152, 339 and 444 days from the beginning of the simulation, as in Fig. \ref{fig:fiducial_dense}. Zero latitude is the equatorial plane and zero longitude (at the center) is the initial location of the companion, namely the positive $x$ axis. The companion is moving towards higher angles, namely from right to left.
}
\label{fig:fiducial_mass_outflow_map}
\end{figure*}

\begin{figure*}
\centering
\includegraphics[trim= 1cm 0.4cm 0.2cm 0.3cm,clip=true,width=0.45\textwidth]{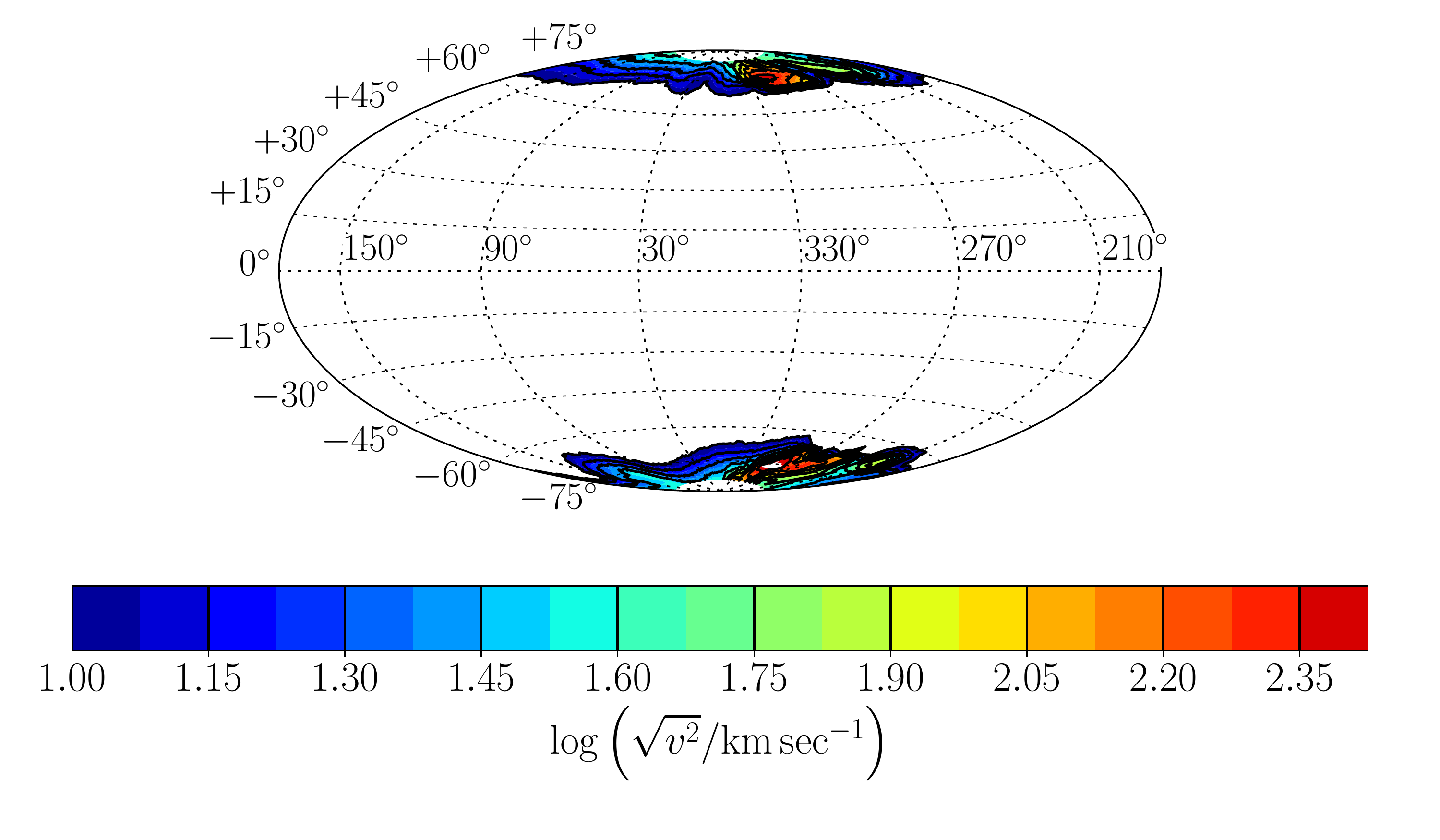}
\includegraphics[trim= 1cm 0.4cm 0.2cm 0.3cm,clip=true,width=0.45\textwidth]{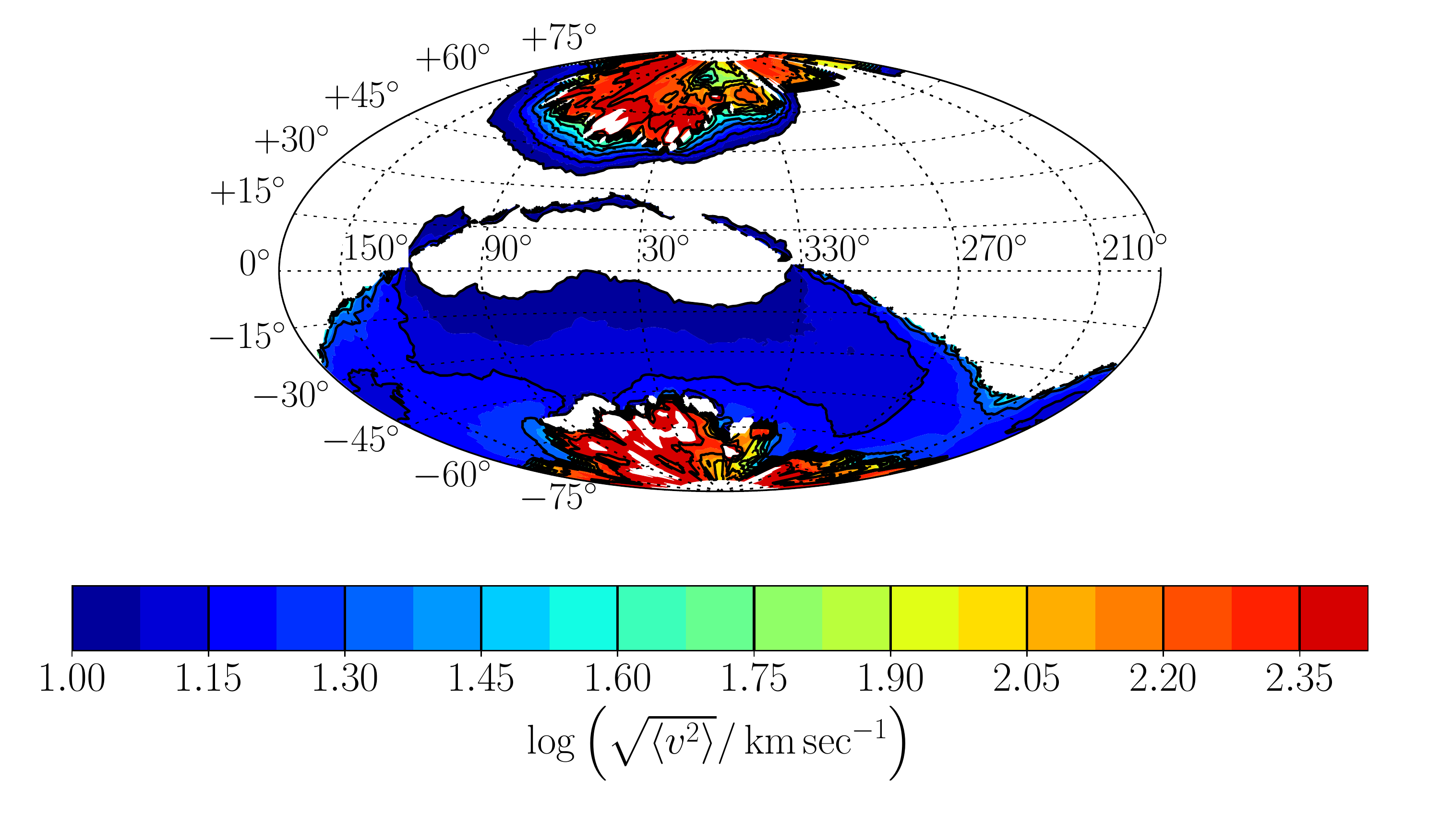}
\hspace*{0.4cm}
\vspace*{0.4cm}
\includegraphics[trim= 1cm 0.4cm 0.2cm 0.3cm,clip=true,width=0.45\textwidth]{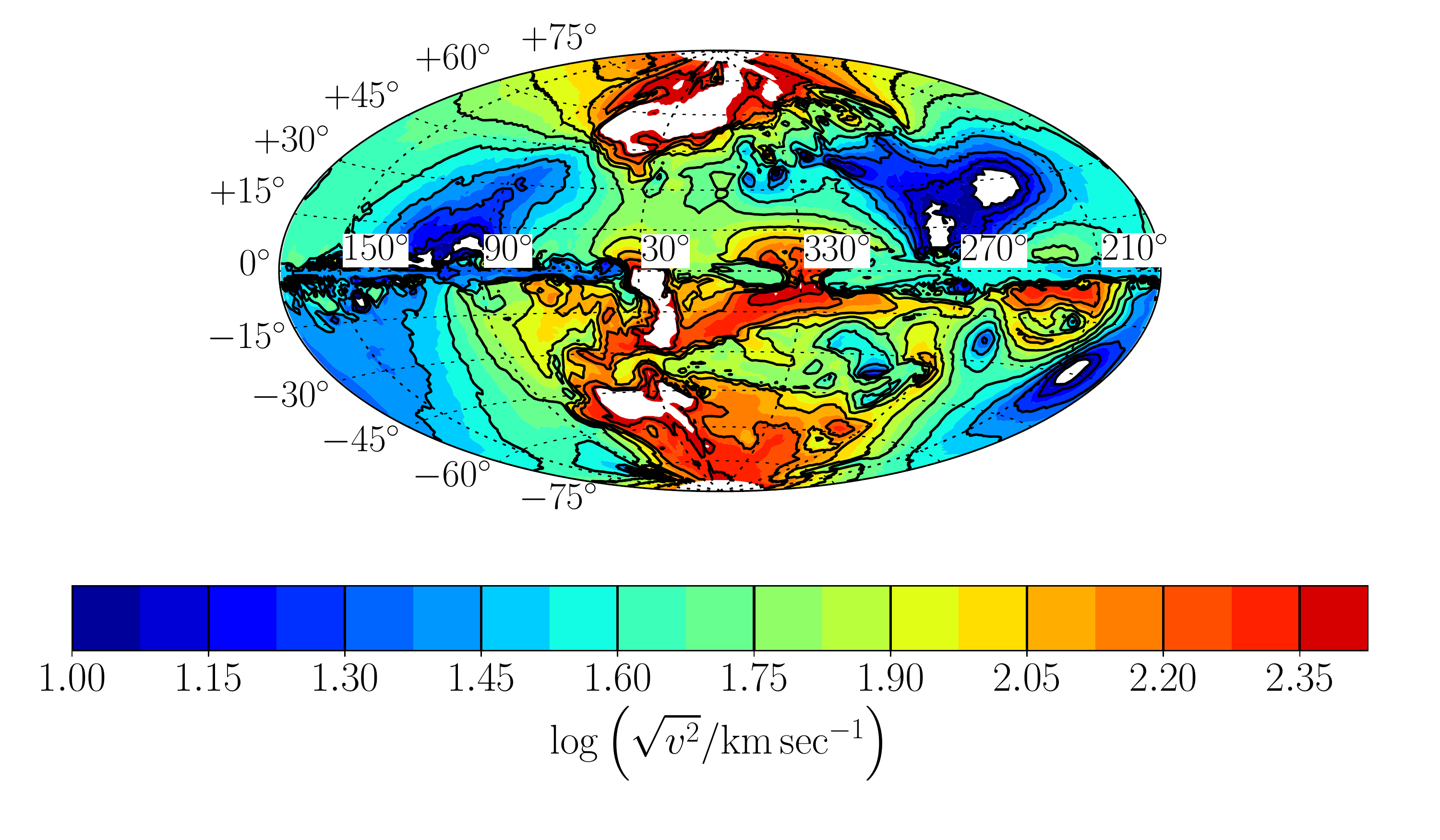}
\includegraphics[trim= 1cm 0.4cm 0.2cm 0.3cm,clip=true,width=0.45\textwidth]{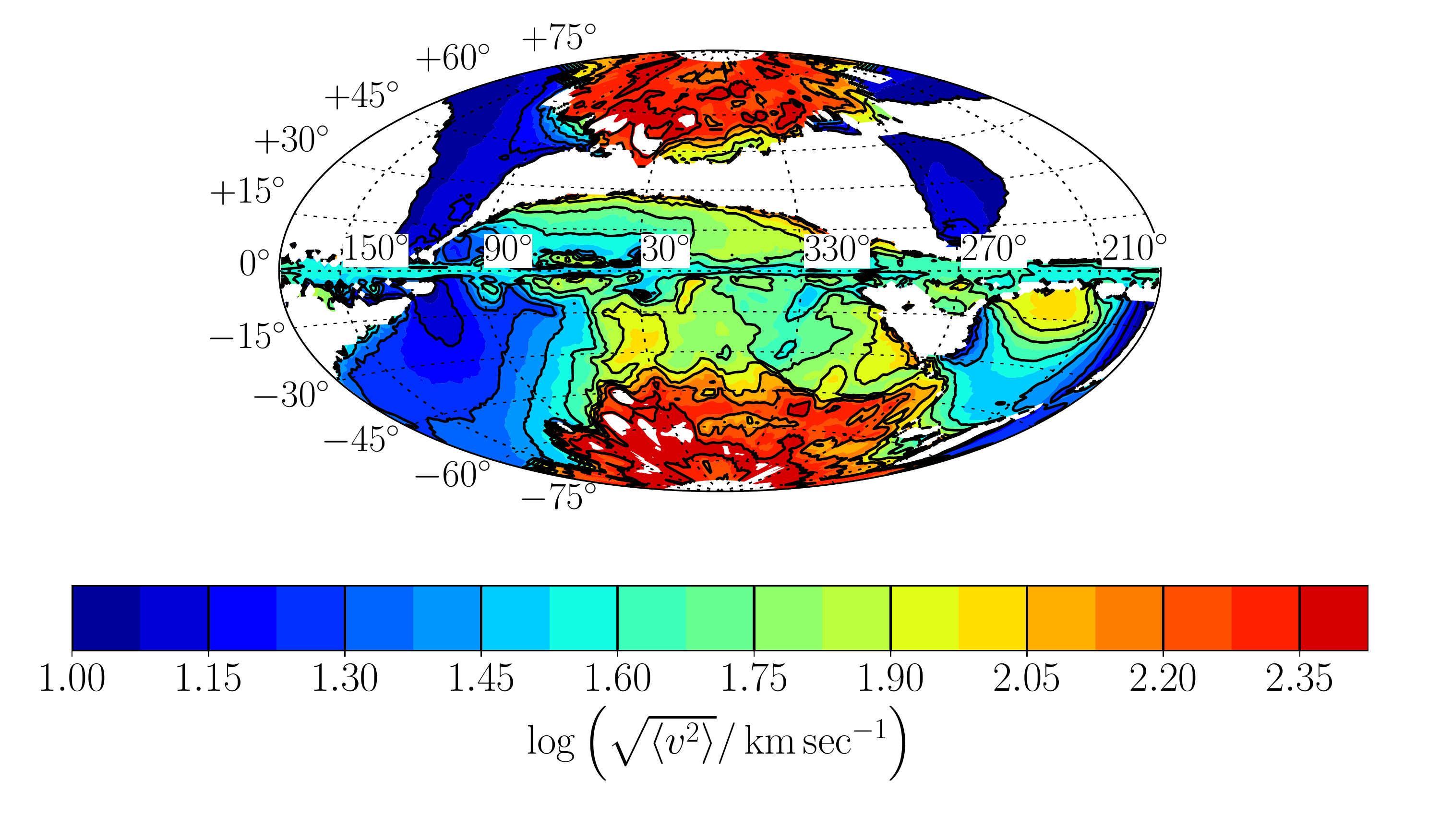}
\hspace*{0.4cm}
\includegraphics[trim= 1cm 0.4cm 0.2cm 0.3cm,clip=true,width=0.45\textwidth]{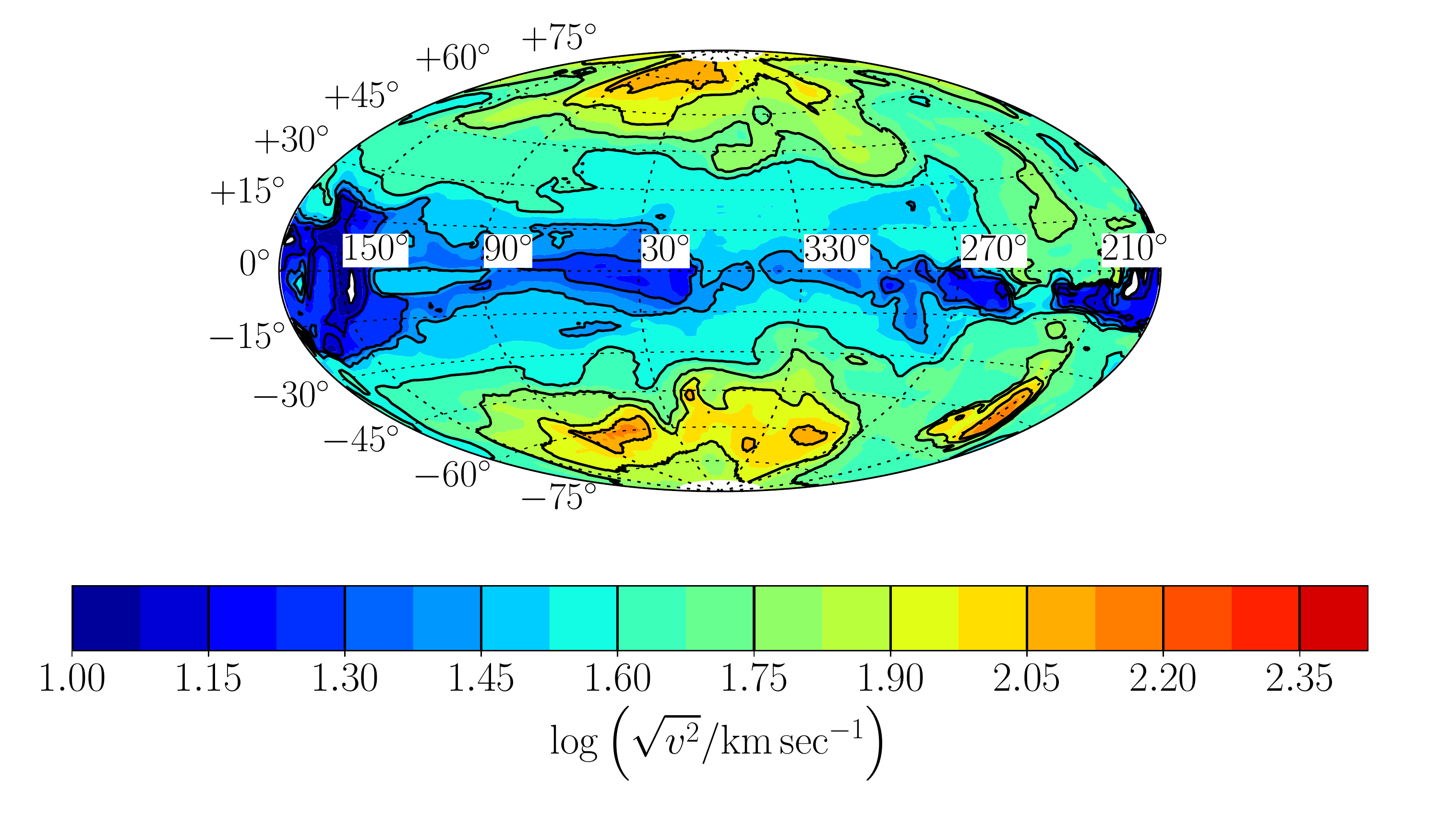}
\includegraphics[trim= 1cm 0.4cm 0.2cm 0.3cm,clip=true,width=0.45\textwidth]{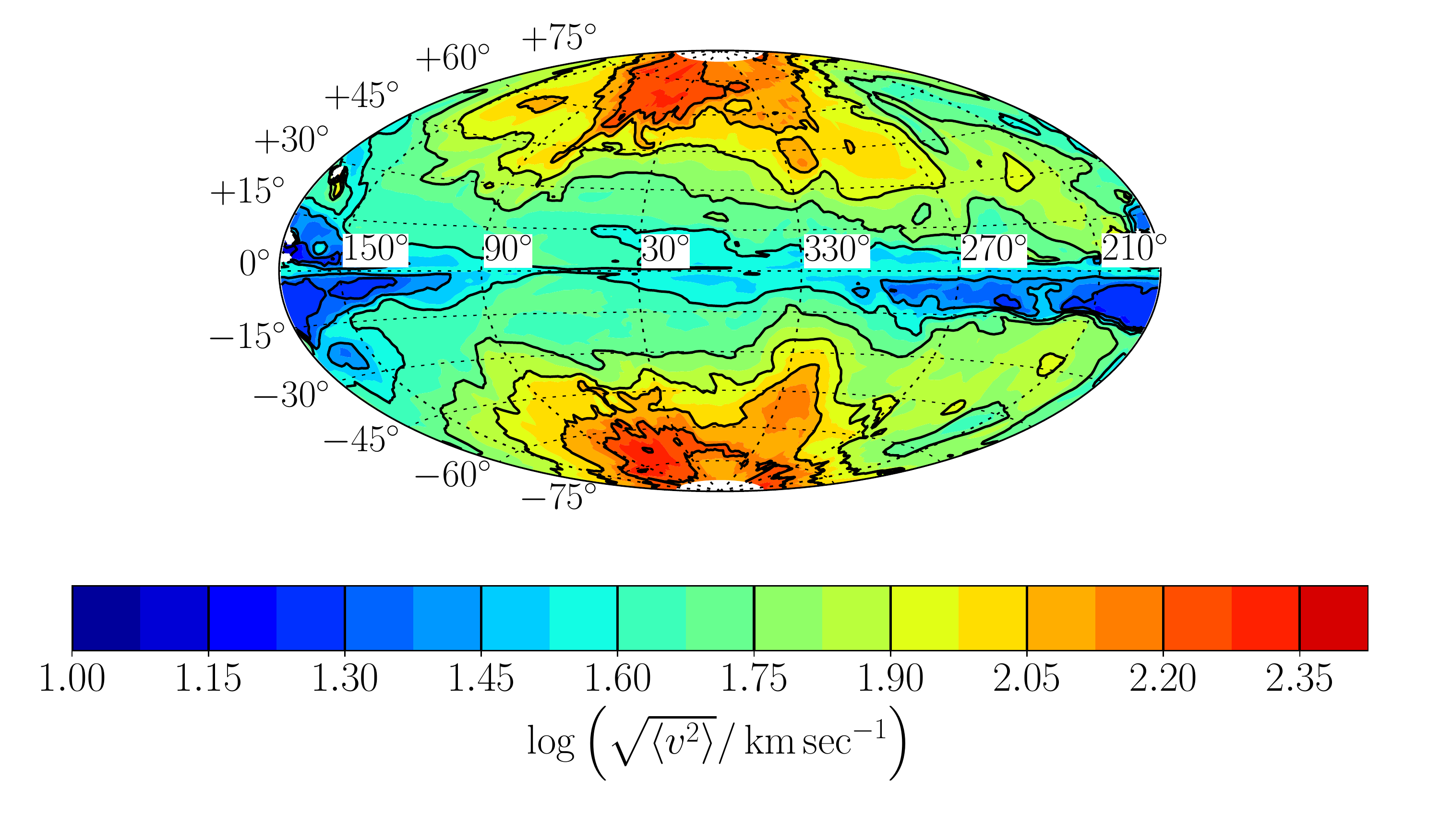}
\hspace*{0.4cm}
\caption{Maps of momentary outflow velocity magnitude (left panels) and of outflow velocity averaged over time (right panels) at a spherical shell with a radius of $R_{\rm out} = 2 \AU$ of simulation \#13 at three times as in Fig. \ref{fig:fiducial_mass_outflow_map}. Grid as in Fig. \ref{fig:fiducial_mass_outflow_map}.  
}
\label{fig:fiducial_velocity_outflow_map}
\end{figure*}

We present in Fig. \ref{fig:fiducial_mass_outflow_map} the maps of the mass loss rate (left panels) and of the total mass loss until the indicated time (right panels) at three times as indicated. In Fig. \ref{fig:fiducial_velocity_outflow_map} we present the outflow velocity maps (left panels) and average velocity maps (right panels) at the same times of Fig. \ref{fig:fiducial_mass_outflow_map}. These maps emphasize the differences between the outflow in the three phases.

 At the beginning of the in-spiral before the jets graze the envelope a small amount of mass with high velocity outflows in the polar directions. This mass comes directly from the jets.
Later, the in-spiral produces an equatorial outflow appearing as a thin red stripe at zero latitude in the second row of Fig. \ref{fig:fiducial_mass_outflow_map}. Ejection along the equatorial plane during the dynamic in-spiral is observed in most CE simulations and it is a direct result of angular momentum conservation (see \citealt{Iaconietal2017} for an estimate of the angle containing the ejecta). In these simulations most of the unbound mass comes from the equatorial outflow. In our simulations with jets the equatorial outflow is slower than the polar outflow and is only mildly unbound. The escape velocity from the system at $R_{\rm out} = 2 \AU$ is approximately $30\;\kms$ while the equatorial ring outflows at slighter higher velocities of $40\; \kms$. At later times, once the jets succeed to penetrate through the dense envelope they cause a substantial mass ejection. In this phase the mass loss rate becomes more isotropic, but the outflow velocity is still larger along the polar directions. The smooth red color in the bottom left panel of Fig. \ref{fig:fiducial_mass_outflow_map} indicates the smooth mass loss rate. The bottom left panel of Fig. \ref{fig:fiducial_velocity_outflow_map} shows the faster polar outflow.

When the companion enters the envelope the jets and gravitational interaction eject mass in a highly non-spherical geometry, with a preferred outflow direction. Momentum conservation implies that the core and companion (the `particles') and the bound gas move in the other direction. In the middle-left panel of Fig. \ref{fig:fiducial_dense} we see dense gas that escapes to the right. The bound system starts moving to the left. 

\begin{figure*}
\centering
\includegraphics[trim= 0.4cm 0.4cm 0.2cm 0.3cm,clip=true,width=0.99\textwidth]{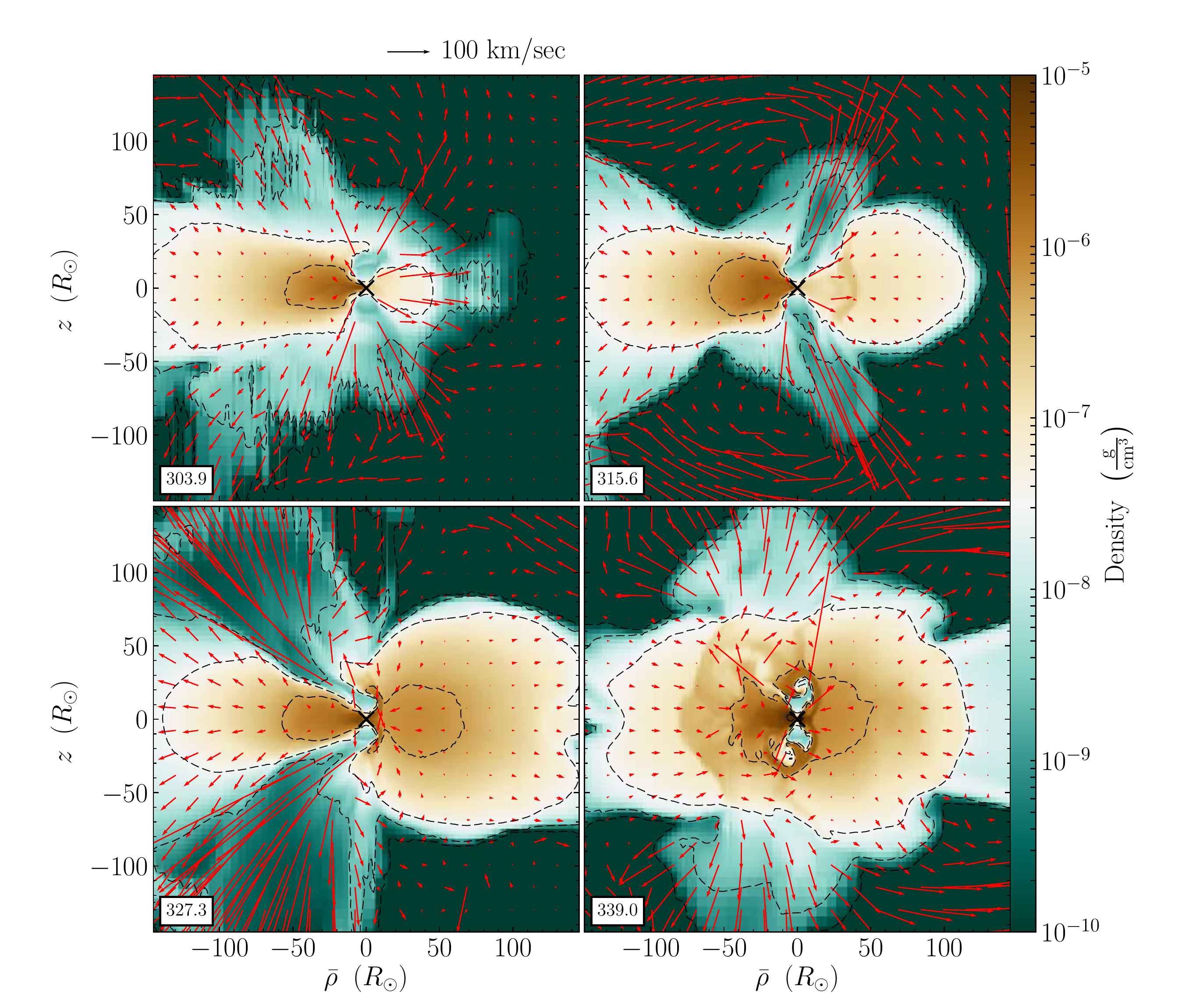} 
\caption{Density maps and velocity vectors in the $\bar{\rho}-z$ plane that is perpendicular to the momentary radius vector of the secondary star relative to the core position and perpendicular to the equatorial plane of simulation \#13 at four times, given in days. We term this plane the tangential plane. The horizontal axis is $\bar{\rho}=\pm \sqrt {(x-x_s)^2+(y-y_s)^2}$, where $(x_s,y_s)$ is the momentary position on the orbital plane of the secondary star. The companion is at the center of each panel and is moving to the right. At early times the jets bend forward (to the right) then to backward (to the left) and then they are choked inside the envelope (last panel). 
}
\label{fig:jets_bending}
\end{figure*}

To elaborate further on the time that the companion enters the envelope we present the flow at four times in Fig. \ref{fig:jets_bending}. The panels show density maps and velocity vectors in the tangential, $\bar{\rho}-z$, plane of simulation \#13. The time of the bottom right panel of Fig. \ref{fig:jets_bending} is the same as in the middle row of Fig. \ref{fig:fiducial_dense}, Fig. \ref{fig:fiducial_mass_outflow_map}, and Fig. \ref{fig:fiducial_velocity_outflow_map}. The tangential plane cuts through the companion and is perpendicular to the equatorial plane as well as to the line joining the secondary star to the core of the primary (the two particles). In the panels the companion is in the center of the frame (black 'X') and moves to the right. At early times (upper left panel), the massive fan that trails the secondary star from behind diverts the jets stream forward (to the right in the panel). The bending forward of the jets pushes material in the opposite direction. The first three panels of Fig. \ref{fig:jets_bending} show a complicated interaction between the jets and envelope, i.e., the bending of the jets and the formation of vortices. The last panel of Fig. \ref{fig:jets_bending} presents two properties of the interaction, the ejection of high velocity gas to the right (the long arrows at the outskirts of the panel) and the chocking of the jets inside the envelope (two bubbles near the center). 

From Fig. \ref{fig:fiducial_dense} and others we see the binary system to move to the left of the grid. The reason is that gas is ejected to the right of the grid and leaves the grid, as seen for example in Fig. \ref{fig:fiducial_dense}. Some of the gas that leaves the grid is unbound and so the motion of the binary system, the kick velocity, is real. However, due to the limited size of the grid some of the gas that leaves the grid is still bound. So the motion of the binary system is overestimated here for numerical reasons. We crudely estimate its value at several $\km \s^{-1}$ in our simulations. Our estimates are similar to, or somewhat larger than, the kick velocity that  \cite{Chamandyetal2019} find in their CEE simulations. 
     
\subsection{Comparison of simulations with and without jets}
\label{subsec:compare_simulations}

\subsubsection{Orbital separation and mass loss rate}
\label{subsubsec:Orbital}

We compare the results of simulations \#6 and \#9 that include jets with the results of simulations \#5 and \#8 that have the same initial conditions but that do not include jets. In simulations \#5 and \#6 the companion starts on the giant surface $a_0=R_1$, while in simulations \#8 and \#9 the companion starts at $a_0=2R_1$.

\begin{figure}
\includegraphics[trim= 0.0cm 0.6cm 0.0cm 0.4cm,clip=true,width=0.99\columnwidth]{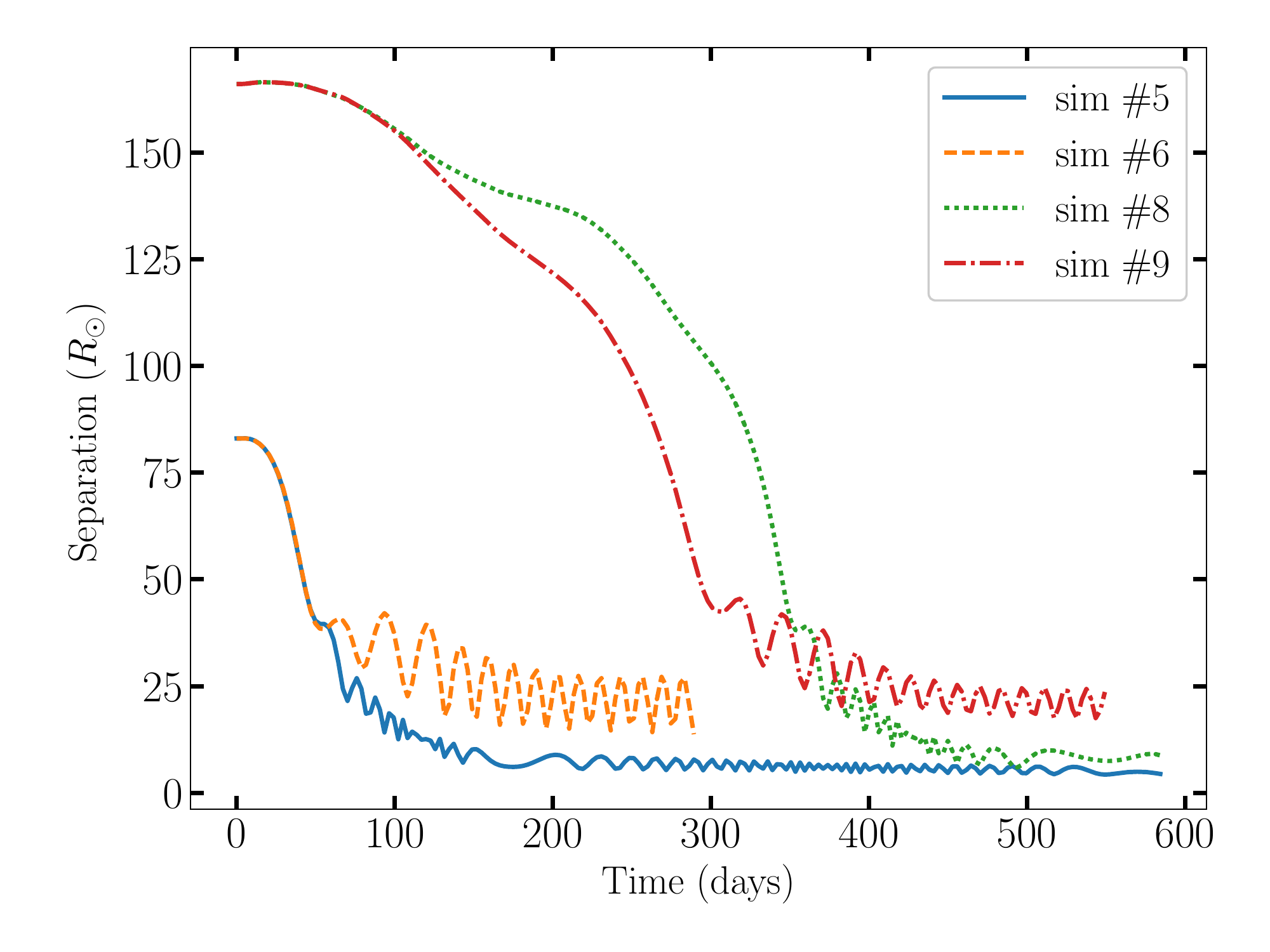}
\caption{ The orbital separation between the giant's core and the companion as a function of time. The launching of jets in simulations \#6 (orange dashed line) and \#9 (red dashed-dotted line) stops the spiraling-in earlier and at a larger and more eccentric orbit compared with simulations with no jets \#5 (blue solid line) and \#8 (green dotted line).}
\label{fig:comparing_separation_vs_time}
\end{figure}

\begin{figure*}
\includegraphics[trim= 0.0cm 4.8cm 0.0cm 6cm,clip=true,width=0.99\textwidth]{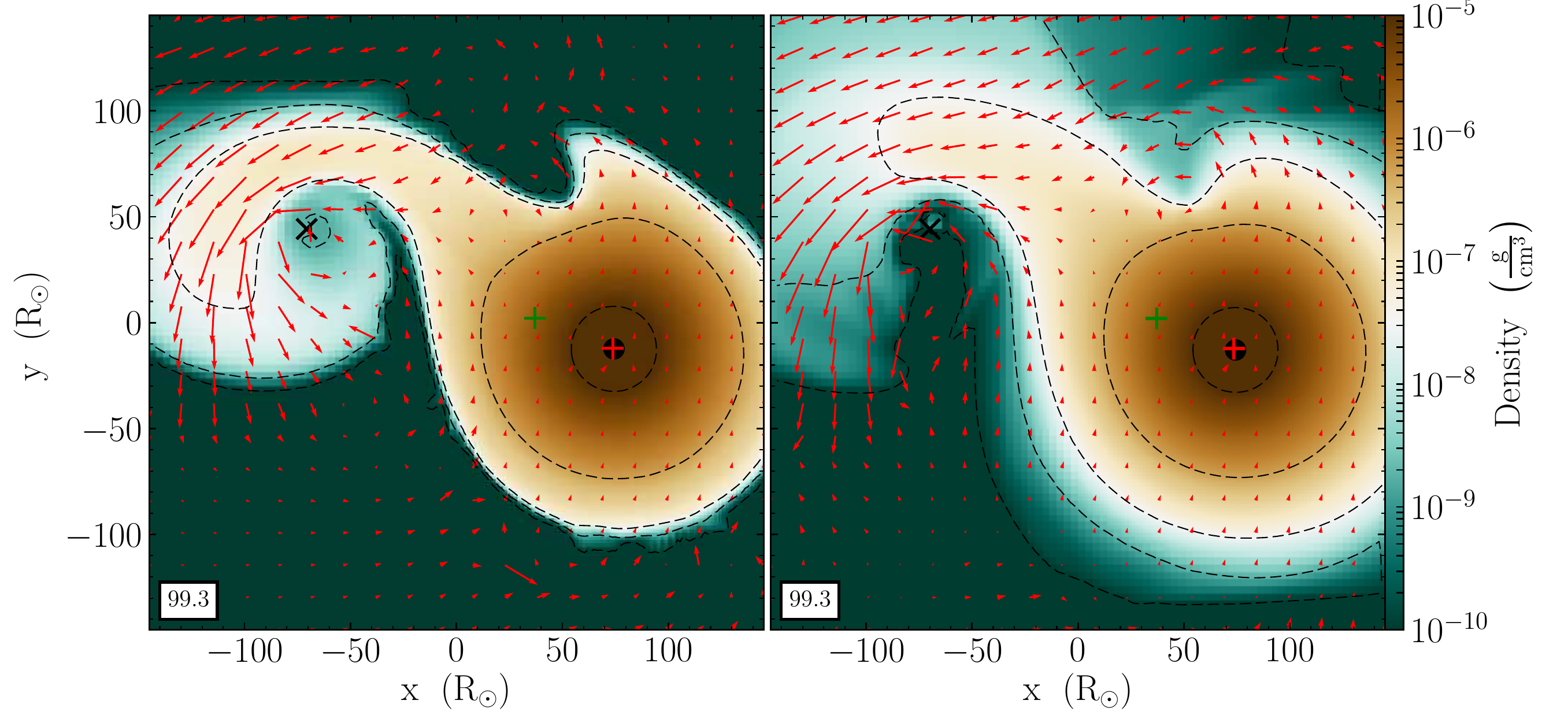}
\caption{ Density maps and velocity vectors in the equatorial plane $z=0$ at $t=99.3$ days, of simulation \#8 without jets (left panel) and simulation \#9 with jets (right panel). The companion orbits counterclockwise, and symbols and scaling are as in Fig. \ref{fig:fiducial_dense}.}
\label{fig:comparing_early}
\end{figure*}

In Fig. \ref{fig:comparing_separation_vs_time} we plot the orbital separation between the giant's core and the companion as a function of time for all four simulations. In simulation \#9, that includes jets, the spiraling-in outside the envelope is faster than in simulation \#8 that has no jets. To find an explanation we present in Fig. \ref{fig:comparing_early} the density and velocity maps at $t=99.3$ days. We see the gas of the fan flowing around the companion. In the jetted simulation (right panel) there is less gas in front of the companion because the jets deflect the gas. The gas behind the companion slows it down, and the gas in front of the companion pulls it. With less gas in front of the companion in the simulation with jets, a stronger slowing-down acts on the companion. 
However, we must recall that for numerical reasons we include no accretion, and so although the effect is real, its  magnitude is not accurate in our simulations. 

After the companion enters the envelope, the main effect of the jets is removal of mass. This further removal of mass slows down the spiraling-in rate, and the companion with jets in simulation \#9 ends at an orbital separation of $21 \rmRodot$  compared with $8 \rmRodot$ in simulation \#8. Simulations \#5 and \#6 show the same effect. 
Overall, the final orbital separation in simulations with jets are more eccentric and a few times larger than the orbits in simulations without jets. We attribute the higher eccentricity to rapid removal of mass by jets: such mass removal in a time much shorter than the orbital period can counteract the tidal circularizing effect (e.g., \citealt{KashiSoker2018}). 

\begin{figure}
\includegraphics[trim= 0.0cm 0.6cm 0.0cm 0.4cm,clip=true,width=0.99\columnwidth]{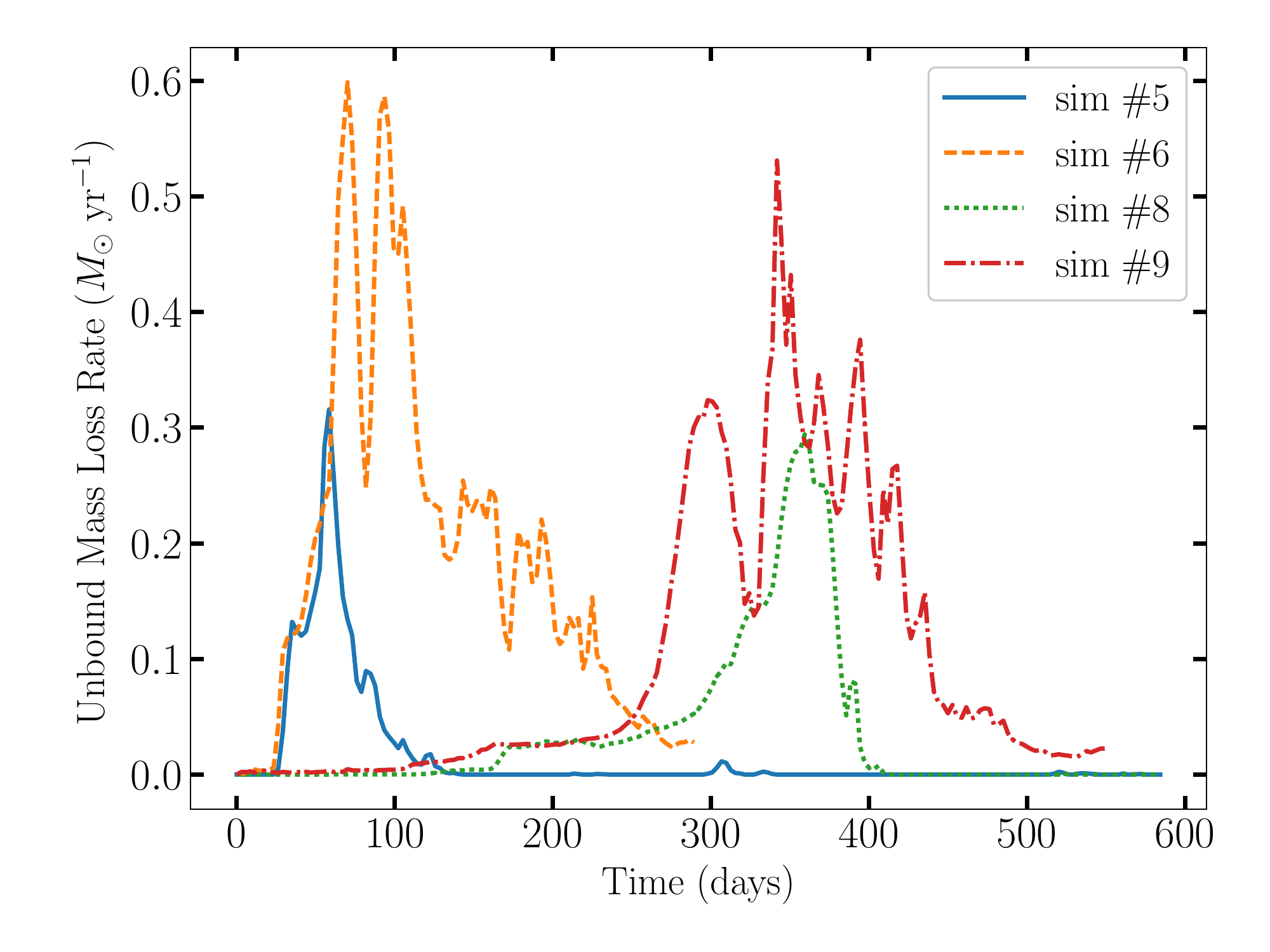}
\caption{ The mass loss rate of unbound mass, $\dot M^{\rm unbound}_{\rm out}$, i.e., with a total positive energy, from a sphere of radius $1 \AU$ as a function of time. The jets of simulations \#6 and \#9 produce higher mass loss rates than in the corresponding simulations \#5 and \#8 that have no jets. }
\label{fig:comparing_massloss_unbind}
\end{figure}

In Fig. \ref{fig:comparing_massloss_unbind} we plot the mass loss rate through a sphere of radius $1 \AU$ of the unbound mass, i.e., having a total (kinetic + thermal + gravitational) positive energy, $\dot M_{\rm out}^{\rm unbound}$.
The simulations without jets suffer mass ejection during the fast in-spiral phase that lasts for approximately a hundred days, and most of the unbound mass outflows at the equator. Simulations with jets experience a phase of higher unbinding that lasts for longer periods. For example, the second unbinding phase lasting from about 340 days to about 450 days in simulation \#9 (the red dashed-dotted line) corresponds to the approximately uniform outflow mass per unit solid angle that the jets produce around the time the orbital separation settles to its final value.  

\subsubsection{In-spiral and outflow properties}
\label{subsubsec:Inspiral}

\begin{figure*}
\centering
\includegraphics[trim= 4cm 0.4cm 0.2cm 0.3cm,clip=true,height=0.9\textheight]{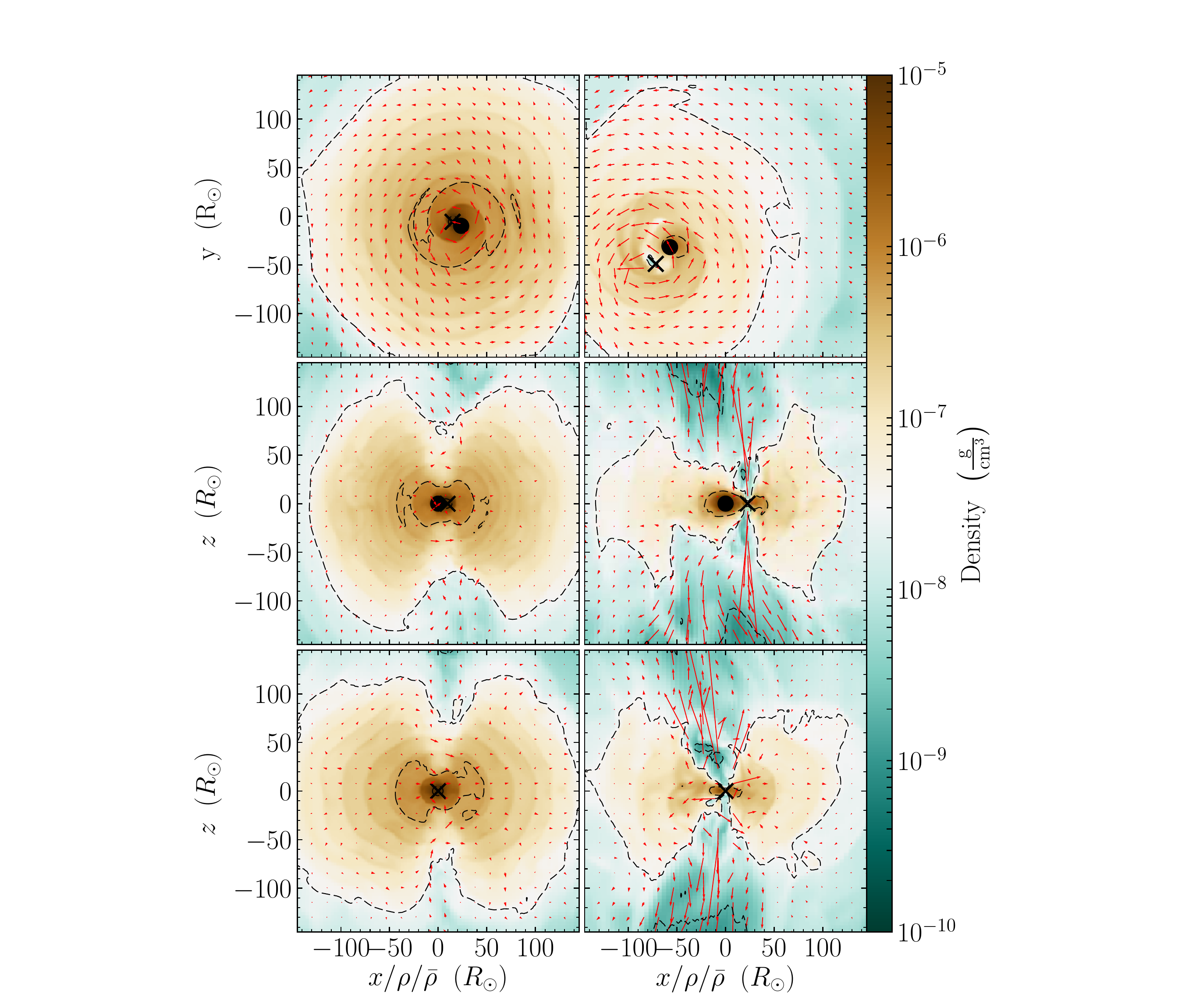} 
\caption{Density maps and velocity vectors in three planes all at $t=452.9$~day. The left column presents simulation \#8 without jets and the right column presents simulation \#9 with jets. 
The upper, middle, and lower rows show the equatorial plane ($x-y$), the meridional plane that is perpendicular to the equatorial plane and contains the momentary positions of both the companion and the core ($\rho-z$), and the tangential plane that is perpendicular to the momentary radius vector of the secondary star relative to the core position and perpendicular to the equatorial plane ($\bar \rho - z$). Other symbols are as in earlier figures. 
}
\label{fig:comparing_dense89}
\end{figure*}

Here we focus only on comparing simulations \#8 (no jets) and \#9 (jets), both started at twice the stellar radius. Fig. \ref{fig:comparing_dense89} shows density maps and velocity vectors in the equatorial, meridional, and tangential planes, from top to bottom, of simulation \#8 (left panels) and simulation \#9 (right panels) at $t=452.9$~day. 

From Fig. \ref{fig:comparing_dense89} and some figures we do not show we note the following differences: (1) The offset of the binary from the center of the grid is larger in the simulation with jets. (2) As we already discussed, simulations with jets end with a larger core-companion separation. (3) The jets make the outflow less symmetric and much more complicated. In particular, after the jets are chocked at earlier times when the companion enters the deep envelope, they eventually break out along the polar directions. 

\begin{figure*}
\centering
\includegraphics[trim= 0.0cm 4.8cm 0.0cm 5cm, clip=true,width=0.99\textwidth]{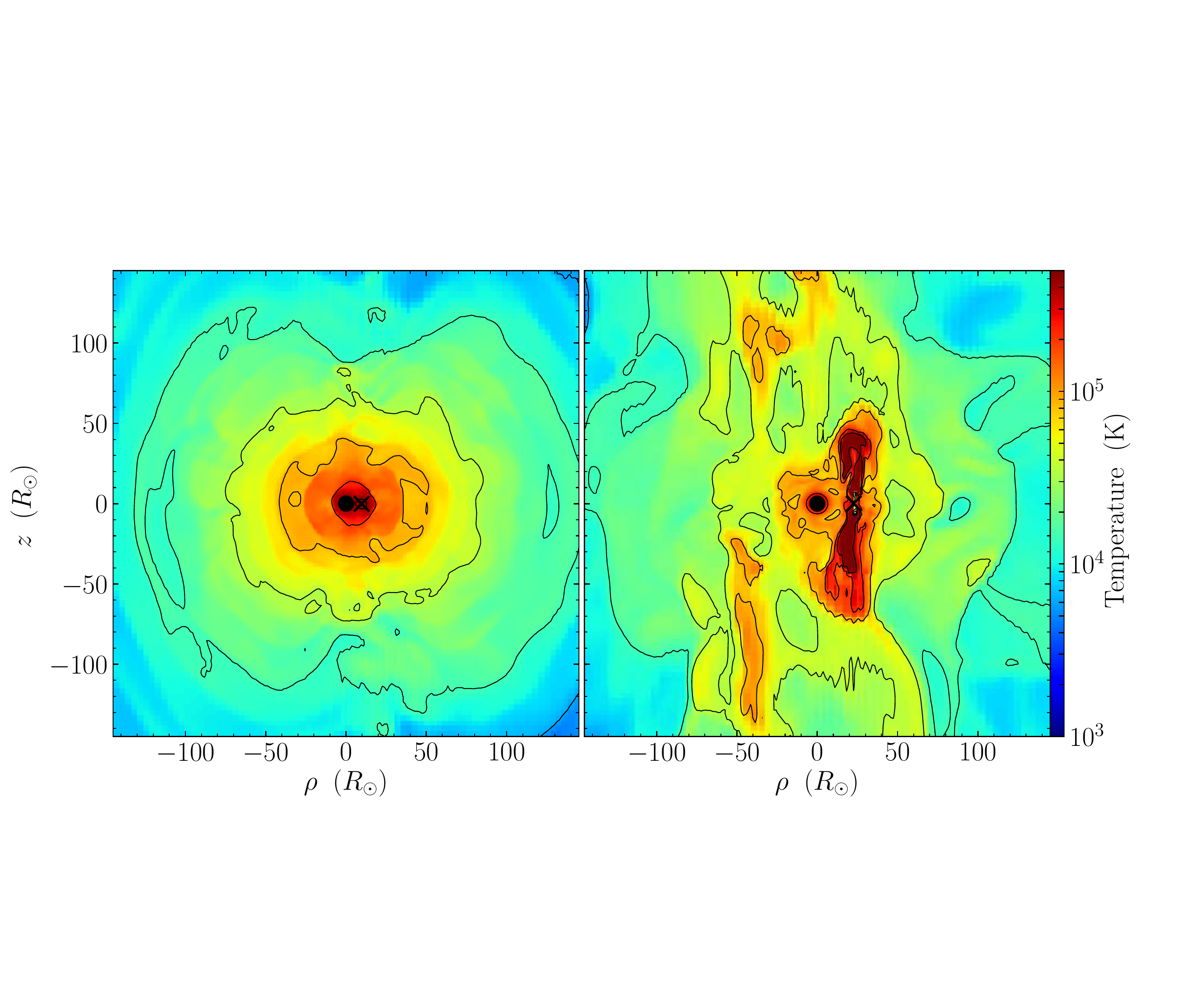} 
\caption{Like middle row of Fig. \ref{fig:comparing_dense89} but presenting the temperature maps.
}
\label{fig:comparing_center_tempFinal}
\end{figure*}

Fig.  \ref{fig:comparing_center_tempFinal} presents the temperature maps in the meridional plane (i.e., perpendicular to the equatorial plane) that contains the companion and core, at $t=452.9$~day. The temperature distribution at the end of simulation \#8, with no jets, has a large scale spherical symmetry with a number of fluctuations. In the equatorial plane (that we do not show here) there is a spiral pattern. In simulation \#9, with jets, the interaction of the jets with the envelope creates two shocks, of the jets and of the envelope, that heat the gas to high temperatures. The right panel in Fig. \ref{fig:comparing_center_tempFinal} emphasizes the bipolar structure that the jets form around the companion that launches the jets. 

\begin{figure}
\includegraphics[trim= 0.0cm 0.6cm 0.0cm 0.4cm,clip=true,width=0.99\columnwidth]{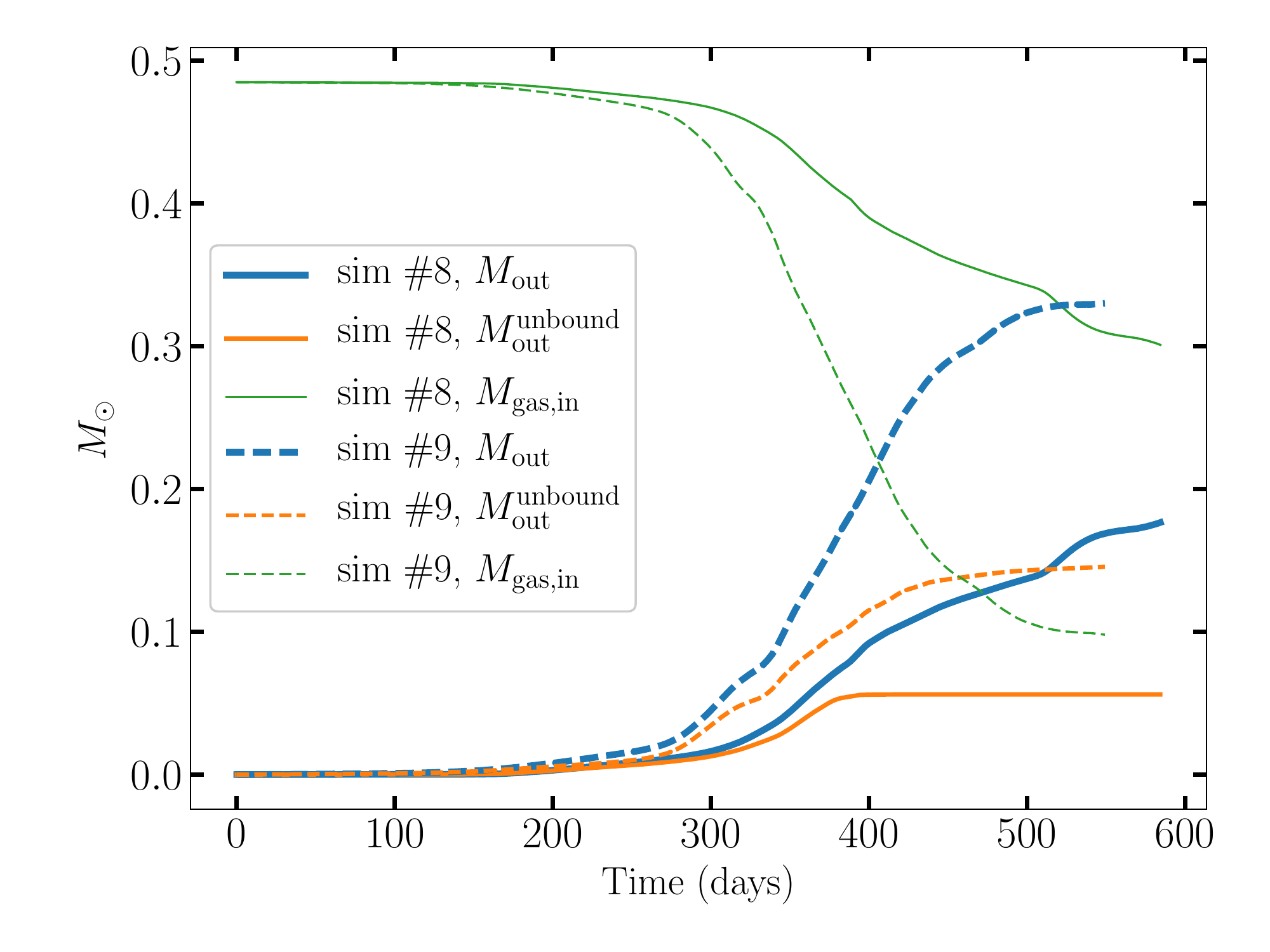}
\caption{ The variation of the envelope gas mass, i.e., gas inside a radius of $R_{\rm out}=1 \AU$, $M_{\rm gas,\;in}$ (thin green lines), of the total mass (bound + unbound) that flowed out through $R_{\rm out}=1 \AU$, $M_{\rm out}$ (thick blue lines), and of the unbound mass  that flowed out through a sphere of  $R_{\rm out}=1 \AU$, $M_{\rm out}^{\rm unbound}$ (orange lines). 
Solid lines are for simulation \#8 and dashed lines are for  simulation \#9 (with jets).}
\label{fig:comparing_mass_plot}
\end{figure}

In Fig. \ref{fig:comparing_mass_plot} we show the contribution of the jets to mass removal. The curves show the variation with time of the mass in the envelope, $M_{\rm gas,in}$, that we define as the gas that resides inside $R_{\rm out}=1 \AU$ (thin green lines), of the total mass (bound + unbound) $M_{\rm out}$ (thick blue lines) that flowed out through a sphere of radius $R_{\rm out}=1 \AU$, and of the unbound mass $M_{\rm out}^{\rm unbound}$ (orange lines) that flowed out through a sphere of radius $R_{\rm out}=1 \AU$.
Solid lines are for simulation \#8 with no jets, and dashed lines are for simulation \#9 with jets. In simulation \#8, with no jets, $\approx 12 \%$ of the ejected mass is unbound, in agreement with simulation SIM3 of \cite{Iaconietal2018} and with simulations Enzo3 and Enzo8 of \cite{Passyetal2012} that have similar initial parameters and were also performed with {\sc Enzo}. In simulation \#9, with jets, $\approx 33 \%$ of the ejected mass is unbound.

We note here that there is another similarity between the results of \citet{Iaconietal2018} and our results in that our simulation \#5 has the same level of energy and angular momentum conservation as their very similar simulation SIM3. Similar level of conservation exists in the other simulations without jets (\#1, \#3, \#8). We discuss the energy conservation level in simulations with jets in section \ref{subsec:JetsEnergetic}.
   
\begin{figure}
\includegraphics[trim= 0.0cm 0.6cm 0.0cm 0.4cm,clip=true,width=0.99\columnwidth]{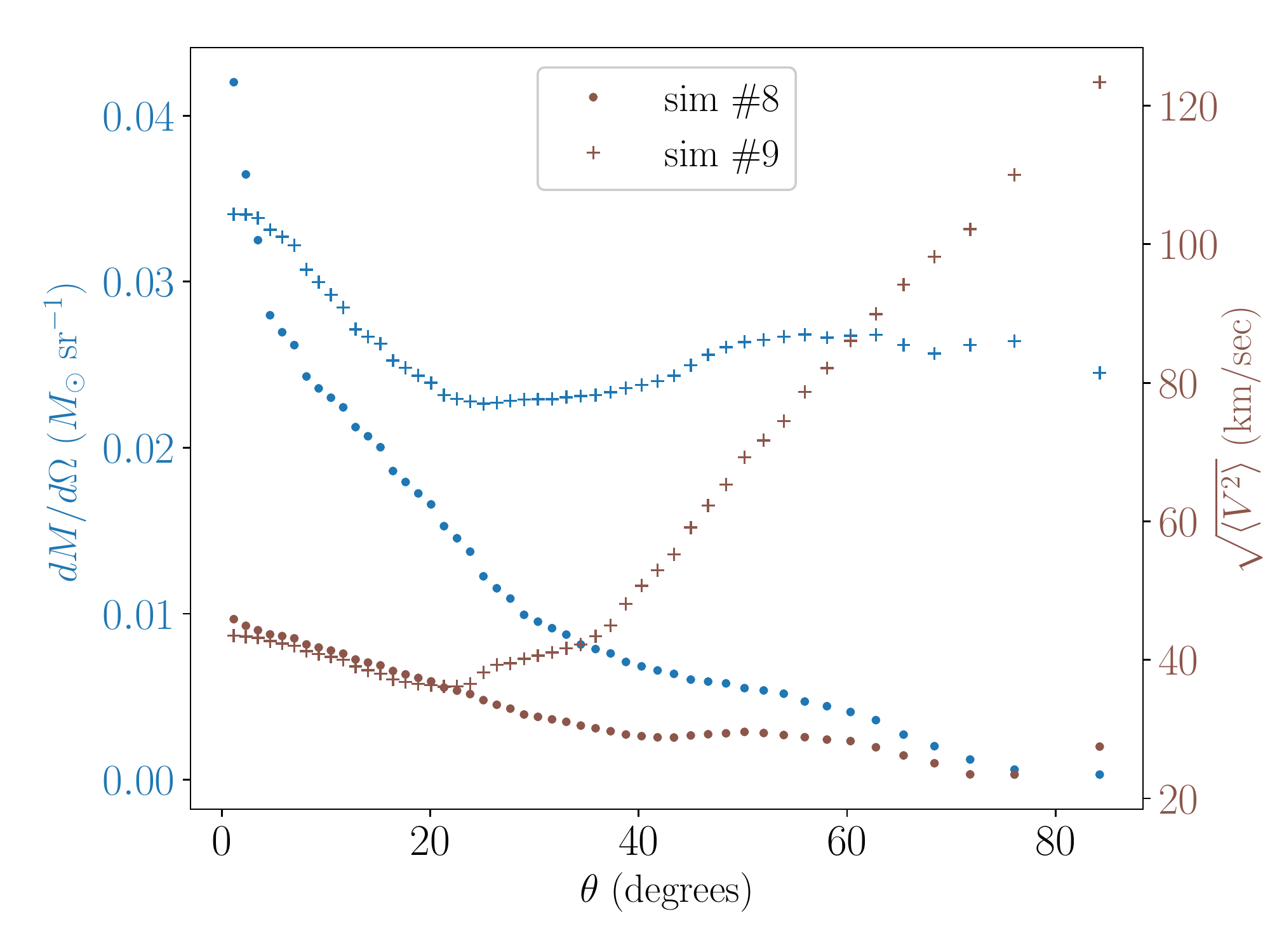}
\caption{ The total mass per unit solid angle that flows out from a sphere of radius $R_{\rm out}=1 \AU$ (blue) and its average velocity (brown) as a function of the angle $\theta$ from the equatorial plane, for simulation \#8 (no jets; O symbols) and for simulation \#9 (with jets; $+$ symbols). The equator is at $\theta=0$ and the poles are at $\theta=90$. Jets produce more evenly distributed mass ejection with much faster outflows at high latitudes. }
\label{fig:comparing_mass_vrms_angle}
\end{figure}

To emphasize the role of jets in shaping the outflow we plot in Fig. \ref{fig:comparing_mass_vrms_angle} the total mass loss (bound and unbound) per unit solid angle that flows out from a sphere of radius $R_{\rm out}=1 \AU$ (blue) and its average velocity (brown), as function of angle from the equatorial plane.  
The main difference is that in the simulation with jets ($+$ symbols) the jets eject more mass relative to the simulation with no jets (O symbols) along the polar directions, and at much higher velocities. 

\subsection{Energetic of jets and mass ejection}
\label{subsec:Energy}

\begin{figure}
\centering
\includegraphics[trim= 0.0cm 0.6cm 0.0cm 0.4cm,clip=true,width=0.99\columnwidth]{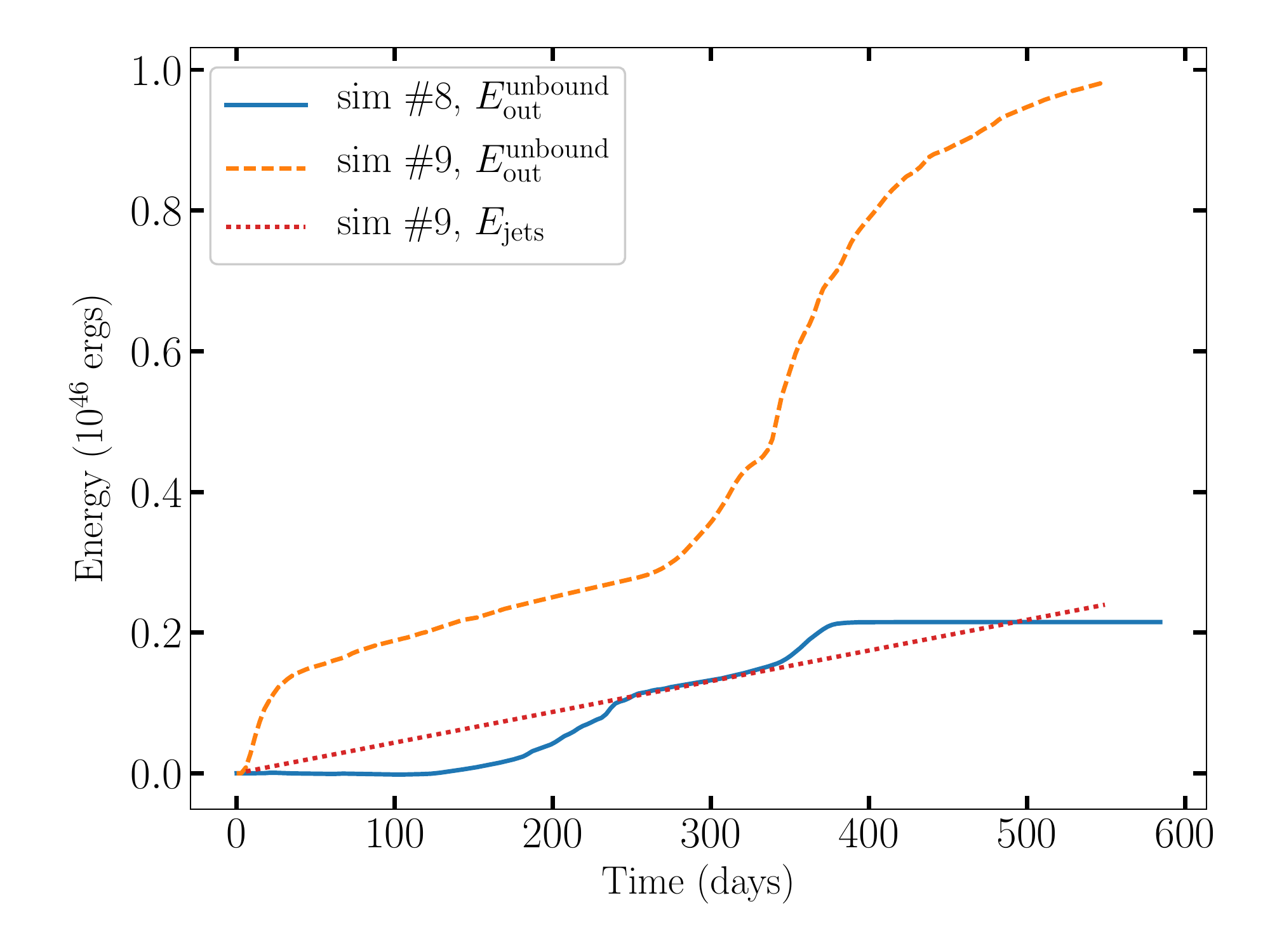}
\caption{The energy that the unbound mass carries with it as it leaves the system in simulation \#9 with jets (dashed-orange line) and in simulation \#8 with no jets (solid-blue line). The dotted-red line presents the kinetic energy injected in the jets in simulation \#9. 
}
\label{fig:comparing_eout}
\end{figure}

From Fig. \ref{fig:comparing_mass_plot} we learned that the simulation with jets (\#9) ejects much more mass than the simulation with no jets (\#8). We here examine the energy of the unbound mass that leaves the system. In Fig. \ref{fig:comparing_eout} we present the energy that the unbound mass carries with it as it leaves the system in both the simulation with jets (\#9; dashed-orange line) and in the simulation with no jets  (\#8; solid-blue line). The dotted-red line represents the kinetic energy injected by the jets in simulation \#9. In section \ref{subsec:JetsEnergetic} we will discuss the effects that replacing envelope gas with jet gas has on the energetics. (We term this `mass removal', or numerical accretion, as we remove mass from the grid that the companion supposedly accretes.)   

The gravitational energy that the companion-core binary system releases by the end of the simulations is $\approx 1 \times 10^{46}  \erg$ in simulation \#9, and $\approx 3 \times 10^{46}  \erg$ in simulation \#8.
These values are more than an order of magnitude larger than the energy carried by the unbound mass in simulation $\#8$ ($\sim 0.2 \times 10^{46}$ erg). This means that the efficiency of the orbital energy in ejecting mass from the system is very low. This is a results shared by all numerical simulations of the CEE that do not include recombination energy (see list of papers in section \ref{sec:intro}). 
 
In Fig. \ref{fig:comparing_eout} we see that the jets that inject a kinetic energy of $E_{\rm jets}(\#9)= 2.4 \times 10^{45} \erg$ by the end of the simulation increase the energy of the unbound gas by $\Delta E^{\rm unbound}_{\rm out}(\#9) \approx 7.4 \times 10^{45} \erg$ with respect to the simualtion without jets. 
{{{{ The fact that $\Delta E^{\rm unbound}_{\rm out}(\#9) > E_{\rm jets}(\#9)$ shows that the jets are very efficient in removing mass, not only by adding their energy to the unbound mass, but also by increasing the fraction of the released orbital energy that goes to the unbound gas. \cite{WilsonNordhaus2019} argue that convection, via mixing, can distribute the released orbital energy in the CE, and by that the convection increases mass removal efficiency. \cite{Chamandyetal2019} further point out that it is crucial to include convection as energy transport since convection might redistribute energy such that a larger fraction of mass is unbound.  
It is possible that the jets in our simulations act to redistribute the released orbital energy in the envelope, and therefore contribute to the unbinding of gas, similarly to what \cite{WilsonNordhaus2019} suggested to be the role of convection. The suggestion that jets and convection both could act as energy distribution agents and could play an important role in the CEE would require a study by itself.   
 }}}} 
 
\begin{figure}
\centering
\includegraphics[trim= 0.0cm 0.6cm 0.0cm 0.4cm,clip=true,width=0.99\columnwidth]{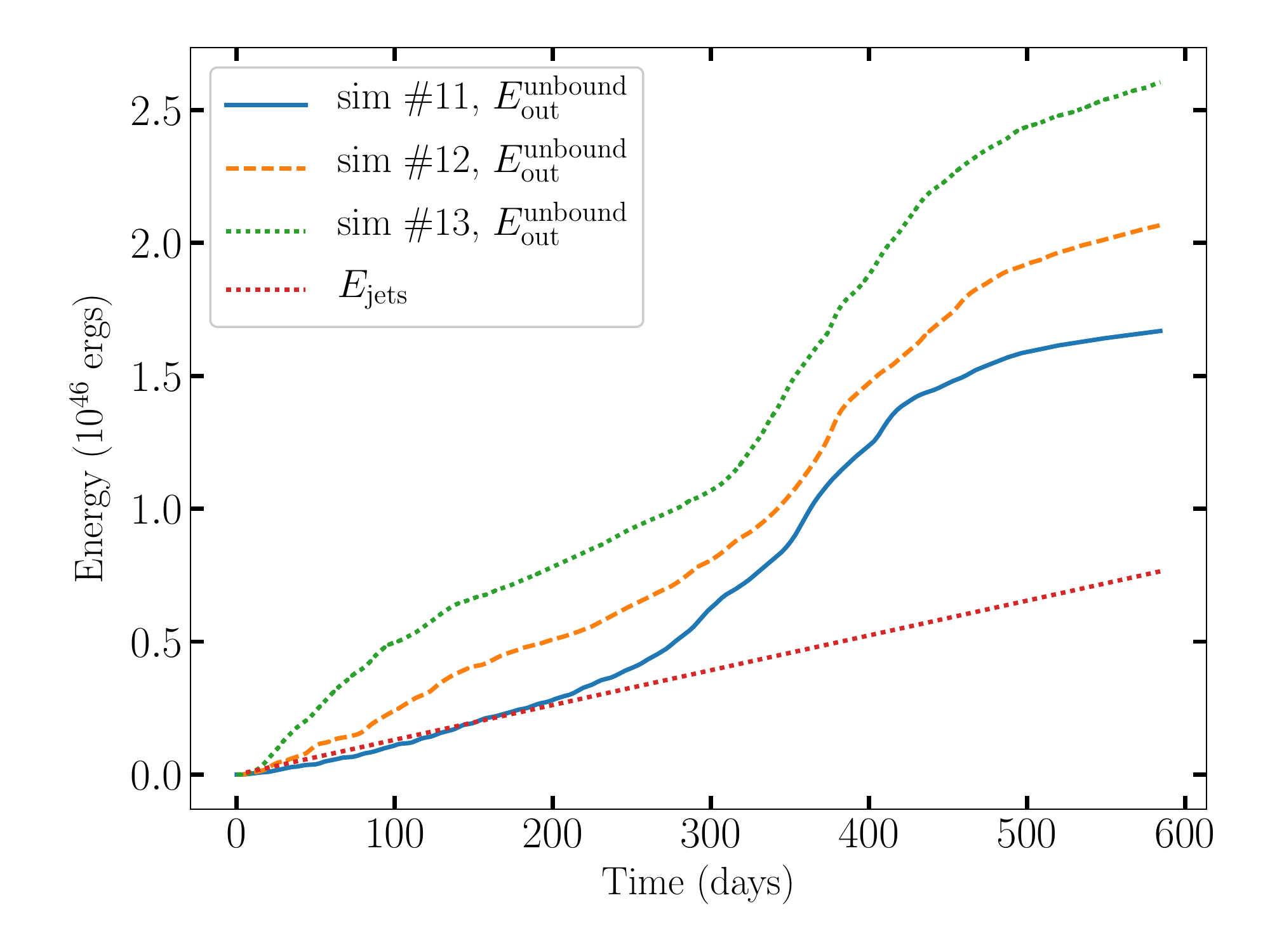}
\caption{The energy that the unbound mass carries with it as it leaves the system in simulation \#11 (third from top; solid blue line), in simulations \#12 (second from top; dashed orange line), and in simulations \#13 (upper line; dotted green line), that differ only in the length of the cones into which we inject the jets, $L_{\rm jets}= 3.6, 7.2, 14.4$~R$_\odot$, respectively. The dotted-red line presents the kinetic energy injected in the jets in each of these simulations. 
}
\label{fig:comparing_eout_J3}
\end{figure}

However, the uncertainties in the exact efficiency of mass removal by jets are large because of the numerical complexity in launching jets. We examine here the influence of one numerical parameter in launching jets. 
In Fig. \ref {fig:comparing_eout_J3} we present the energy that the unbound mass carries in three simulations with a jets mass loss rate of $\dot M_{\rm jets}=0.003 M_\odot \yr^{-1}$ but different lengths of the cones into which we inject the jets, $L_{\rm jets}= 3.6$~R$_\odot$ in simulation \#11, $L_{\rm jets}= 7.2$~R$_\odot$ in simulation \#12 (like in simulation \#9), and  $L_{\rm jets}= 14.4$~R$_\odot$ in simulation \#13. As in all simulations the jets have the same velocity and the same mass loss rate, the kinetic energy power of the jets is the same in the three simulations $\dot E_{\rm jets} = 1.5 \times 10^{38} \erg \s^{-1}$. At the end of the  three simulations the jets injected a total kinetic energy of $E_{\rm jets}=7.6 \times 10^{45} \erg$. The energy of the unbound mass that left the grid is quite different in the three simulations, $E^{\rm unbound}_{\rm out}(\#11)=1.7 \times 10^{46} \erg = 2.2 E_{\rm jets}$ , $E^{\rm unbound}_{\rm out}(\#12)=2.1 \times 10^{46} \erg = 2.7 E_{\rm jets}$, $E^{\rm unbound}_{\rm out}(\#13)=2.6 \times 10^{46} \erg = 3.4 E_{\rm jets}$.

{{{{ We can compare the energy that the jets carry with the relevant binding energy of the envelope. The initial binding energy of the envelope layer from $r=20$~R$_\odot$, i.e., the typical final orbital separation of our simulations, to the giant surface is $2.2\times10^{46} \erg $. This is larger than the energy that the jets carry, and might further point to the role of jets not only in carry energy, but in redistribution of the orbital energy.   }}}}

Our conclusion from this section and from Fig. \ref{fig:comparing_mass_plot} is that despite the large numerical uncertainties, the jets are quite efficient in removing mass from the common envelope. 

\section{NUMERICAL CONSIDERATIONS}
\label{sec:numerical_limitations}

In this section we examine our numerical method limitations. We start discussing how the numerical parameter $L_{\rm jets}$, the length of the cone within which we numerically inject the jets, affects our results. We then check energy conservation, and end by discussing numerical resolution effects.

\subsection{The effect of the injection length, $L_{\rm jets}$, on the simulation results}
\label{ssec:lej}

For each of the two simulations \#6 and \#9 we perform a second simulation, \#7 and \#10 respectively, that differs only in the value of $L_{\rm jets}$. All four simulations have $\dot{M}_{\rm jets} = 0.001 \msyr$.  We also performed 3 simulations with a higher  $\dot{M}_{\rm jets} = 0.003 \msyr$ and $L_{\rm jets}$ = 3.6 (\#11), 7.2 (\#12) or 14.4~R$_\odot$ (\#13). 

Although $L_{\rm jets}$ is a numerical expedient to inject the jets into the grid, we can try and give it a physical meaning. We assume that an accretion disk around the companion launches the jets, so we assert that $L_{\rm jets}$ should be of the same order of magnitude as the radius of the accretion disk, $R_{\rm acc}$. Assuming accretion from a Roche lobe overflow, we use equation (4.20) from \cite{Franketal2002}, and obtain:
\begin{equation}
L_{\rm jets} \simeq R_{\rm acc} = a\left(1+q\right)\left(0.5-0.227\log_{10}q\right)^4,
\label{eq:jets_length}
\end{equation} 
where $a$ is the orbital separation and $q=M_{\rm donor}/M_{\rm accretor}$ is the mass ratio between the donor star and the accretor star. For our simulations $q=0.88/0.3$ and
\begin{equation}
L_{\rm jets} \simeq 7.86 \left(\frac{a}{83 \rmRodot} \right) \rmRodot. 
\label{eq:jets_length_values}
\end{equation} 
The chosen values of $L_{\rm jets}=3.6, 7.2$ and $14.4 \rmRodot$ bracket this value. These values are also several times the radius of the putative main sequence star companion.

\begin{figure*}
\centering
\includegraphics[trim= 0.0cm 0.6cm 0.0cm 0.4cm,clip=true,width=0.99\columnwidth]{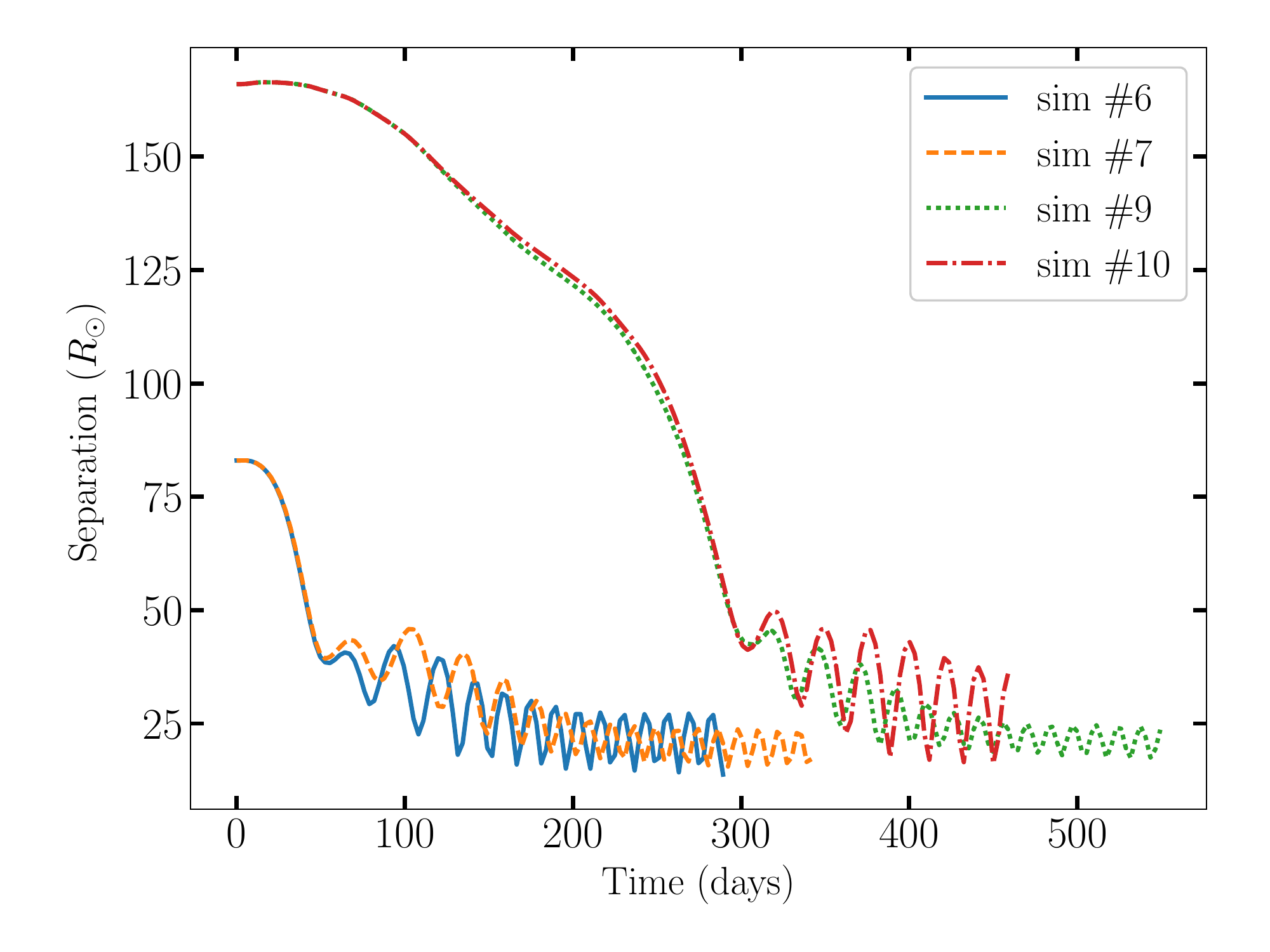}
\includegraphics[trim= 0.0cm 0.6cm 0.0cm 0.4cm,clip=true,width=0.99\columnwidth]{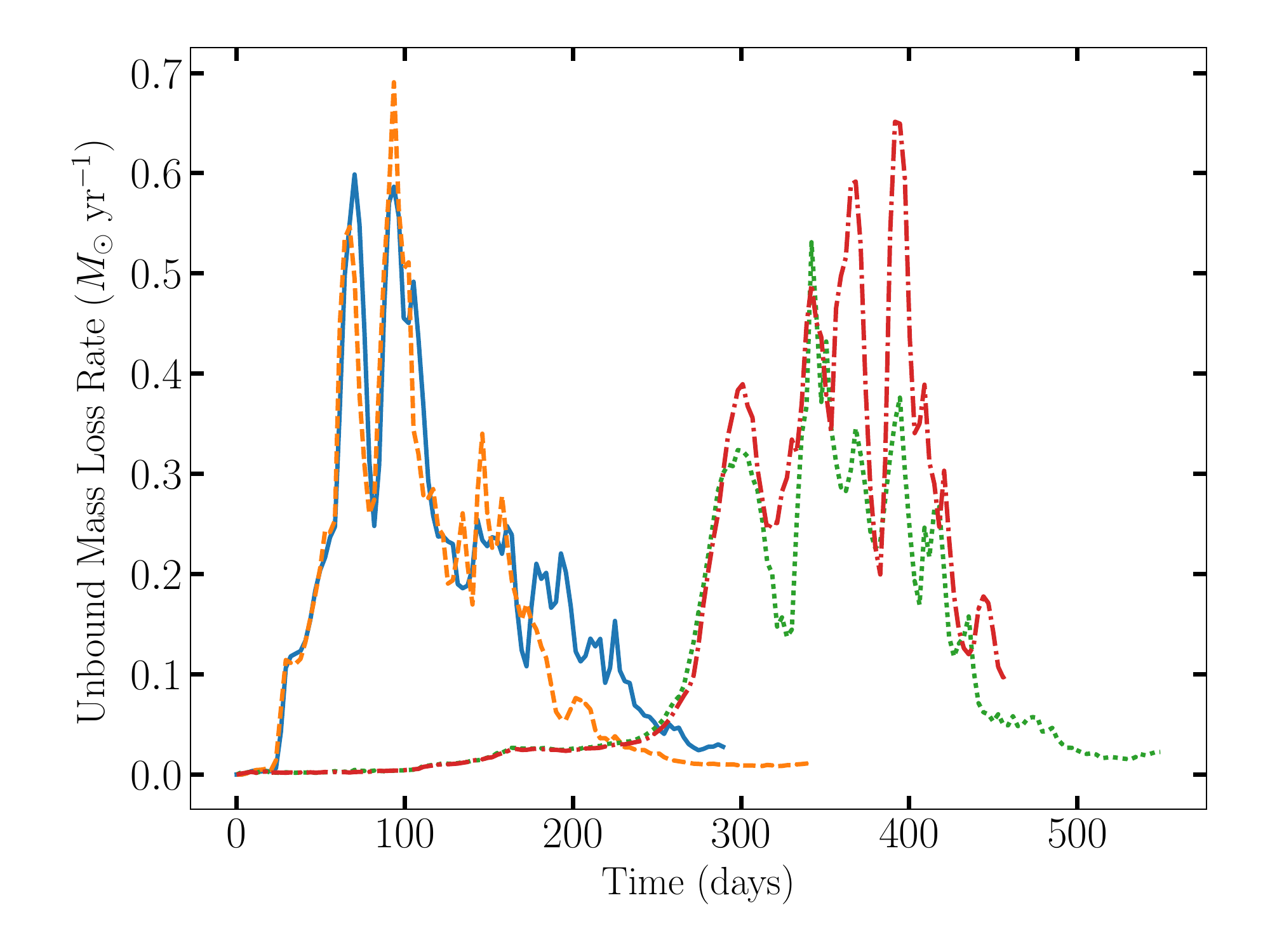}
\includegraphics[trim= 0.0cm 0.6cm 0.0cm 0.4cm,clip=true,width=0.99\columnwidth]{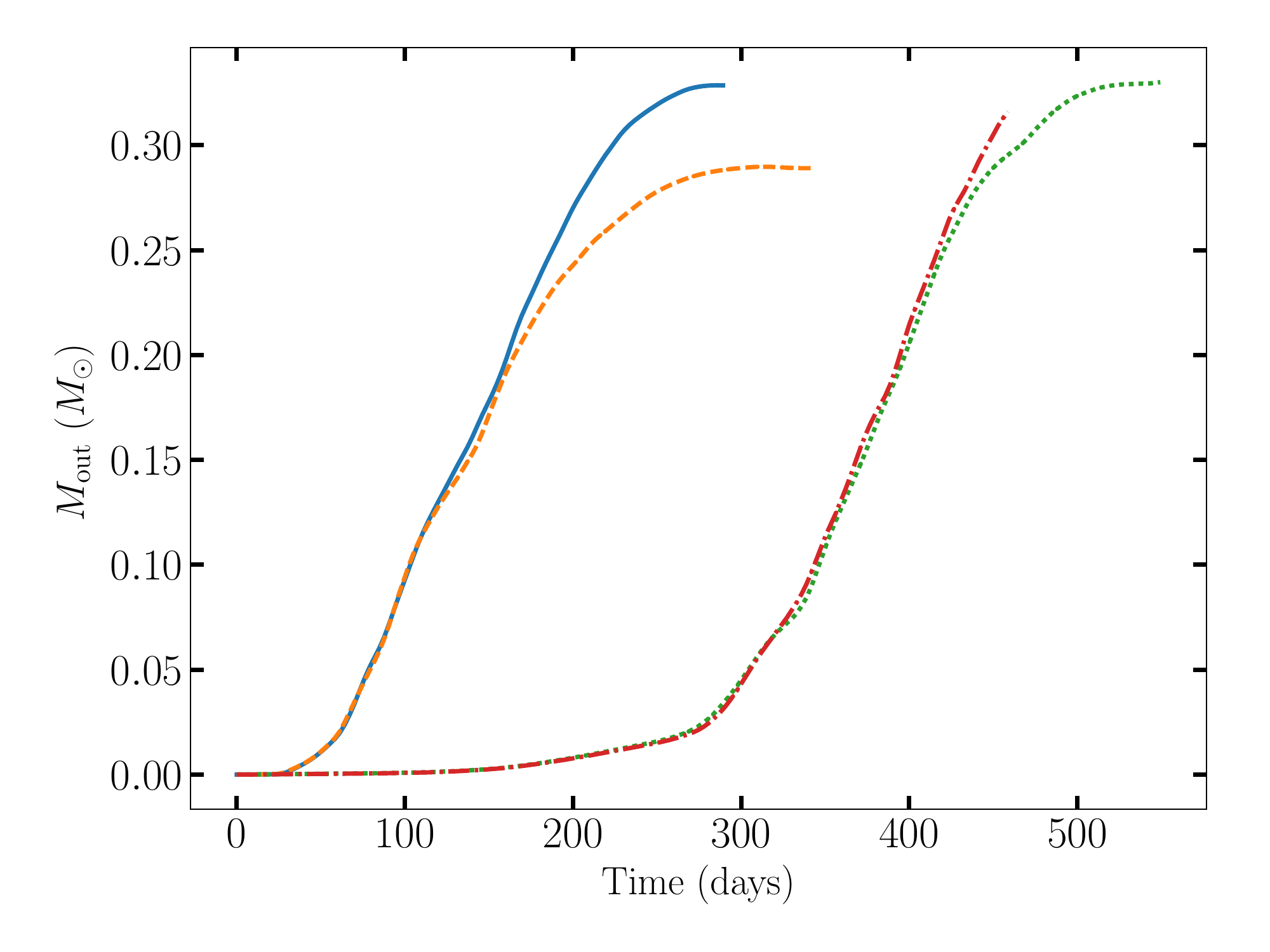}
\includegraphics[trim= 0.0cm 0.6cm 0.0cm 0.4cm,clip=true,width=0.99\columnwidth]{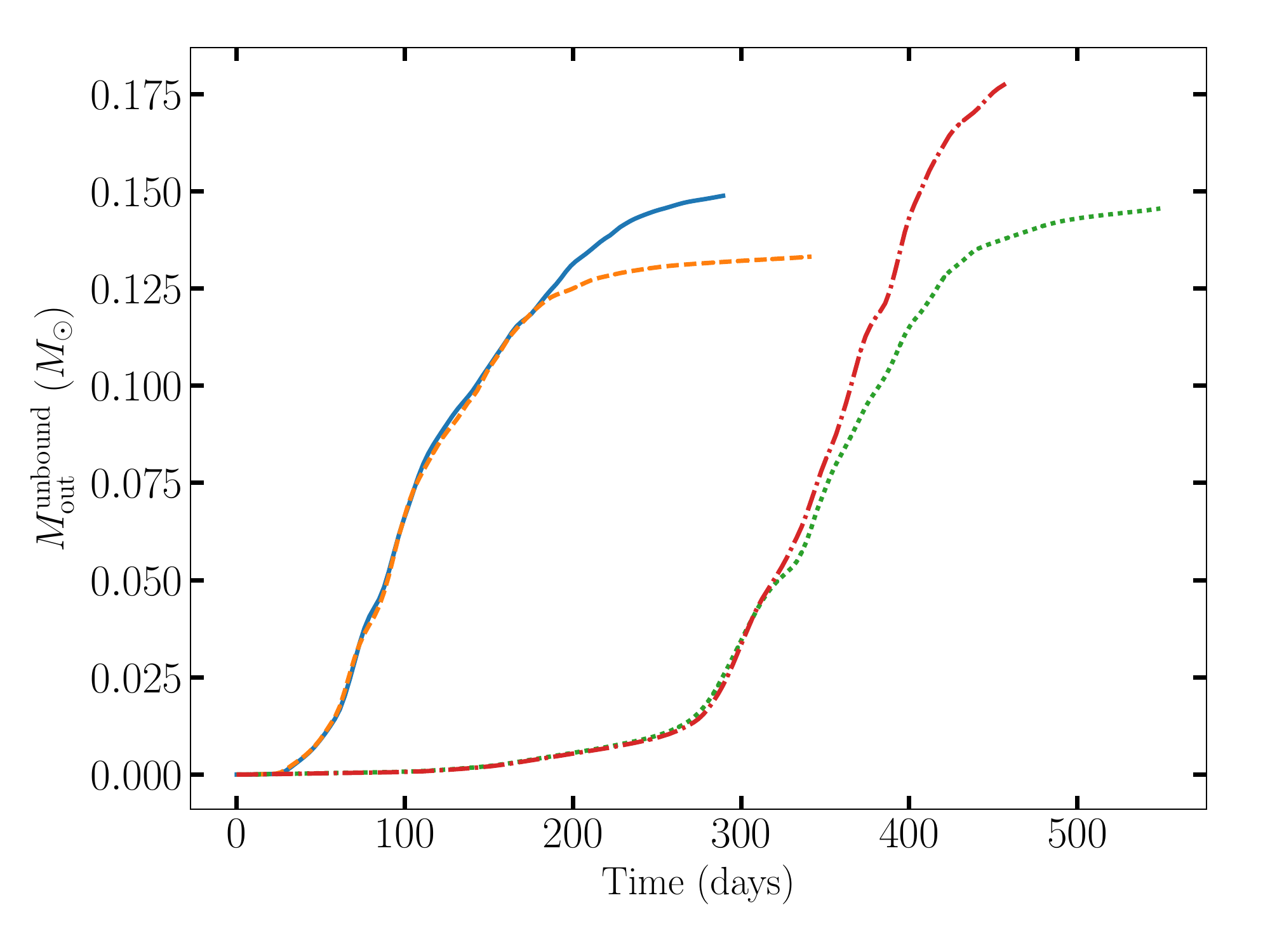}
\caption{ 
The separation (upper left panel), unbound mass loss rate, $\dot M^{\rm unbound}_{\rm out}$, (upper right panel), total mass loss (bottom left panel), and the unbound total mass loss (bottom right panel) of simulations \#6 (solid blue line), \#7 (dashed yellow line), \#9 (green dotted line), and \#10 (red dashed-dotted line).
}
\label{fig:numerical_limitation_J1}
\end{figure*}

In Fig. \ref{fig:numerical_limitation_J1} we compare simulations \#6 (solid blue line; smaller $L_{\rm jets}$) with \#7 (yellow dashed line; larger $L_{\rm jets}$), both starting on the surface of the giant and \#9 (green dotted line; smaller $L_{\rm jets}$) with \#10 (red dashed-dotted line; larger $L_{\rm jets}$), both starting at twice the giant's radius. The separation (upper left panel), unbound mass loss rate ($\dot M^{\rm unbound}_{\rm out}$; upper right panel), total mass loss ($M_{\rm out}$; bottom left panel) and the total unbound mass loss ($M_{\rm out}^{\rm unbound}$; bottom right panel) are as described in subsection \ref{subsec:compare_simulations}. 

From Fig. \ref{fig:numerical_limitation_J1} we identify the following numerical properties. 
Significant differences appear only when the companion is deep inside the envelope. We can partially understand this as at smaller separations the length $L_{\rm jets}= 7.2 R_\odot$ (and more so $L_{\rm jets}= 14.4 R_\odot$) is only slightly smaller than the final orbital separation of $a \la 25 R_\odot$. However, the differences in mass loss properties between the simulation with jet \#9 and that without, simulation \#8, appear before day 300 (Fig. \ref{fig:comparing_mass_plot}), and well before the effects caused by the injection length $L_{\rm jets}$ come into play (approximately at 400 in the comparable simulations \#9 and \#10). From this we conclude that the effect of jets in removing mass is not due to the injection length expedient. 
We also see that the real differences between simulations with jets and without jets (comparing simulations \#6 with \#5 and simulations \#9 with \#8) appear before the numerical effects of $L_{\rm jets}$ (comparing simulations \#6 with \#7 and simulations \#9 with \#10) by comparing Fig. \ref{fig:comparing_massloss_unbind} and the upper right panel of Fig. \ref{fig:numerical_limitation_J1}, respectively.

The upper left panel of Fig. \ref{fig:numerical_limitation_J1} shows that simulations with larger $L_{\rm jets}$ have similar final orbital separation, but the eccentricity is larger. This is particularly so for simulations starting with the companion at twice the giant's radius. The comparison between simulations \#11-13 shows the same trend.

In discussing the dependence of the mass ejection with the value of $L_{\rm jet}$, we need to consider separately the simulations starting with the companion on the surface, i.e, simulations \#6 and \#7 from simulations \#9 and \#10 that start with the companion at $a_0=2R_1$. 
The jets with larger $L_{\rm jets}$ in simulation \#7 eject and unbind less mass. On the other hand, when the companion starts well outside the envelope, $a_0=2R_1$, larger $L_{\rm jets}$ leads to {\it higher} mass loss rates (i.e., more mass is lost in simulation \#10 than in \#9). Similarly, simulations with higher mass injection rates into the jets, all starting at twice the primary's radius (simulations \#11, \#12 and \#13) show the same trend: simulation \#13 ($L_{\rm jets}=14.4$)  unbinds  48\% of the initial envelope, while simulations \#12 ($L_{\rm jets}=7.2$) unbinds 33\% of the envelope, and \#11 ($L_{\rm jets}=3.6$) unbinds 29\% of the envelope. The different trends between the influence of $L_{\rm jets}$ in simulations starting with $a_0=R_1$ (\#6 and \#7) and those starting with $a_0=2R_1$ come from the different envelope structure when the companion finally enters the envelope. This should be better explored in future studies.  

\begin{figure}
\includegraphics[trim= 0.0cm 0.6cm 0.0cm 0.4cm,clip=true,width=0.99\columnwidth]{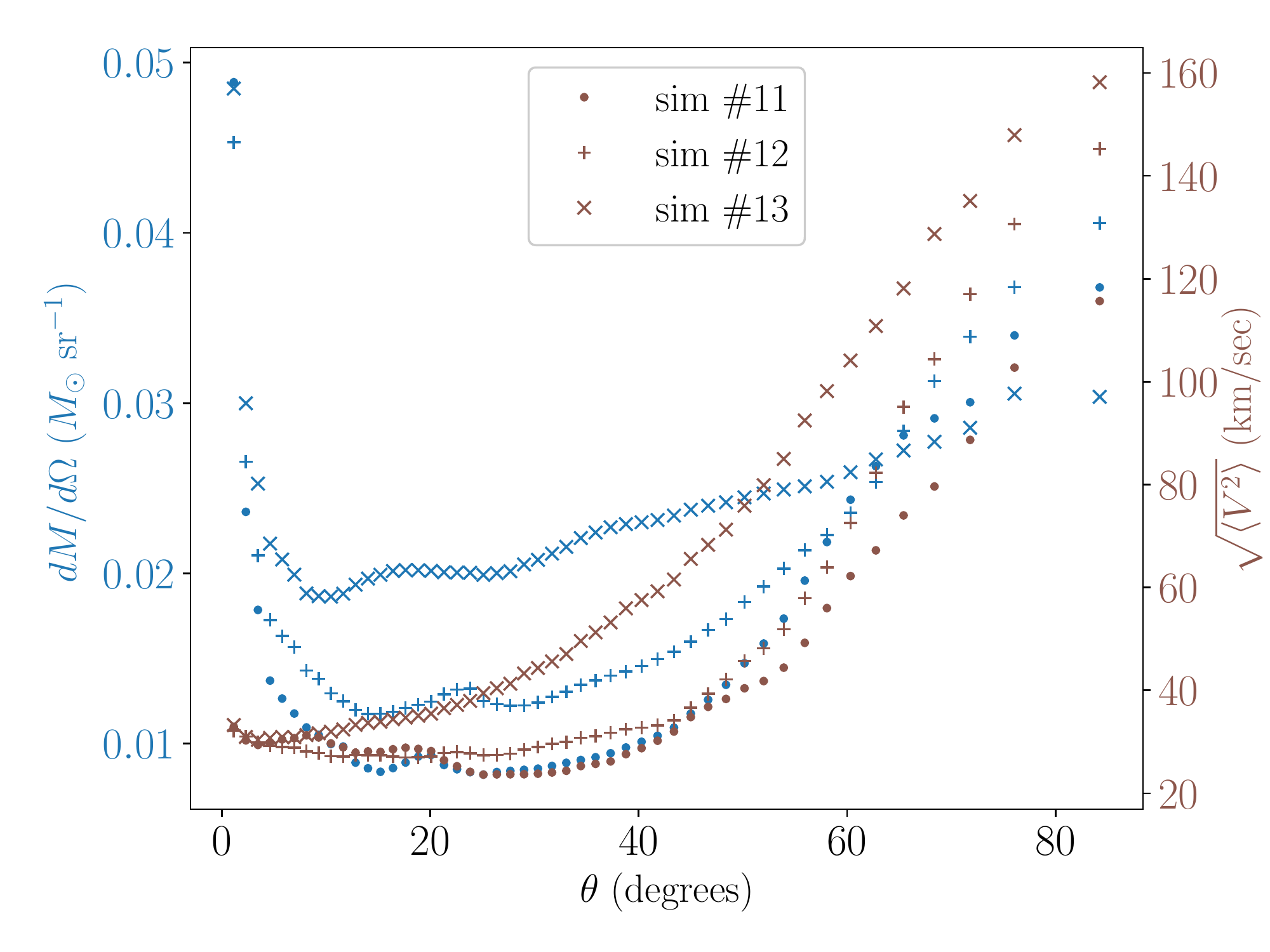}
\caption{ 
Like Fig. \ref{fig:comparing_mass_vrms_angle} but for simulations \#11 (circles), \#12 (pluses), \#13 (xs) having $L_{\rm jets}=3.6,\;7.2,\;14.4\rmRodot{}$, respectively.
}
\label{fig:mass_angle_limitation_J3}
\end{figure}

The variation of the mass loss rate per unit solid angle with latitude is more complicated as we see in Fig. \ref{fig:mass_angle_limitation_J3}. Low $L_{\rm jets}$ simulation have a much stronger dependence of this value with latitude, while high $L_{\rm jets}$ simulations show a much flatter behaviour.
As for the velocity, a higher value of $L_{\rm jets}$ leads to a higher outflow velocity almost in all directions.  

A second numerical aspect of the injection length, $L_{\rm jets}$ is that by injecting the jet over a certain number of cells near the companion, we are overriding the density in those cells, thus altering their mass. The density we impose on these cells is
\begin{equation}
\begin{aligned}
    \rho_{\rm jets} =3\times10^{-7}  
    \left( \frac{\dot{M}_{\rm jets}}{0.001 \msyr} \right)
    \left(\frac{\Delta x}{0.8 \rmRodot}\right)^{-2} \\ \times \left(\frac{v_{\rm jets}}{400\; \kms}\right)^{-1}
    \left(\frac{1-\cos{\theta_{\rm jets}}}{1-\cos{30^{\circ}}}\right)^{-1} \g \cm^{-3},
\end{aligned}
\label{eq:max_density}
\end{equation}
where $\Delta x$ is the distance from the cell center to the companion. These density values can be substantially lower than the density they substitute, particularly when the companion is deep inside the envelope. In the dense regions the result is that we subtract mass from the grid. 
We can consider this mass deficit as an artificial accretion that the numerical injection of the jets produces.
  
\begin{figure}
\includegraphics[trim= 0.0cm 0.6cm 0.0cm 0.4cm,clip=true,width=0.99\columnwidth]{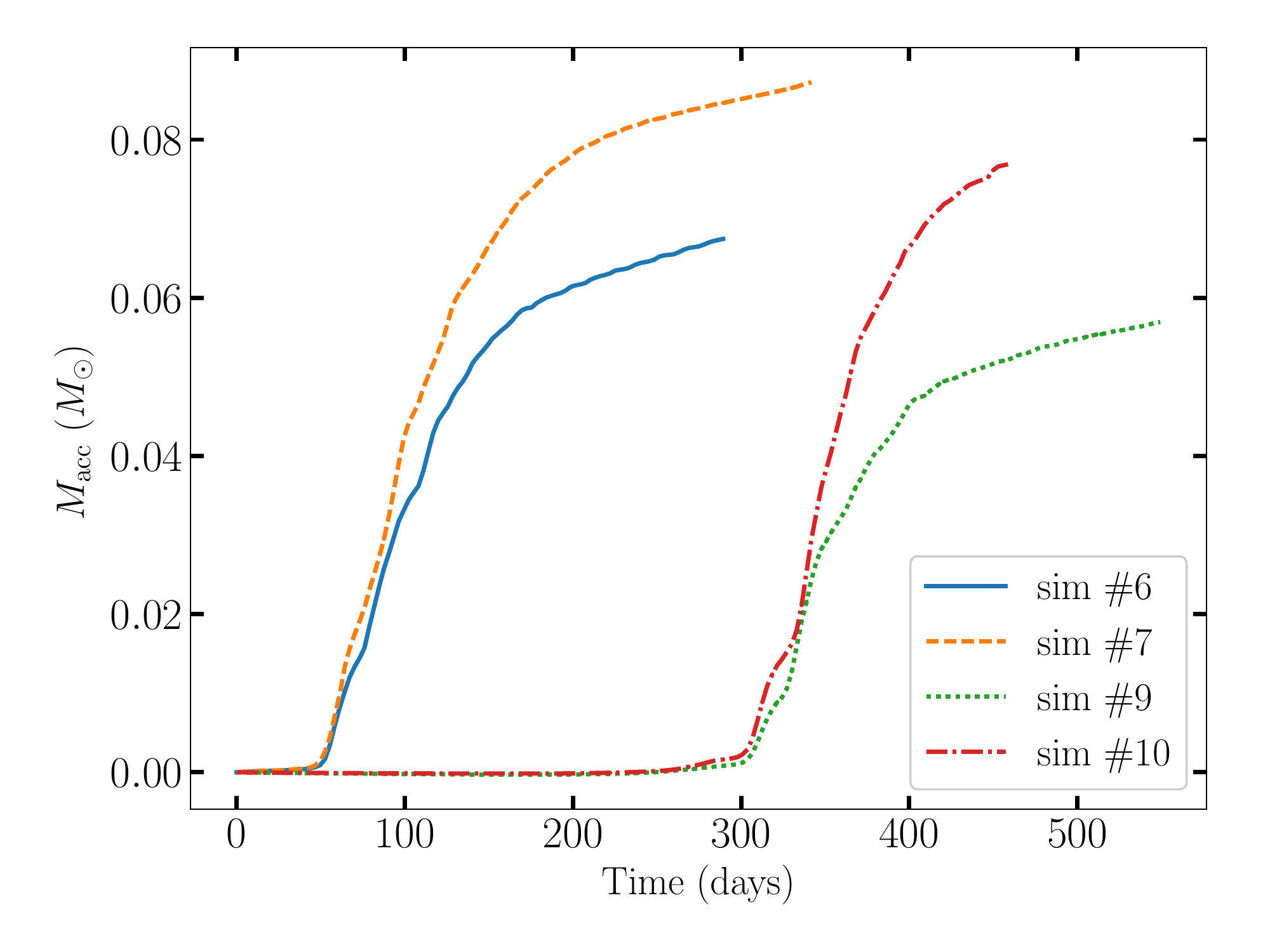}
\caption{ The `accreted mass' as function of time. Here the accreted mass is not really accreted by the companion, but rather the mass we remove from the vicinity of the companion when we inject the jets.  }
\label{fig:diff_mass_limitation_J1}
\end{figure}

In the knowledge that our code has excellent mass conservation, we can  calculate this `accreted mass'.
We track the mass subtracted in this way from each simulation, and we plot it as a function of time, $M_{\rm acc}(t)$, in Fig. \ref{fig:diff_mass_limitation_J1}. We see, as expected, that higher values of $L_{\rm jets}$ that imply large cone volumes remove more mass from the grid. We find the same trend in simulations \#11 to \#13. 
  
The values of the numerical accretion mass that we find in the different simulations imply that the total mass loss rate in the jets amounts to $\simeq 1-14 \%$ of the accreted mass, i.e., $M_{\rm jets} \simeq (0.01-0.14) M_{\rm acc}$. This is compatible with known values of jets launching, e.g., in young stellar objects \citep{Pudritzetal2012}. 
    
In conclusion, the behaviour of the orbital decay does not  depend on the value of $L_{\rm jets}$, while the amount of ejected mass changes by a factor of up to 2 when $L_{\rm jets}$ is altered. The removal of mass by the jets is real though, as the differences between different values of $L_{\rm jets}$ appear only after the effects of the jets, compared with simulations without jets, appear. Future numerical studies will have to better explore the numerical scheme to launch jets. 
 
The numerical effects of $L_{\rm jets}$ become larger as the orbital separation decreases. One of the challenges of future CEE simulations with jets is to better handle the spiraling-in and the launching of jets at small orbital separations, when the spiraling-in process has settled to a very slow rate.
  
\subsection{The energetics of jets launching}
\label{subsec:JetsEnergetic}
\subsubsection{Numerically removed energy}
\label{subsubsec:Eremoval}
When we inject the jets into the grid, we remove from the simulation the gas that was residing there before. By that we remove also its energy, mainly gravitational energy and thermal energy, as its kinetic energy is small. We insert the jets with the same pressure as the environment, which implies the same total thermal energy as that of the gas we remove. The jets have much lower density than that of the gas we remove, so that their gravitational energy is much smaller. 
Overall, the net energetic effect of jets injection is that we remove the gravitational energy of the mass we remove, and we inject the kinetic energy of the jets.

Let us demonstrate this for simulation \#9. 
In that run we remove a total mass (during the entire simulation) of $M_{\rm acc}=0.057M_\odot$ (Fig. \ref{fig:diff_mass_limitation_J1}). We calculate the gravitational energy of this mass as it resides in the two jet injection cones and in the gravitational potential of the secondary star, 
$E_{\rm acc,g,2} = - (3/2) G M_2 M_{\rm acc}/ L_{\rm jets} = - 1.35 \times 10^{46} \erg$. We can only estimate the gravitational energy of the removed mass in the gravitational potential of the envelope and the core because of the motion of all components relative to each other during the simulation. We approximate $E_{\rm acc,g,1} \simeq -G M_1 M_{\rm acc} /(0.5 R_1) \simeq -0.46  \times 10^{46} \erg$. Overall we remove a negative energy of $E_{\rm acc,g} \simeq -1.8 \times 10^{46} \erg$. Indeed, when we count all the energy in the grid and the energy that is carried by the mass outflowing from the grid, we find that at the end of simulation \#9 we have an excess of about $1.8 \times 10^{46} \erg$. This shows that in that simulation we conserve energy to within 10\%, a level similar to what \cite{Sandquistetal1998} or \cite{Passyetal2012} have in their simulations. We next try to answer the question of what effect this energy injection may have on the simulation.

\subsubsection{The accretion energy}
\label{subsubsec:Accretion}

Is it possible to justify the numerically injected energy on a physical basis? When gas accretes onto the companion, accretion energy is liberated. In reality, accretion energy accounts for the jets' kinetic energy, while the rest is liberated locally as radiation. In our simulation we therefore need to determined whether the artificially-injected energy plus the jets' kinetic energy are approximately similar to the amount of energy that would be expected by accreting mass onto the companion.

The jets in simulation \#9 carry a total kinetic energy of $E_{\rm jets} = 2.4 \times 10^{45} \erg$. Adding the amount of energy that is numerically removed  ($E_{\rm acc,g} \simeq -1.8 \times 10^{46} \erg$), we see that the accretion process should supply an energy of $\vert E_{\rm acc,g} \vert + E_{\rm jets} \simeq 2 \times 10^{46} \erg$. 
We have already discussed that by injecting the jets we are artificially deleting mass from the grid ($M_{\rm acc} = 0.057$~M$_\odot$). The accretion of this mass would release an energy of $\simeq 0.5 G M_2 M_{\rm acc} /R_2 \simeq 6 \times 10^{46} \erg$, for $R_2=0.5 R_\odot$. This is three times the energy that we need to account for, showing that our artificial injection is not unreasonable enough to make us question the result. If we consider the likely possibility that the companion swells as it accretes mass, say to a radius of $\simeq 1 R_\odot$, then the accretion process would justify the amount of energy we are artificially injecting. {{{{ For comparison, the binding energy of the envelope above $r=20 R_\odot$ is $2.2\times10^{46} \erg $. (section \ref{subsec:Energy}). }}}} 
 
\subsubsection{Very slow jets}
\label{subsubsec:SlowJets}
As described above, there are two kinds of energy we inject into the grid: the kinetic energy of the jets and the removal of gas, (which is assumed to be accreted) that has a negative energy. In simulations \#9 the total energies are $E_{\rm jets} \simeq 0.24 \times 10^{46} \erg$
and $ E_{\rm acc,g}  \simeq - 1.8 \times 10^{46} \erg$, respectively. 
Despite the fact that the kinetic energy of the jets is much smaller (in magnitude), we find that it is the kinetic energy of the jets that plays the major role in envelope ejection. 

To show that, we conducted a small experiment. We computed a simulation identical to \#9, but where we launched the jets with an initial velocity of $10 \km \s^{-1}$ instead of $400 \km \s^{-1}$. In this slow-jets simulation the kinetic energy of the jets is negligible, but not their thermal energy as it is equals to that of the mass we remove. In addition, the jets launching in the slow-jets simulation removes $27\%$ more mass ($M_{\rm acc}$) than in simulation \#9. With that, the slow-jets simulation removes more negative energy. Despite this, the total unbound mass that leaves the grid in the slow-jets simulation is {\it half} the total unbound mass that leaves the grid in simulation \#9, indicating that it is the kinetic energy to play a large role in mass unbinding, not the artificially-injected (accretion) energy (i.e., negative energy removal). 
   
To quantify the effect we compare the unbound mass in the slow-jet simulation with the no-jet simulation (\#8; see Fig. \ref{fig:comparing_mass_plot} for mass loss in simulations \#8 and \#9): $M_{\rm out}^{\rm unbound}({\rm slow-jets})-M_{\rm out}^{\rm unbound}(\#8)=0.022M_\odot$. The same comparison for simulations \#8 (no jets) and \#9 (regular jets) is 
$M_{\rm out}^{\rm unbound}(\#9)-M_{\rm out}^{\rm unbound}(\#8)=0.089M_\odot$, more than 4 times larger.
Since the slow-jet case mimics liberation of accretion energy as local heat, rather than kinetic energy, we concluude that the kinetic energy of the jets, namely the propagation of jets through the envlope, plays a decisive role in removing mass from the envelope and unbind it. 

\subsection{Resolution study}
\label{ssec:resolution}
  
\begin{figure*}
\centering
\includegraphics[trim= 0.0cm 0.6cm 0.0cm 0.4cm,clip=true,width=0.99\columnwidth]{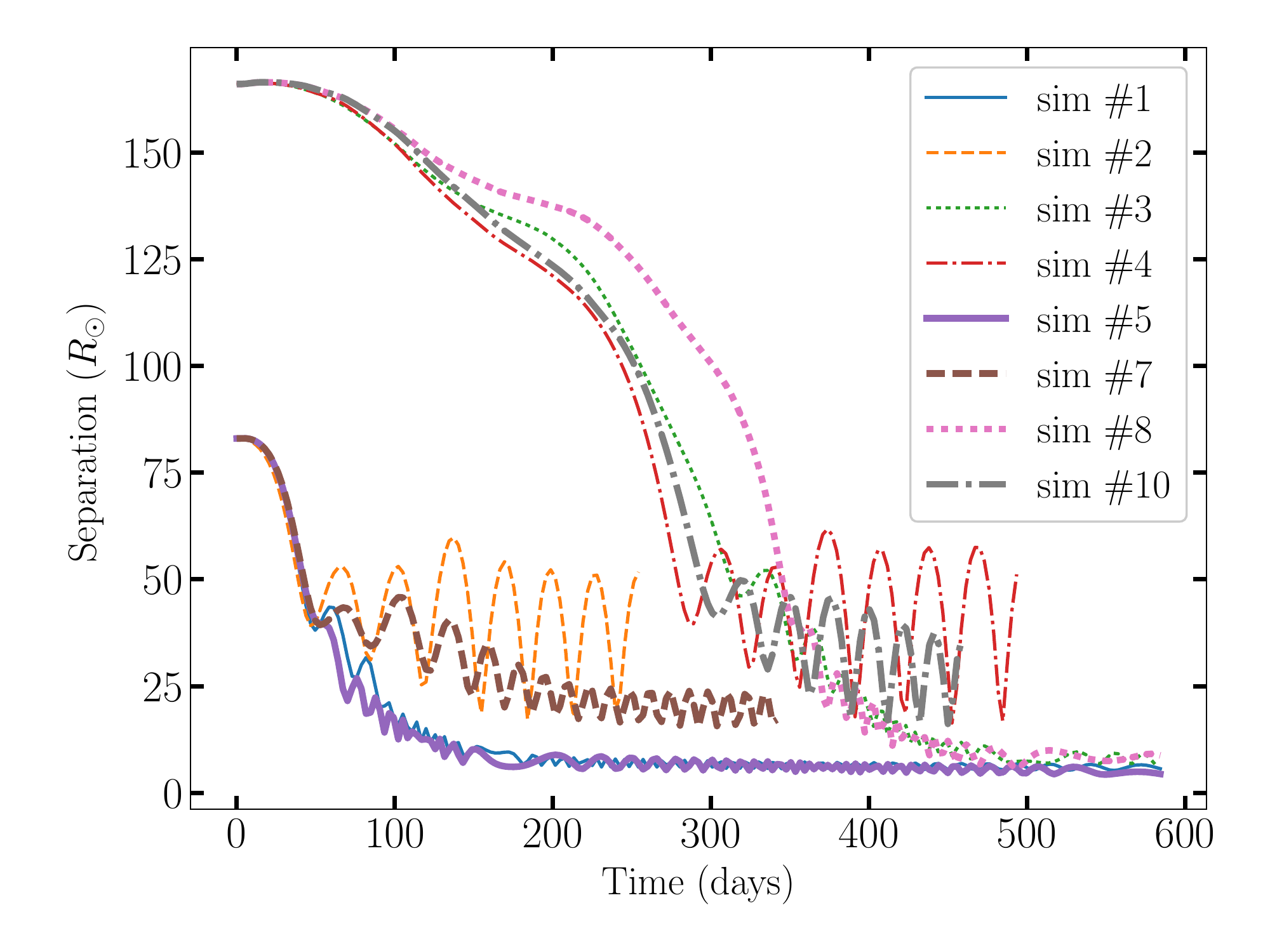}
\includegraphics[trim= 0.0cm 0.6cm 0.0cm 0.4cm,clip=true,width=0.99\columnwidth]{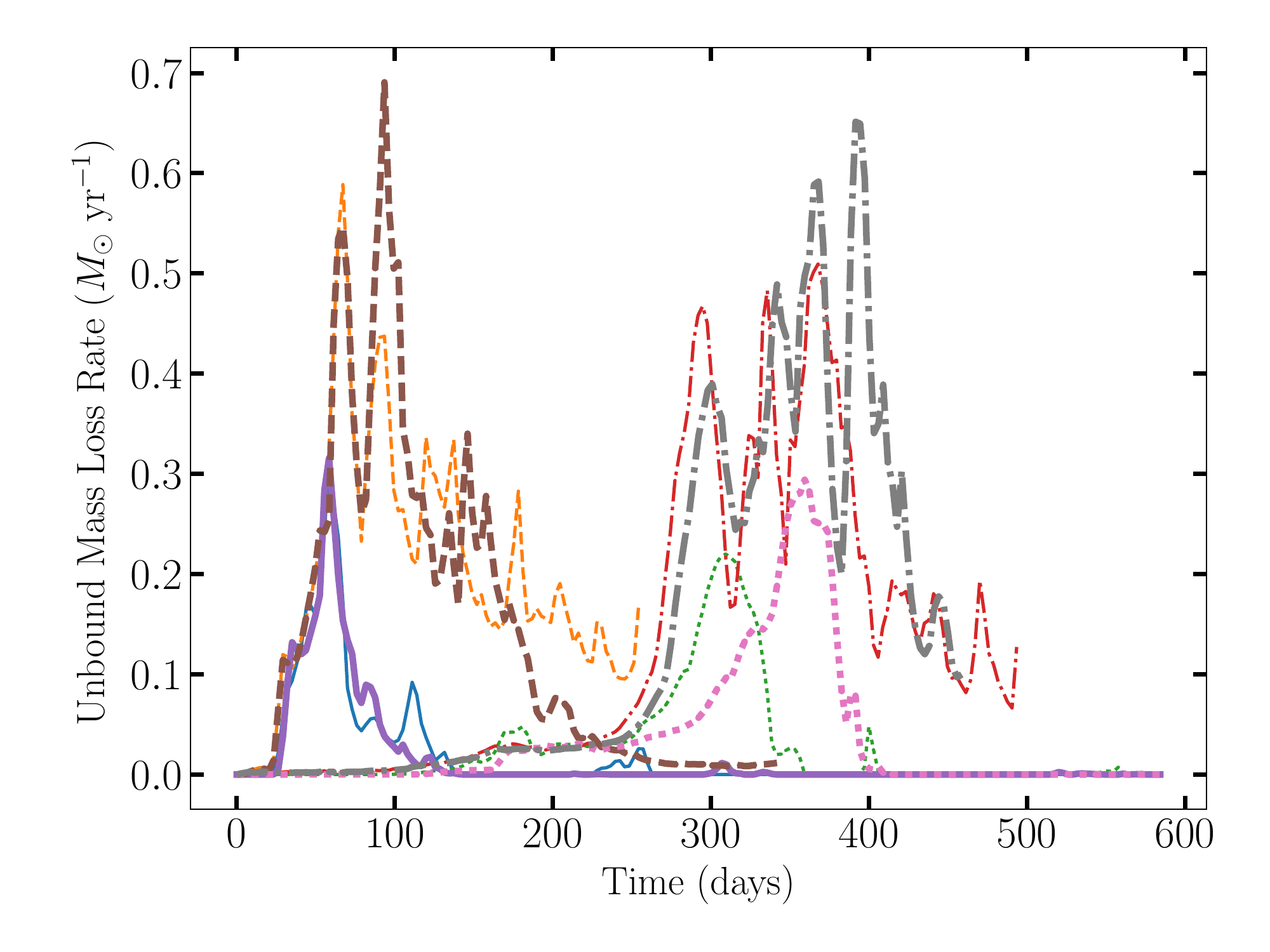}
\includegraphics[trim= 0.0cm 0.6cm 0.0cm 0.4cm,clip=true,width=0.99\columnwidth]{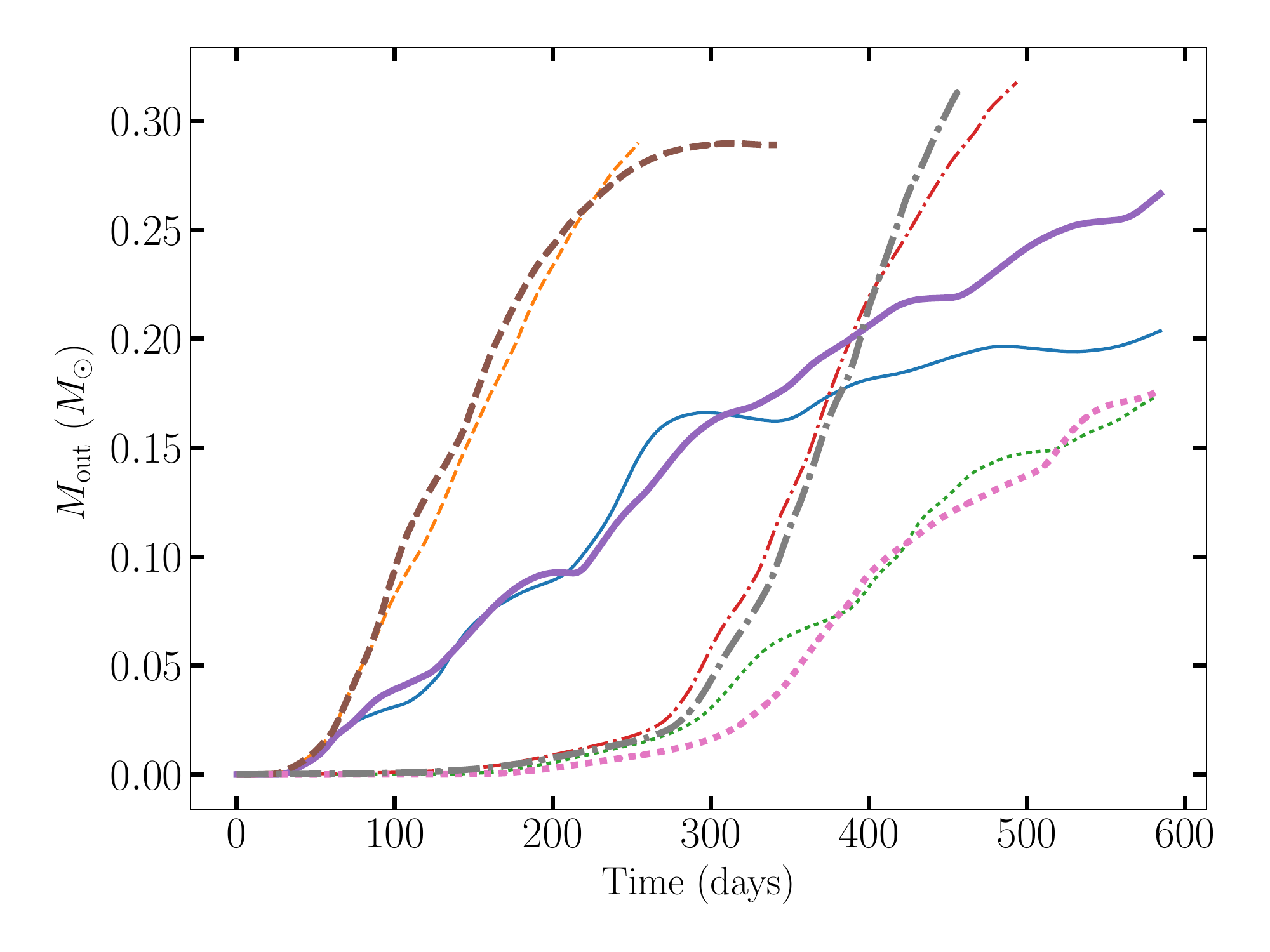}
\includegraphics[trim= 0.0cm 0.6cm 0.0cm 0.4cm,clip=true,width=0.99\columnwidth]{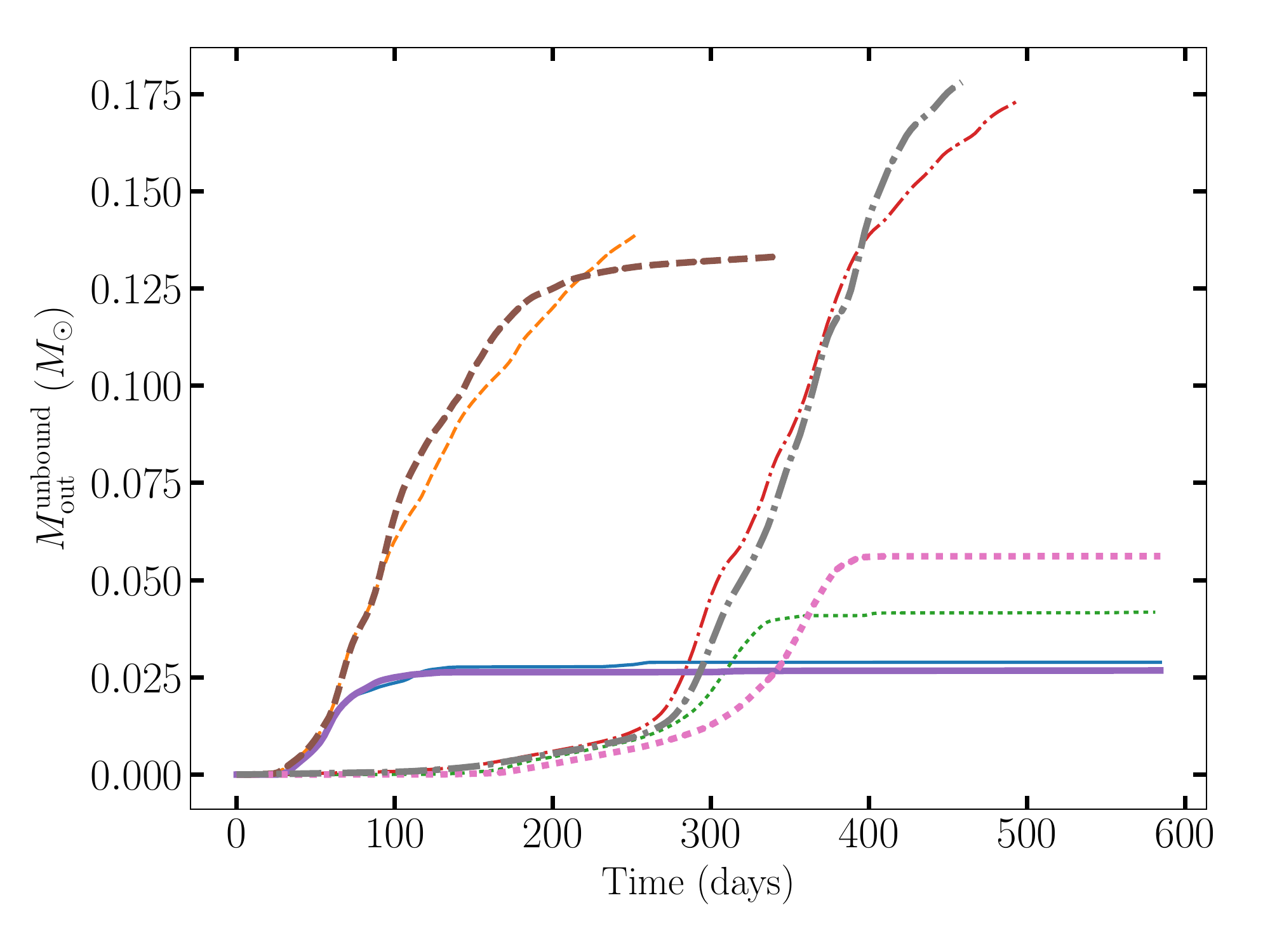}
\caption{ The separation (upper left panel), unbound mass loss rate, $\dot M^{\rm unbound}_{\rm out}$, (upper right panel), total mass loss (bottom left panel), and the unbound total mass loss (bottom right panel) of simulations \#1-4 (thin lines) and their higher resolution counterparts (thick lines). We plot simulations \#1 and \#5 in solid lines, \#2 and \#7 in dashed lines, \#3 and \#8 with dotted lines, and \#4 and \#10 with dashed-dotted lines.
}
\label{fig:numerical_resolution_study}
\end{figure*}

We perform four simulation, \#5, \#7, \#8, \#10, identical to simulations \#1-4, respectively, but with double the linear resolution of the top grid (128 instead of 64 cells per side).
In Fig. \ref{fig:numerical_resolution_study} we compare the same properties that we presented in Fig. \ref{fig:numerical_limitation_J1} for the high and low resolution pairs

The low resolution at the surface of the giant in the low resolution simulations maintains less accurately the hydro-static equilibrium. As a result, the envelope tends to inflate more and hence produces a bigger fan when the companion starts at $a_0=2R_1$. Therefore the companion spirals-in earlier in simulations \#3 and \#4 compared to simulations \#8 and \#10 respectively. The earlier spiraling-in 
results in unbinding eventually less material (see right panels). Another effect of the resolution is that the final orbits in the lower resolution  simulations with jets are more eccentric and have larger separations.
Despite these differences, the results of the two resolutions are sufficiently close to conclude that, at the order-of-magnitude level of precision afforded by this experiment, the resolution is not impacting the main conclusion, namely that jets may lead to substantially more unbound mass.

\section{Summary and discussion}
\label{sec:summary}

We present a first study of the CE interaction including the effect of jets, using hydrodynamic simulations that include self gravity. The strength of our approach is that the feedback of the jet activity on the in-spiral is included. A second advantage is the use of a stellar structure which has been extremely well quantified in the context of CE simulations with three distinct codes. This anchors the current, quite experimental setup in well quantified numerical simulations of the CE interaction.

Despite some numerical uncertainties (section \ref{subsec:Energy}  and \ref{sec:numerical_limitations}, and Fig. \ref{fig:comparing_eout_J3}), we confidently reach the following conclusions. 

\begin{enumerate}
\item The jets are quite efficient in removing mass from the envelope, removing approximately 30\% of the envelope, about 3 times more than the equivalent simulation with no jets (Figs.
\ref{fig:comparing_massloss_unbind}, \ref{fig:comparing_mass_plot} and \ref{fig:comparing_eout}).
\item The jets lead to higher mass loss rates and much higher velocities in the polar directions (Figs. \ref{fig:fiducial_mass_outflow_map}, \ref{fig:fiducial_velocity_outflow_map},  \ref{fig:comparing_dense89}, and \ref{fig:comparing_mass_vrms_angle}).
\item In simulations with jets the core-companion binary systems reach the very slow spiraling-in phase with much wider orbits (20-30~R$_{\odot}$) and much higher eccentricity. (e.g., Fig.  \ref{fig:comparing_separation_vs_time}; we discuss this further below).
\item 
By vigorously removing mass the jets introduce into the outflowing gas some morphological features that lack axisymmetrical structures (Figs. \ref{fig:fiducial_dense} and \ref{fig:jets_bending}). The extent to which the morphology of a subsequent PN is affected would have to await a treatment similar to that carried out by \citet{Franketal2018}. It is interesting that post-CE PN jets can be kinematically clocked to have been launched just before or just after the CE main ejection \citep{Tocknelletal2014}. Given that there are some uncertainties on these measurements it is possible that some of these jets may be due to an ejection during some part of the CE in-spiral \citep{BlackmanLucchini2014}.
%

\end{enumerate}

The simulations with jets remove mass more efficiently than simulations without jets. This in turn leads to larger orbital separations (20-30~R$_\odot$), too large to fit the majority of post-CE binaries ($\la 10R_\odot$; e.g., \citealt{Zorotovicetal2011}). Only a handful of post-RGB and post-AGB post-CE binaries exists at separations larger than $\sim 10$~$R_\odot$. Those, are more consistent with a prolonged phase of Roche lobe overflow before the CE due to a large companion-to-primary mass ratio, or can be explained by a particularly large primary mass \citep{IaconiDeMarco2019}. 

We think that the solution to this tension could be  that the core-companion binary system experiences further inspiralling due to gravitational interaction with the flat envelope around the binary system (or circumbinary disk) as suggested by \cite{KashiSoker2011MN};(see also \citealt{ChenPodsiadlowski2017}).  However, the interaction with a circumbinary disk increases the eccentricity \citep{KashiSoker2011MN}. As many post-CE binary have circular orbits, we think that there must be as well tidal interaction with the leftover envelope near and around the core. Such a tidal interaction tends to circularize the orbit. If this tidal interaction lasts for a longer time than the interaction with the circumbinary disk, then the orbit becomes circular. 

\section*{Acknowledgments}
{{{{ We thank an anonymous referee for useful comments. }}} }
SaS was supported in part by the Pazi Foundation. 
NS acknowledges support from the Israel Science Foundation.
RI is grateful for the financial support provided by the Postodoctoral Research Fellowship of the Japan Society for the Promotion of Science and by the International Macquarie University Research Excellence Scholarship. OD acknowledges support from the Australian Research Council Future Fellowship scheme (FT120100452).
Computations described in this work were performed using the ENZO code (http://enzo-project.org), which is the product of a collaborative effort of scientists at many universities and national laboratories. This work also required the use and integration of a Python package for astronomy, yt (http://yt-project.org, \citealt{Turk2011}). This project was undertaken with the assistance of resources and services from the National Computational Infrastructure (NCI), which is supported by the Australian Government.


\label{lastpage}

\begin{thebibliography}{0}

\bibitem[Abbott et al.(2016)]{Abbottetal2016} Abbott, B.~P., Abbott, R., Abbott, T.~D., et al.\ 2016, \prl, 116, 61102

\bibitem[Ablimit et al.(2016)]{Ablimitetal2016} Ablimit, I., Maeda, K., \& Li, X.-D.\ 2016, \apj, 826, 53

\bibitem[Ablimit, \& Maeda(2018)]{AblimitMaeda2018} Ablimit, I., \& Maeda, K.\ 2018, \apj, 866, 151

\bibitem[Akashi \& Soker(2008)]{AkashiSoker2008} Akashi, M., \& Soker, N.\ 2008, \na, 13, 157

\bibitem[Armitage \& Livio(2000)]{ArmitageLivio2000} Armitage, P.~J., \& Livio, M.\ 2000, \apj, 532, 540

\bibitem[Balick, \& Frank(2002)]{BalickFrank2002} Balick, B., \& Frank, A.\ 2002, Annual Review of Astronomy and Astrophysics, 40, 439.

\bibitem[Blackman \& Lucchini(2014)]{BlackmanLucchini2014} Blackman, E.~G., \& Lucchini, S.\ 2014, \mnras, 440, L16

\bibitem[Bondi(1952)]{Bondi1952} Bondi, H.\ 1952, \mnras, 112, 195.

\bibitem[Bryan et al.(2014)]{Bryanetal2014} Bryan, G.~L., Norman, M.~L., O'Shea, B.~W., et al.\ 2014, The Astrophysical Journal Supplement Series, 211, 19.

\bibitem[Chamandy et al.(2018)]{Chamandyetal2018a} Chamandy, L., Frank, A., Blackman, E.~G., et al.\ 2018, \mnras, 480, 1898

\bibitem[Chamandy et al.(2019)]{Chamandyetal2019} Chamandy L., Tu Y., Blackman E.~G., Carroll-Nellenback J., Frank A., Liu B., Nordhaus J., 2019, MNRAS, 486, 1070

\bibitem[Chen \& Podsiadlowski(2017)]{ChenPodsiadlowski2017} Chen, W.-C., \& Podsiadlowski, P.\ 2017, \apjl, 837, L19

\bibitem[Chevalier(2012)]{Chevalier2012} Chevalier, R.~A.\ 2012, \apj, 752, L2

\bibitem[Eggleton(1983)]{Eggleton1983} Eggleton, P.~P.\ 1983, \apj, 268, 368

\bibitem[Frank et al.(2002)]{Franketal2002} Frank, J., King, A., \& Raine, D.~J.\ 2002, Accretion Power in Astrophysics.

\bibitem[Frank et al.(2018)]{Franketal2018} Frank, A., Chen, Z., Reichardt, T., et al.\ 2018, ArXiv e-prints , arXiv:1807.05925.

\bibitem[Garc{\'{\i}}a-Segura et al.(2018)]{Garciaetal2018} Garc{\'{\i}}a-Segura, G., Ricker, P.~M., \& Taam, R.~E.\ 2018, \apj, 860, 19 

\bibitem[Glanz \& Perets(2018)]{GlanzPerets2018} Glanz, H., \& Perets, H.~B.\ 2018, \mnras, 478, L12 

\bibitem[Iaconi \& De Marco(2019)]{IaconiDeMarco2019} Iaconi, R., \& De Marco, O.\ 2019, arXiv:1902.02039

\bibitem[Iaconi et al.(2017)]{Iaconietal2017} Iaconi, R., Reichardt, T., Staff, J., De Marco, O., Passy, J.-C., Price, D., Wurster, J., \& Herwig, F.\ 2017, \mnras, 464, 4028

\bibitem[Iaconi et al.(2018)]{Iaconietal2018} Iaconi, R., De Marco, O., Passy, J.-C., \& Staff, J.\ 2018, mnras, 477, 2349 

\bibitem[Ivanova(2018)]{Ivanova2018} Ivanova, N.\ 2018, \apjl, 858, L24

\bibitem[Ivanova et al.(2013b)]{Ivanovaetal2013} Ivanova, N., Justham, S., Chen, X., et al.\ 2013b, \aapr, 21, 59

\bibitem[Ivanova \& Nandez(2016)]{IvanovaNandez2016} Ivanova, N., \& Nandez, J.~L.~A.\ 2016, \mnras, 462, 362 

\bibitem[Jones et al.(2015)]{Jonesetal2015} Jones, D., Boffin, H.~M.~J., Rodr{\'\i}guez-Gil, P., et al.\ 2015, \aap, 580, A19.

\bibitem[Kashi \& Soker(2011)]{KashiSoker2011MN} Kashi, A., \& Soker, N.\ 2011, \mnras, 417, 1466

\bibitem[Kashi \& Soker(2018)]{KashiSoker2018} Kashi, A., \& Soker, N.\ 2018, \mnras, 480, 3195 

\bibitem[Kuruwita et al.(2016)]{Kuruwitaetal2016}  Kuruwita, R.~L., Staff, J., \& De Marco, O.\ 2016, \mnras, 461, 486

\bibitem[Livio \& Soker(1988)]{LivioSoker1988} Livio, M., \& Soker, N.\ 1988, \apj, 329, 764

\bibitem[L{\'o}pez-C{\'a}mara et al.(2019)]{Lopezetal2019} L{\'o}pez-C{\'a}mara, D., De Colle, F., \& Moreno M{\'e}ndez, E.\ 2019, \mnras, 482, 3646 

\bibitem[MacLeod et al.(2018a)]{MacLeodetal2018a} MacLeod, M., Ostriker, E.~C., \& Stone, J.~M.\ 2018a, \apj, 863, 5 

\bibitem[MacLeod et al.(2018b)]{MacLeodetal2018b} MacLeod, M., Ostriker, E.~C., \& Stone, J.~M.\ 2018b, \apj, 868, 136 

\bibitem[Miszalski et al.(2013)]{Miszalskietal2013} Miszalski, B., Boffin, H.~M.~J., \& Corradi, R.~L.~M.\ 2013, \mnras, 428, L39.

\bibitem[Montez et al.(2010)]{Montezetal2010} Montez, R., De Marco, O., Kastner, J.~H., et al.\ 2010, \apj, 721, 1820.

\bibitem[Montez et al.(2015)]{Montezetal2015} Montez, R., Kastner, J.~H., Balick, B., et al.\ 2015, \apj, 800, 8.

\bibitem[Moreno M{\'e}ndez et al.(2017)]{MorenoMendezetal2017} Moreno M{\'e}ndez, E., L{\'o}pez-C{\'a}mara, D., \& De Colle, F.\ 2017, \mnras, 470, 2929 

\bibitem[Murguia-Berthier et al.(2017)]{Murguiaetal2017} Murguia-Berthier, A., MacLeod, M., Ramirez-Ruiz, E., Antoni, A., \& Macias, P.\ 2017, \apj, 845, 173 

\bibitem[Nandez \& Ivanova(2016)]{NandezIvanova2016} Nandez, J.~L.~A., \& Ivanova, N.\ 2016, \mnras, 460, 3992 

\bibitem[Nandez et al.(2014)]{Nandezetal2014} Nandez, J.~L.~A., Ivanova, N., \& Lombardi, J.~C., Jr.\ 2014, \apj, 786, 39

\bibitem[Ohlmann et al.(2016)]{Ohlmannetal2016} Ohlmann, S.~T., R{\"o}pke, F.~K., Pakmor, R., \& Springel, V.\ 2016, \apjl, 816, L9

\bibitem[O'Shea et al.(2004)]{O'Sheaetal2004} O'Shea, B.~W., Bryan, G., Bordner, J., et al.\ 2004, ArXiv e-prints , astro–ph/0403044.

\bibitem[Papish et al.(2015)]{Papishetal2015} Papish, O., Soker, N., \& Bukay, I.\ 2015, \mnras, 449, 288

\bibitem[Passy \& Bryan(2014)]{PassyBryan2014} Passy, J.-C., \& Bryan, G.~L.\ 2014, \apjs, 215, 8

\bibitem[Passy et al.(2012)]{Passyetal2012} Passy, J.-C., De Marco, O., Fryer, C.~L., et al.\ 2012, \apj, 744, 52

\bibitem[Paczynski(1976)]{Paczynski1976} Paczynski, B.\ 1976, Structure and Evolution of Close Binary Systems, 75.

\bibitem[Pudritz et al.(2012)]{Pudritzetal2012} Pudritz, R.~E., Hardcastle, M.~J., \& Gabuzda, D.~C.\ 2012, \ssr, 169, 27.

\bibitem[Rasio, \& Livio(1996)]{RasioLivio1996} Rasio, F.~A., \& Livio, M.\ 1996, \apj, 471, 366.

\bibitem[Reichardt et al.(2018)]{Reichardtetal2018} Reichardt, T.~A., De Marco, O., Iaconi, R., Tout, C.~A., \& Price, D.~J.\ 2018, arXiv:1809.02297

\bibitem[Ricker \& Taam(2012)]{RickerTaam2012} Ricker, P.~M., \& Taam, R.~E.\ 2012, \apj, 746, 74

\bibitem[Ruffert (1993)]{Ruffert1993} Ruffert M., 1993, \aa, 280, 141

\bibitem[Sandquist et al.(1998)]{Sandquistetal1998} Sandquist, E.~L., Taam, R.~E., Chen, X., Bodenheimer, P., \& Burkert, A.\ 1998, \apj, 500, 909

\bibitem[Sandquist et al.(2000)]{Sandquistetal2000} Sandquist, E.~L., Taam, R.~E., \& Burkert, A.\ 2000, \apj, 533, 984.

\bibitem[Shiber(2018)]{Shiber2018} Shiber, S.\ 2018, Galaxies, 6, 96

\bibitem[Shiber et al.(2017)]{Shiberetal2017} Shiber, S., Kashi, A., \& Soker, N.\ 2017, \mnras, 465, L54

\bibitem[Shiber et al.(2016)]{Shiberetal2016} Shiber, S., Schreier, R., \& Soker, N.\ 2016, Research in Astronomy and Astrophysics, 16, 117

\bibitem[Shiber, \& Soker(2018)]{ShiberSoker2018} Shiber, S., \& Soker, N.\ 2018, \mnras, 477, 2584

\bibitem[Soker(1994)]{Soker1994} Soker, N.\ 1994, \mnras, 270, 774.

\bibitem[Soker(2004)]{Soker2004} Soker, N.\ 2004, \na, 9, 399

\bibitem[Soker(2015)]{Soker2015} Soker, N.\ 2015, \apj, 800, 114

\bibitem[Soker(2016a)]{Soker2016Rev} Soker, N.\ 2016a, \nar, 75, 1

\bibitem[Soker(2017b)]{Soker2017final} Soker, N.\ 2017b, \mnras, 471, 4839 

\bibitem[Soker et al.(2018)]{Sokeretal2018} Soker N., Grichener A., \& Sabach E.,\ 2018, \apj, 863, L14

\bibitem[Staff et al.(2016a)]{Staffetal2016MN} Staff, J.~E., De Marco, O., Macdonald, D., Galaviz, P., Passy, J.C., Iaconi, R., \& Mac Low, M.-M\ 2016a, \mnras, 455, 3511

\bibitem[Staff et al.(2016b)]{Staffetal2016MN8} Staff, J.~E., De Marco,
 O., Wood, P., Galaviz, P., \& Passy, J.-C.\ 2016b, \mnras, 458, 832

\bibitem[Taam \& Ricker(2010)]{TaamRicker2010} Taam, R.~E., \& Ricker, P.~M.\ 2010, \nar, 54, 65

\bibitem[Tocknell et al.(2014)]{Tocknelletal2014} Tocknell, J., De Marco, O., \& Wardle, M.\ 2014, \mnras, 439, 2014 

\bibitem[Toonen, \& Nelemans(2013)]{ToonenNelemans2013} Toonen, S., \& Nelemans, G.\ 2013, \aap, 557, A87.

\bibitem[Turk et al.(2011)]{Turk2011} Turk, M.~J., Smith, B.~D., Oishi, J.~S., et al.\ 2011, \apjs, 192, 9 

\bibitem[Webbink(1984)]{Webbink1984} Webbink, R.~F.\ 1984, \apj, 277, 355.

\bibitem[Wilson, \& Nordhaus(2019)]{WilsonNordhaus2019}  Wilson, E.~C., \& Nordhaus, J.\ 2019, \mnras, 485, 4492. 

\bibitem[Witt et al.(2009)]{Wittetal2009} Witt, A.~N., Vijh, U.~P., Hobbs, L.~M., et al.\ 2009, \apj, 693, 1946.

\bibitem[Zorotovic et al.(2011)]{Zorotovicetal2011} Zorotovic, M., Schreiber, M.~R., \& G{\"a}nsicke, B.~T.\ 2011, \aap, 536, A42

\end{thebibliography}
\end{document}